\documentclass[11pt,reqno]{amsart}

\usepackage[T1]{fontenc}
\usepackage[utf8]{inputenc}
\usepackage[english]{babel}

\usepackage[a4paper,margin=1.15in]{geometry}
\usepackage{microtype}
\usepackage{newpxtext}
\usepackage{mathtools}
\usepackage{amssymb}
\usepackage{newpxmath}
\usepackage{bm}
\usepackage{braket}
\usepackage{extpfeil}
\input{Ustmry.fd}
\DeclareFontShape{U}{stmry}{m}{n}{
  <5><6><7><8><9><10> gen * stmary
  <5.5> stmary5
  <10.95><12><14.4><17.28><20.74><24.88> stmary10
}{}
\allowdisplaybreaks
\numberwithin{equation}{section}

\usepackage{array}
\usepackage{tabularx}

\usepackage[shortlabels]{enumitem}
\usepackage[title]{appendix}
\usepackage[all,poly]{xy}
\usepackage{tikz}
\usetikzlibrary{arrows.meta,positioning}

\usepackage{xcolor}
\definecolor{linkblue}{RGB}{0,72,128}
\usepackage[
colorlinks=true,
linkcolor=linkblue,
citecolor=linkblue,
urlcolor=linkblue
]{hyperref}
\usepackage[nameinlink,noabbrev]{cleveref}

\theoremstyle{plain}
\newtheorem{thm}{Theorem}[section]

\newtheorem{lemma}[thm]{Lemma}
\newtheorem{corollary}[thm]{Corollary}
\newtheorem{corollary*}[thm]{Corollary*}
\newtheorem{proposition}[thm]{Proposition}
\newtheorem{proposition*}[thm]{Proposition*}

\theoremstyle{definition}

\newtheorem{definition}[thm]{Definition}
\newtheorem{example}[thm]{Example}

\newtheorem{defn-thm}[thm]{Definition-Theorem}

\theoremstyle{remark}
\newtheorem{remark}[thm]{Remark}

\newcommand{\Rmnum}[1]{\uppercase\expandafter{\romannumeral #1}}

\tikzset{
	solid black dot/.style={
		circle,
		fill=black,
		inner sep=1.5pt
	}
}

\def\d{\mathrm{d}}

\usepackage{color}

\definecolor{yellow}{rgb}{1,1,0}
\definecolor{orange}{rgb}{1,.7,0}
\definecolor{red}{rgb}{1,0,0}
\definecolor{white}{rgb}{1,1,1}
\definecolor{green}{rgb}{0,1,0}

\begin{document}
	
\raggedbottom

	\title[Topological recursion, Virasoro constraints, CohFTs]{Eynard-Orantin Topological Recursion, Moduli-Space Virasoro Constraints, and Semisimple CohFTs}
	\author{Dingkang Xiong}
	\address{Department of Mathematical Sciences, Tsinghua University, Beijing 100084, China}
	\email{xdk22@mails.tsinghua.edu.cn}
	\author{Jian Zhou}
	\address{Department of Mathematical Sciences, Tsinghua University, Beijing 100084, China}
	\email{jianzhou@mail.tsinghua.edu.cn}
	
	\begin{abstract}
		We use Virasoro constraints to prove Eynard's intersection-number formula for Eynard--Orantin topological recursion with several simple ramification points. The key construction couples the higher Weil--Petersson partition functions associated with the individual ramification points through a quadratic differential operator determined by the Bergman kernel.
		
		We identify this operator with the quadratic part of Givental’s quantized \(R\)-matrix, thereby obtaining a Virasoro-theoretic proof of the DOSS identification of topological recursion with Givental’s formula for semisimple cohomological field theories. 
		
		Furthermore, while the DOSS identification fixes the distinguished \(y\)-function, we keep the Bergman kernel/\(R\)-matrix fixed and allow \(y\) to vary, showing that this amounts to shifting the
		evaluation point of the corresponding ancestor potential. 
	\end{abstract}
	
	\maketitle
		\setcounter{tocdepth}{1}
		\tableofcontents

		\section{Introduction}
		
		\subsection{Background}
		The Eynard--Orantin topological recursion \cite{eynardInvariantsAlgebraicCurves2007} associates to a spectral curve
		\[
		(\Sigma,x,y,B)
		\]
		a sequence of symmetric multidifferentials \(W_{g,n}\) on \(\Sigma^n\), indexed by the stable range \(2g-2+n>0\). It has proved to be a universal mechanism in enumerative geometry.  The Airy curve recovers the Witten--Kontsevich intersection numbers of \(\psi\)-classes on \(\overline{\mathcal M}_{g,n}\) \cite{eynardInvariantsAlgebraicCurves2007,zhouTopologicalRecursionsEynard2013}; the Kontsevich curve gives higher Weil--Petersson volumes \cite{eynardIntersectionNumbersSpectral2011,zhouTopologicalRecursionsEynard2013}; the Lambert curve encodes simple Hurwitz numbers and yields the ELSV formula \cite{borotMatrixModelSimple2011a,eynardLaplaceTransformCutandJoin2011,ekedahlHurwitzNumbersIntersections2001,dunin-barkowskiPolynomialityHurwitzNumbers2015}; and the framed mirror curve of \(\mathbb C^3\) gives the Mari\~no--Vafa formula for cubic Hodge integrals \cite{marinoFramedKnotsLarge2002,liuProofConjectureMarinoVafa2003a,zhouLocalMirrorSymmetry2009,zhouLocalMirrorSymmetry2009a,eynardInvariantsSpectralCurves2014}.  
		Further applications include the remodeling conjecture of Bouchard–Klemm–Mariño–Pasquetti (BKMP) \cite{bouchardRemodelingBModel2009,bouchardTopologicalOpenStrings2010}, which relates topological recursion to mirror symmetry for toric Calabi-Yau threefolds and orbifolds \cite{chenBouchardKlemmMarinoPasquettiConjectureC32018,eynardComputationOpenGromov2015,liuAllgenusOpenclosedMirror2020,fangRemodelingConjectureToric2019}.
		These examples suggest that topological recursion is a local residue formalism which nevertheless packages global intersection-theoretic and Gromov--Witten type invariants.
		
		One explanation of this phenomenon is Eynard's intersection-number formula \cite{eynardIntersectionNumbersSpectral2011,eynardInvariantsSpectralCurves2014}.  For spectral curves with simple ramification points, Eynard expresses \(W_{g,n}\) in terms of intersection numbers on moduli spaces of curves decorated by the local Taylor coefficients of \(y\) and \(B\).  
		Thus the topological recursion admits a moduli-space interpretation in which the local geometry of the spectral curve is translated into tautological intersection theory.  
		
		A complementary structural explanation comes from Givental’s quantization formalism \cite{giventalGromovWittenInvariantsQuantization2001} and semisimple cohomological field theories in the sense of Kontsevich--Manin \cite{kontsevichGromovWittenClassesQuantum1994}.  Givental's formula reconstructs the total partition function of a semisimple CohFT by applying quantized symplectic transformations to a product of Witten--Kontsevich tau-functions \cite{giventalGromovWittenInvariantsQuantization2001}, while Teleman's theorem shows that this reconstruction is universal for semisimple CohFTs \cite{telemanStructure2DSemisimple2012}. 
		 Dunin-Barkowski, Orantin, Shadrin and Spitz (DOSS) identified the Givental formula with the spectral-curve topological recursion for local spectral curves with simple ramification points under some compatibility conditions \cite{dunin-barkowskiIdentificationGiventalFormula2014} (see \cite{milanovEynardOrantinRecursion2014} for a related formulation in the singularity-theory setting).
		As an application, DOSS proved the Norbury--Scott conjecture \cite{norburyGromovWittenInvariants2014} relating a spectral curve to the stationary Gromov--Witten theory of \(\mathbb P^1\). 
		
		Eynard’s intersection-number formula and the DOSS correspondence therefore give two complementary descriptions of topological recursion: one in terms of tautological intersection theory and the other in terms of Givental’s action on semisimple CohFTs.
		
		The purpose of this paper is to revisit these bridges from the viewpoint of Virasoro constraints, generalizing the second named author's work \cite{zhouTopologicalRecursionsEynard2013,zhouTopologicalRecursionVirasoro2024}.  Eynard's original proof of the intersection-number formula uses the Rauch variational formula, while DOSS proved the comparison with Givental's formula through graph-sum descriptions.  We give a different proof of Eynard's formula for spectral curves with several simple ramification points by comparing the residue recursion directly with Virasoro constraints on moduli spaces of curves.  We also compare the resulting partition function with Givental quantization, obtaining a Virasoro-theoretic proof of the DOSS identification and a generic-background extension.
		
		\subsection{Main results and strategy}
		Suppose that \(x\) has finitely many simple ramification points \(c_1,\ldots,c_N\).  Near \(c_i\), choose a local Airy coordinate \(z_i\) such that
		\[
		x(p)=x(c_i)+\frac{1}{2}z_i^2(p), \qquad p\in U_i.
		\]
		Write the local expansions
		\[
		y(p)=\sum_{m\ge0}y_m^i z_i(p)^m,\qquad y_1^i\ne0,
		\]
		and, for $p\in U_i$, $q\in U_j$,
		\[
		B(p,q)
		=
		\delta_{ij}\frac{\d z_i(p)\d z_j(q)}{(z_i(p)-z_j(q))^2}
		+\sum_{k,l\ge0}B^{i,j}_{k,l}z_i(p)^kz_j(q)^l\d z_i(p)\d z_j(q).
		\]
		The numbers $y_m^i$ and $B^{i,j}_{k,l}$ are the local coefficients of the spectral curve at the ramification points. Eynard's formula \cite{eynardIntersectionNumbersSpectral2011,eynardInvariantsSpectralCurves2014} says that $W_{g,n}$ can be expanded in a basis of differentials $\d\xi^i_m$ built from $B$, and that the coefficients are intersection numbers on moduli spaces of curves, decorated by these local data. In the one-ramification-point case, this formula can be related to the Virasoro constraints for higher Weil-Petersson volumes, following Liu and Xu \cite{liuRecursionFormulaeHigher2008}, and to the topological recursion type reformulation in \cite{zhouTopologicalRecursionsEynard2013}.  The second-named author used this approach in \cite{zhouTopologicalRecursionVirasoro2024} to give a proof of Eynard's formula for a spectral curve with one ramification point.
		
		The first aim of this paper is to extend this Virasoro-constraint approach to spectral curves with several simple ramification points.  We keep all computations local at the ramification points, so the global geometry of $\Sigma$ enters only through the coefficients just displayed.  The odd part of $y$ at $c_i$ is encoded by the Schur transform \cite{eynardIntersectionNumbersSpectral2011} $\mathbf s^i=(s^i_1,s^i_2,\ldots)$ defined by
		\[
		\exp\left(-\sum_{a\ge1}s^i_aw^a\right):=\sum_{m\ge0}(2m+1)!!\,\frac{y^i_{2m+1}}{y^i_1}w^m,
		\]
		while the even regular coefficients of $B$ enter through
		\[
		\widehat B^{i,j}_{k,l}:=(2k-1)!!(2l-1)!!B^{i,j}_{2k,2l}.
		\]
		By expanding \(1/(y(z_i)-y(-z_i))\) in the local Airy coordinate, we expand the recursion kernel explicitly in the local Airy coordinate \eqref{kernel expansion} and introduce the normalized basis $\d\xi^i_m$ in \eqref{normalized dxi}.   
		Following Eynard's expansion \cite{eynardIntersectionNumbersSpectral2011,eynardInvariantsSpectralCurves2014}, with respect to this basis we write
		\begin{equation}
			W_{g,n}(p_1,\ldots,p_n)
			=
			\sum_{(i_a,m_a)}
			\left\langle \tau_{i_1,m_1}\cdots \tau_{i_n,m_n}\right\rangle_g^{\mathrm{EO}}
			\prod_{a=1}^n \d\xi^{i_a}_{m_a}(p_a),
		\end{equation}
		and compute directly the recursion satisfied by the correlators. 
		Following \cite{zhouTopologicalRecursionVirasoro2024}, the residue calculation is reduced to two elementary operations.  These two operations give the splitting and merging constants $\alpha$ \eqref{alpha} and $\mu$ \eqref{mu}, expressed explicitly in terms of $y^i_1$, the Schur variables $\mathbf s^i$, and the normalized Bergman coefficients $\widehat B^{i,j}_{k,l}$. 
		At the level of correlators, Proposition \ref{brief expansion of TR} gives the following form of the Eynard--Orantin recursion:
			\begin{equation}
			\begin{aligned}
				&\left\langle
				\tau_{i_0,m_0}\tau_{i_1,m_1}\cdots\tau_{i_n,m_n}
				\right\rangle_g^{\mathrm{EO}} \\
				=&
				\sum_{\substack{1\le j\le n\\ i_j=i_0}}
				\sum_{i,m}
				\alpha^{i_0,m_0;i_j,m_j}_{i,m}
				\left\langle
				\tau_{i,m}
				\prod_{\substack{1\le k\le n\\ k\ne j}}
				\tau_{i_k,m_k}
				\right\rangle_g^{\mathrm{EO}}
				\\
				&+\frac{1}{2}
				\sum_{j_1,k_1,j_2,k_2}
				\mu^{i_0,m_0}_{j_1,k_1;j_2,k_2}
				\Bigg[
				\left\langle
				\tau_{j_1,k_1}\tau_{j_2,k_2}
				\prod_{j=1}^{n}\tau_{i_j,m_j}
				\right\rangle_{g-1}^{\mathrm{EO}}
				\\
				&\qquad\qquad\qquad\qquad
				+
				\sum_{\substack{g_1+g_2=g\\ I_1\sqcup I_2=[n]}}^{\mathrm{st}}
				\left\langle
				\tau_{j_1,k_1}
				\prod_{j\in I_1}\tau_{i_j,m_j}
				\right\rangle_{g_1}^{\mathrm{EO}}
				\left\langle
				\tau_{j_2,k_2}
				\prod_{j\in I_2}\tau_{i_j,m_j}
				\right\rangle_{g_2}^{\mathrm{EO}}
				\Bigg],
			\end{aligned}
		\end{equation}
		The complete result is the recursion relation \eqref{recursion relation for EO TR}, together with the initial values obtained from $W_{0,3}$ and $W_{1,1}$.  Thus the Eynard-Orantin residue recursion is translated into a closed recursion for the correlators in the $\d\xi$-basis.
		
		On the intersection-theoretic side, we use the Virasoro constraints for higher Weil--Petersson volumes due to Liu and Xu \cite{liuRecursionFormulaeHigher2008}, generalizing the result of Mulase and Safnuk \cite{mulaseMirzakhanisRecursionRelations2006}.
		Their partition function is
		\[
		Z(\mathbf t;\mathbf s)=\exp F(\mathbf t;\mathbf s),
		\]
		with
		\begin{equation}
			F(\mathbf t;\mathbf s)
			=
			\sum_{g\ge0}\sum_{\mathbf b,\mathbf n}
			\left\langle
			\prod_{a\ge1}\kappa_a^{b_a}
			\prod_{d\ge0}\tau_d^{n_d}
			\right\rangle_g
			\frac{\mathbf s^{\mathbf b}}{\mathbf b!}
			\frac{\mathbf t^{\mathbf n}}{\mathbf n!},
		\end{equation}
		where $	\left\langle
		\prod_{a\ge1}\kappa_a^{b_a}
		\prod_{d\ge0}\tau_d^{n_d}
		\right\rangle_g$ are the intersection numbers of $\kappa$-classes and $\psi$-classes on moduli spaces of curves.
		
		The Liu--Xu constraints \eqref{LiuXu operator} can be reorganized into standard Virasoro
		operators \eqref{shift LiuXu} and become the Witten--Kontsevich Virasoro constraints after
		the shift
		\[
		F(\mathbf t;\mathbf s)
		=
		F_{\mathrm{WK}}(t_0,t_1,t_2+p_2,t_3+p_3,\ldots),
		\qquad
		1-\sum_{m\ge2}p_mw^{m-1}
		=
		\exp\left(-\sum_{a\ge1}s_aw^a\right).
		\]
		
		For the ramification point \(c_i\), the Schur transform of the odd
		local expansion of \(y\) gives
		\[
		p^i_m
		=
		-(2m-1)!!\,\frac{y^i_{2m-1}}{y^i_1},
		\qquad m\ge2.
		\]
		After the rescaling by \(y^i_1\), set
		\[
		q^i_{\mathrm{WP},m}
		=
		\begin{cases}
			t^i_{\mathrm{WP},0}, & m=0,\\
			t^i_{\mathrm{WP},m}-(2m-1)!!\,y^i_{2m-1},
			& m\ge1.
		\end{cases}
		\]
		The corresponding dilated higher Weil--Petersson partition function is
		\[
		Z_{\mathrm{WP}}
		(\mathbf t^i_{\mathrm{WP}};\mathbf s^i;y^i_1)
		:=
		(y^i_1)^{1/24}Z_{\mathrm{WK}}(\mathbf q^i_{\mathrm{WP}}).
		\]
		
		We take one copy of this partition function for each ramification
		point and couple them through the quadratic differential operator
		determined by the normalized Bergman coefficients:
		\begin{equation}\label{eqn:Z-loc}
			Z_{\mathrm{loc}}
			(\mathbf t_{\mathrm{TR}};\mathbf s;\widehat B;\mathbf y)
			:=
			\exp\left(
			\frac12
			\sum_{i,j=1}^N\sum_{k,l\ge0}
			\widehat B^{i,j}_{k,l}
			\frac{\partial^2}
			{\partial t^i_{\mathrm{TR},k}
				\partial t^j_{\mathrm{TR},l}}
			\right)
			\prod_{i=1}^N
			Z_{\mathrm{WP}}
			(\mathbf t^i_{\mathrm{TR}};\mathbf s^i;y^i_1).
		\end{equation}
		Conjugating the local Virasoro constraints by this gluing operator produces a system of Virasoro constraints for \(Z_{\mathrm{loc}}\).
		The induced recursion \eqref{recursion relation of TR intersection numbers}, together with its initial values, agrees with the Eynard--Orantin	recursion \eqref{recursion relation for EO TR}.
		
		Our first main theorem, Theorem \ref{main thm: relation}, is the equality of these two families of correlators.  Since the two recursions have the same initial data and uniquely determine all stable correlators, the EO correlators agree with those generated by \(Z_{\mathrm{loc}}\). More precisely,
		\begin{thm}[Theorem  \ref{main thm: relation}]
			Expand the local glued partition function $Z_{\mathrm{loc}}$ as
			\begin{equation}
				Z_{\mathrm{loc}}=\exp \big(\sum_{2g-2+n>0} \sum_{(i_a,m_a)} \left\langle \tau_{i_1,m_1}\cdots\tau_{i_n,m_n}\right\rangle_g^{\mathrm{loc}}
				\frac{1}{n!}\prod_{a=1}^n t^{i_a}_{\mathrm{TR},m_a}  \big),
			\end{equation}
			then the two families of correlators agree:
			\begin{equation}
				\left\langle \tau_{i_1,m_1}\cdots\tau_{i_n,m_n}\right\rangle_g^{\mathrm{EO}}=\left\langle \tau_{i_1,m_1}\cdots\tau_{i_n,m_n}\right\rangle_g^{\mathrm{loc}},
			\end{equation}
			i.e.,
			\begin{equation}
				W_{g,n}(p_1,\ldots,p_n)
				=
				\sum_{(i_a,m_a)}
				\left\langle \tau_{i_1,m_1}\cdots\tau_{i_n,m_n}\right\rangle_g^{\mathrm{loc}}
				\prod_{a=1}^n \d\xi^{i_a}_{m_a}(p_a).
			\end{equation}
		\end{thm}
		Equivalently, the local glued partition function is the generating function obtained from the Eynard-Orantin multidifferentials after replacing $\d\xi^i_m$ by the corresponding variables $t^i_{\mathrm{TR},m}$, i.e.,
		\[
			Z_{\mathrm{loc}}(\mathbf t_{\mathrm{TR}};\mathbf s;\widehat B;\mathbf y)=\exp \big( \sum_{2g-2+n>0}\frac{1}{n!} W_{g,n}\big|_{\d \xi^i_m\to t^{i}_{\mathrm{TR},m}}  \big).
		\]
		
		From \eqref{eqn:Z-loc} one easily gets a graph sum expansion of $Z_{\mathrm{loc}}$ (Theorem \ref{thm of local glued partition function} ).  Expanding the quadratic gluing operator expresses $Z_{\mathrm{loc}}$ as a sum over stable decorated graphs: vertices carry the local intersection theories determined by $(\mathbf s^i,y_1^i)$, and edges carry the propagators $\widehat B^{i,j}_{k,l}$. 
		It is equivalent to Eynard's combinatorial-moduli-space description \cite{eynardInvariantsSpectralCurves2014} and also equivalent to the description in \cite{dunin-barkowskiIdentificationGiventalFormula2014}.
		
		This proves Eynard's multi-ramification-point intersection-number formula by a direct comparison with Virasoro constraints.  
		
		According to a private communication of S. Shadrin, our result has some connections to the result in \cite{borotHigherAiryStructures2024a}.
		
		The next main result of this paper relates the above local construction to Givental quantization and semisimple cohomological field theory. 
		We first recall the reconstruction formula of Givental \cite{giventalGromovWittenInvariantsQuantization2001} and Teleman \cite{telemanStructure2DSemisimple2012}. See Section \ref{section Topological Recursion and Givental Quantization} for notation. 
		 
		For the Frobenius manifold underlying a semisimple CohFT, let $u^i$ be canonical coordinates and put $U=\mathrm{diag}(u^1,\ldots,u^N)$.  Let $\Psi$ be the transition matrix from the normalized canonical basis	$f_i=\psi_{i1}^{-1}\partial_{u^i}$ to the flat basis, let $R(z)$ be the Givental $R$-matrix, and let $S(z^{-1})$ be an $S$-calibration.  Then Givental's formula, together with Teleman's reconstruction theorem, gives the total descendant partition function in the form
		\begin{equation}
		Z_{\mathrm{Des}}
		=
		e^{C(u)}\widehat S_u^{-1}\widehat\Psi\,\widehat R_u\,\widehat{e^{U/z}}
		Z_{\mathrm{WK}}^{\otimes N},
		\end{equation}
		with a function $C(u)$.
		Here $\hat{\cdot}$ is the Givental quantization and all the operators act on the dilaton variables $\mathbf{q}$.
		Equivalently, for the total ancestor partition function in flat coordinates one has
		\begin{equation}
		Z_{\mathrm{FA}}
		=
		\left(\prod_{i=1}^N\psi_{i1}\right)^{1/24}
		\widehat\Psi\,\widehat R\,Z_{\mathrm{WK}}^{\otimes N}.
		\end{equation}
		
		We show that, after identifying the series determined by \(B\) with the Givental \(R\)-matrix at a semisimple point and choosing \(y\) to satisfy the corresponding DOSS compatibility condition, the topological-recursion partition function obtained above realizes Givental's ancestor formula.
		
		When the Bergman kernel is normalized by $\mathcal A$-cycles, or when its local coefficients satisfy \eqref{extra condition on B}, the coefficients $\widehat B^{i,j}_{0,k}$ define a matrix-valued series \cite{dunin-barkowskiIdentificationGiventalFormula2014}:
		\begin{equation}
			R(z)=I+R_1z+R_2z^2+\cdots,\qquad
			(R_m)^i{}_j=(-1)^{m-1}\widehat B^{i,j}_{0,m-1} \ \text{for } m\geq 1.
		\end{equation}
		The same assumption implies the symplectic identity $R(-z)^*R(z)=I$ and the quadratic product formula	\eqref{double product formula}. 
		
		To compare this construction with a semisimple CohFT, we fix a semisimple point $v$ of a Frobenius manifold and assume that, in the normalized canonical basis, this series agrees with the Givental $R$-matrix at $v$.
		
		For a symplectic series \(R(z)\), Givental's quantization acts on a
		function \(G\) by \cite{giventalGromovWittenInvariantsQuantization2001}
		\begin{equation}
			(\widehat R \cdot G)(\mathbf q)=[e^{V(\partial_{\mathbf q},\partial_{\mathbf q})/2}G](R^{-1}\mathbf q),
		\end{equation}
		where the quadratic form $V(\partial_{\mathbf{q}},\partial_{\mathbf{q}})$ is determined by the same formula as \eqref{double product formula}.
		In particular, the coefficients of $\hat{B}$ that appear as propagators in the graph expansion are precisely the coefficients of the quadratic form associated with the symplectic transformation $R(z)$.  Thus the gluing operator in $Z_{\mathrm{loc}}$ is the quadratic part of the quantized $R$-matrix in Givental's formalism.
		
		The same assumption also gives the following change of basis of differentials:
		\[
		\d\xi^i_m
		=
		\sum_{l=0}^m\sum_{a=1}^N
		\left(-\widehat B^{a,i}_{0,m-l-1}\right)(-1)^l\d\xi^a_{(l)},
		\]
		where
		\[
		\widehat B^{a,i}_{0,-1}:=-\delta_a^i,\quad \d \xi^a_{(l)}:=\d (\frac{d^l}{d x^l}\xi^a_0).
		\]
		Equivalently, under the identification
		\[
		\d\xi^i_m\longleftrightarrow t^i_{\mathrm{TR},m},
		\qquad
		(-1)^l\d\xi^i_{(l)}
		\longleftrightarrow t^i_{\mathrm{CA},l},
		\]
		the change of basis corresponds to the coordinate transformation
		\eqref{transform of R on t}
		\[
		\mathbf t_{\mathrm{TR}}(z)
		=
		R(z)^{-1}\mathbf t_{\mathrm{CA}}(z).
		\]
		The notation \(\mathbf t_{\mathrm{CA}}\) anticipates its later
		identification with the canonical ancestor variables of the CohFT.
		This change of variables plays a crucial role in the proof.
		
		Applying it to \(Z_{\mathrm{loc}}\), we obtain a transformed topological-recursion partition function, denoted by \(Z_{\mathrm{CA}}\). 
		At this stage, no ancestor-theoretic interpretation is assumed. If \(y\) is chosen compatibly with \(B\) as in DOSS
		\cite{dunin-barkowskiIdentificationGiventalFormula2014},
		\begin{equation}\label{compability condition on y}
			y^i_1=\psi_{i1},
			\qquad
			\sum_{a=1}^{N}\widehat B^{a,i}_{0,l}y^a_1
			=-(2l+3)!!y^i_{2l+3},
		\end{equation}
		then Theorem \ref{Theorem Givental Quantization} identifies \(Z_{\mathrm{CA}}\) with the canonical ancestor partition function and gives
		\begin{equation}
		\begin{aligned}
		Z_{\mathrm{CA}}(\mathbf{t}_{\mathrm{CA}}):=&Z_{\mathrm{loc}}(\mathbf{t}_{\mathrm{TR}}(\mathbf{t}_{\mathrm{CA}}))=Z_{\mathrm{loc}}(R(z)^{-1}\mathbf t_{\mathrm{CA}}(z))\\
		=&
		\exp \big(\sum_{2g-2+n>0}\frac{1}{n!}W_{g,n}\big|_{(-1)^l\d\xi^i_{(l)}\to t^i_{\mathrm{CA},l}}\big)\\
		=&
		\widehat R
		\prod_{i=1}^N
		Z_{\mathrm{WK}}\left(
		(\psi_{i1})^{-1/3}t^i_{\mathrm{CA},0},
		(\psi_{i1})^{-1}t^i_{\mathrm{CA},1},
		(\psi_{i1})^{-5/3}t^i_{\mathrm{CA},2},\ldots
		\right).
		\end{aligned}
		\end{equation}
		This recovers the result of DOSS \cite{dunin-barkowskiIdentificationGiventalFormula2014}.
		After applying the $\Psi$-change of basis this becomes precisely the flat ancestor formula
		\begin{equation}
		Z_{\mathrm{FA}}(\mathbf{t}_{\mathrm{FA}})
		=
		\left(\prod_{i=1}^N\psi_{i1}\right)^{1/24}
		\widehat\Psi\,\widehat R\,Z_{\mathrm{WK}}^{\otimes N}=Z_{\mathrm{CA}}(\mathbf{t}_{\mathrm{CA}})\big|_{\mathbf{t}_{\mathrm{FA}}=\Psi \mathbf{t}_{\mathrm{CA}}}.
		\end{equation}
		Conversely, starting from the $R$-matrix and the normalization factors $\psi_{i1}$ at a semisimple point, the product formula determines the normalized coefficients $\widehat B^{i,j}_{k,l}$ of a local Bergman kernel, while the remaining local coefficients of $B$ are not determined by the $R$-matrix.
		The compatibility condition \eqref{compability condition on y} determines the odd local expansion of the distinguished function \(y_B\) relevant to topological recursion.
		
		The subsequent Corollary \ref{origin point EO} expresses $W_{g,n}$ as the expansion of the canonical ancestor genus-$g$ free energy in the basis $(-1)^m\d\xi^i_{(m)}$ :
		\begin{equation}
		W_{g,n}=\sum_{(i_k,m_k)}\frac{\partial^n F^{\mathrm{CA}}_g }
		 {\partial t^{i_1}_{\mathrm{CA},m_1}\cdots\partial t^{i_n}_{\mathrm{CA},m_n}}
		 \bigg|_{\mathbf t_{\mathrm{CA}}=0}
		 \prod_{k=1}^{n} (-1)^{m_k}\d \xi^{i_k}_{(m_k)},
		 \end{equation}
		where $F^{\mathrm{CA}}_g$ is the free energy of the ancestor partition function.
		
		We also record a generic-background version of this statement.  
		Keeping the Bergman kernel $B$ and the above identification of its associated series $R(z)$ with the Givental $R$-matrix of a fixed semisimple CohFT, denote by $y_B$ the distinguished compatible
		$y$-function defined above.
		For a general $y$-function, we define a difference \eqref{q'}
		\[
		\delta \mathbf{y}(z)=\mathbf{y}_B (z)-\mathbf{y}(z).
		\]
		Consequently, the multidifferentials of the spectral curve with this general \(y\)-function are obtained by evaluating the same canonical ancestor free energies at the shifted point \(\mathbf t_{\mathrm{CA}}=R\delta\mathbf y\) (\textbf{Proposition} \ref{generic background EO}):
		\begin{equation}
		W_{g,n}
		=
		\sum_{(i_a,m_a)}
		\left.
		\frac{\partial^n F^{\mathrm{CA}}_g}
		{\partial t^{i_1}_{\mathrm{CA},m_1}\cdots
			\partial t^{i_n}_{\mathrm{CA},m_n}}
		\right|_{\mathbf t_{\mathrm{CA}}=R(z)\delta\mathbf y(z)}
		\prod_{a=1}^n(-1)^{m_a}\d\xi^{i_a}_{(m_a)}.
		\end{equation}
		Thus, varying \(y\) shifts the evaluation point of the corresponding ancestor potential without changing the CohFT itself.
		
		The rest of the paper is organized as follows.  Section \ref{section Eynard-Orantin topological recursion} develops the local expansion of the Eynard-Orantin recursion and derives the recursion for the EO correlators.  Section \ref{section Virasoro constraints for intersection numbers on moduli spaces of Riemann surfaces} recalls the Virasoro constraints for $\psi$- and $\kappa$-class intersection numbers, introduces the dilaton rescaling, constructs the local glued partition function, and proves its graph expansion.  Section \ref{section The equivalence between Eynard-Orantin topological recursion and Virasoro constraints} proves the equivalence theorem and discusses the Airy, Kontsevich, and deformed Bergman-kernel examples.  Section \ref{section Topological Recursion and Givental Quantization} proves the comparison with Givental quantization, explains the semisimple CohFT interpretation, treats generic backgrounds. And we discusses the $\mathbb P^1$ Gromov-Witten example in Appendix \ref{appendix A}.
		
	\vspace{.1in}
	{\bf Declaration of AI Assistance}.
	No AI tools were used in developing the mathematical ideas, results, or proofs presented in this paper. AI tools were used solely to polish the language of the manuscript.

\vspace{.2in}
{\bf Acknowledgements}. The second-named author  is partly supported by NSFC (No.~12371254, No.~11890662, No.~12061131014).

	\vskip 20pt
	
		\section{Eynard-Orantin topological recursion}\label{section Eynard-Orantin topological recursion}
		In this section, we review the definition and basic ingredients of the Eynard-Orantin topological recursion. We introduce the correlators associated with the recursion, following the works of Eynard \cite{eynardIntersectionNumbersSpectral2011,eynardInvariantsSpectralCurves2014}, and derive a set of recursion relations for them through a detailed computation, following \cite{zhouTopologicalRecursionVirasoro2024} .
		
		\subsection{Basic data of Eynard-Orantin topological recursion}
		There are several variants of the Eynard-Orantin topological recursion. In general, the input consists of a spectral curve comprising the following data:\\
		--A Riemann surface \(\Sigma\);\\
		--A meromorphic function \(x:\Sigma\to \mathbb P^1\), regarded as a branched covering;\\
		--A function \(y\) on \(\Sigma\), which is meromorphic (or, more generally, multi-valued meromorphic) and satisfies \(\d y\ne0\) at the ramification points of \(x\);\\
		--A symmetric bidifferential \(B(p,q)\) on \(\Sigma\) which has only a double pole on the diagonal \(p=q\) with no residue.
		
		We will only consider the case where \(x\) has only simple ramification points; for topological recursion with general ramification points see \cite{bouchardThinkGloballyCompute2013}. For any ramification point \(c\) of \(x\), there exists a unique involution map \(\sigma\) defined in a neighborhood \(U\) of \(c\), such that \(x(\sigma(p))=x(p)\) for all \(p\in U\) and \(\sigma\ne\mathrm{id}_U\). The recursion kernel is
		\begin{equation}\label{def of recursion kenrel}
			K(p,q):=\frac{\int_{q'=\sigma(q)}^{q}B(p,q')}{(y(q)-y(\sigma(q)))\d x(q)},\qquad p\in\Sigma,\ q\in U.
		\end{equation}
		
		The Eynard-Orantin topological recursion produces a sequence of multi-differentials \(W_{g,n}\) from
		\begin{equation}
			\begin{aligned}
				W_{0,1}(p)&=y(p)\d x(p),\\
				W_{0,2}(p_1,p_2)&=B(p_1,p_2),
			\end{aligned}
		\end{equation}
		and, for \(2g-2+n>0\),
		\begin{equation}\label{def of EO TR}
			\begin{aligned}
				W_{g,n+1}(p_0,p_1,\ldots,p_n)
				:=&-\frac12\sum_{i=1}^N\operatorname{Res}_{p=c_i}K(p_0,p)
				\Big[
				W_{g-1,n+2}(p,\sigma_i(p),p_{[n]})
				\\
				&\qquad\qquad+
				\sum_{\substack{g_1+g_2=g\\ I_1\sqcup I_2=[n]}}'
				W_{g_1,1+|I_1|}(p,p_{I_1})
				W_{g_2,1+|I_2|}(\sigma_i(p),p_{I_2})
				\Big],
			\end{aligned}
		\end{equation}
		where \(c_1,\ldots,c_N\) are all ramification points of \(x\), \(\sigma_i\) is the local involution near \(c_i\), \([n]=\{1,\ldots,n\}\), and \(p_{[n]}=(p_1,\ldots,p_n)\). The summation \(\sum'\) excludes the unstable terms involving \(W_{0,1}\). We ignore double poles of \(x\), since under our assumptions they do not contribute to the recursion.
		
		\subsection{Local Airy coordinate and local topological recursion}
		For each simple ramification point \(c_i\), choose a local coordinate chart \((U_i,z_i)\) centered at \(c_i\), unique up to sign, such that
		\begin{equation}\label{expansion of x}
			x(p)=x(c_i)+\frac12z_i(p)^2,\qquad p\in U_i.
		\end{equation}
		Then \(z_i(\sigma_i(p))=-z_i(p)\). We call \(z_i\) the local Airy coordinate and write \(z_i\) for \(z_i(p)\) when no confusion arises.
		
		In this coordinate, in this work we assume that one can expand \(y\) around each \(c_i\) as
		\begin{equation}\label{expansion of y}
			y(p)=\sum_{m=0}^{\infty}y_m^i z_i^m,\qquad y_1^i\ne0.
		\end{equation}
		The even-order terms do not appear in the recursion kernel.
	When there is only one ramification point and when the Bergman kernel is 
	$B(z_1,z_2) = \frac{dz_1dz_2}{(z_1-z_2)^2}$,
	the recursion produces the $n$-point correlation functions of the Witten-Kontsevich partition function \cite{eynardInvariantsAlgebraicCurves2007,zhouTopologicalRecursionsEynard2013}. 

		\begin{remark}
			In general, \(y\) may have a singularity at some ramification point \(c_i\):
			\begin{equation}
				y=y_{-k}^i z_i^{-k}+y_{-k+1}^i z_i^{-k+1}+\cdots+y_{-1}^i z_i^{-1}+\text{non-negative power terms}.
			\end{equation}
			A direct computation shows that if \(y_{-k}^i\ne0\) for some \(k\ge3\), then this ramification point does not contribute to the recursion. Since \(y_{-2}^i\) does not enter the recursion kernel either, the only admissible singular behavior is
			\begin{equation}
				y=y_{-1}^i z_i^{-1}+\text{non-negative power terms}.
			\end{equation}
			When the spectral curve is the Bessel curve:
			\begin{equation}
				x=\frac{1}{2}z^2,\quad y=\frac{1}{z},\quad B(z_1,z_2) = \frac{dz_1dz_2}{(z_1-z_2)^2},
			\end{equation}
		by a result due to \cite{doTopologicalRecursionBessel2018}
  		the recursion produces the $n$-point correlation functions of the   Br\'ezin-Gross-Witten (BGW) partition function \cite{brezinExternalFieldProblem1980,grossPossibleThirdorderPhase1980}. 
   		For the general case when $y$ is regular at some ramification points and is irregular at some other ramification, see \cite{chekhovTopologicalRecursionHard2019}. 
		We intend to deal with this case in subsequent work using the method developed in the present paper.
		\end{remark}
		
		For \(p\in U_i\) and \(q\in U_j\), write
		\begin{equation}\label{expansion of B}
			B(p,q)=\delta_{ij}\frac{\d z_i(p)\d z_j(q)}{(z_i(p)-z_j(q))^2}
			+\sum_{k,l\ge0}B^{i,j}_{k,l}z_i(p)^kz_j(q)^l\d z_i(p)\d z_j(q).
		\end{equation}
		The coefficient of the diagonal double pole is assumed to be \(1\). We note that the coefficients are symmetric, i.e., 
		\begin{equation}
			B^{i,j}_{k,l}=B^{j,i}_{l,k}.
		\end{equation}
		
		Since topological recursion uses only the local information near the ramification points, the local data \eqref{expansion of x}, \eqref{expansion of y}, and \eqref{expansion of B} define the local topological recursion.
		
		\subsection{Schur transform}
		The function \(y\) appears in the denominator of the recursion kernel \(K\), so the residue computation uses the inverse of the odd part of \(y\). 
	To make contact with the work \cite{liuRecursionFormulaeHigher2008} on higher Weil-Petersson volumes, let us recall the Schur transform \cite{eynardIntersectionNumbersSpectral2011} of a sequence of formal variables. 
	For a sequence
		\[
		\mathbf u=(u_1=1,u_3,u_5,\ldots),
		\]
		define its Schur transform \(\mathbf s=(s_1,s_2,\ldots)\) by
		\begin{equation}\label{Schur transform}
			\exp\left(-\sum_{a=1}^{\infty}s_aw^a\right)
			:=\sum_{m=0}^{\infty}(2m+1)!!\,u_{2m+1}w^m.
		\end{equation}
		Some initial relations are
		\begin{equation}
			s_1=-3u_3,\qquad
			s_2=-15u_5+\frac92u_3^2,\qquad
			s_3=-105u_7+45u_5u_3-9u_3^3.
		\end{equation}
		
		Following \cite{liuRecursionFormulaeHigher2008}, let \(\mathbb N^\infty\) be the set of sequences \(\mathbf b=(b_1,b_2,\ldots)\) with only finitely many nonzero entries, and define
		\begin{equation}
			|\mathbf b|=\sum_{a\ge1}ab_a,\qquad
			\Vert\mathbf b\Vert=\sum_{a\ge1}b_a,\qquad
			\mathbf b!=\prod_{a\ge1}b_a!.
		\end{equation}
		Define \(\beta_{\mathbf b}\) recursively by \(\beta_{\mathbf0}=1\) and
		\begin{equation}
			\beta_{\mathbf b}
			=\sum_{\substack{\mathbf L+\mathbf L'=\mathbf b\\ \mathbf L'\ne\mathbf0}}
			\frac{(-1)^{\Vert\mathbf L'\Vert-1}\beta_{\mathbf L}}
			{\mathbf L'!(2|\mathbf L'|+1)!!}.
		\end{equation}
		Then
		\begin{equation}
			\exp\left(-\sum_{a=1}^{\infty}s_aw^a\right)
			=\sum_{\mathbf b\in\mathbb N^\infty}
			(-1)^{\Vert\mathbf b\Vert}\frac{\mathbf s^{\mathbf b}}{\mathbf b!}w^{|\mathbf b|},
			\qquad
			\mathbf s^{\mathbf b}=\prod_{a\ge1}s_a^{b_a},
		\end{equation}
		and
		\begin{equation}
			u_{2m+1}
			=\frac1{(2m+1)!!}\sum_{|\mathbf b|=m}
			\frac{(-1)^{\Vert\mathbf b\Vert}}{\mathbf b!}\mathbf s^{\mathbf b}
			=\sum_{|\mathbf b|=m}
			\frac{(-1)^{\Vert\mathbf b\Vert}}{(2|\mathbf b|+1)!!\mathbf b!}\mathbf s^{\mathbf b}.
		\end{equation}
		The inverse series is
		\begin{equation}
			\frac1{1+u_3z^2+u_5z^4+\cdots}
			=\sum_{\mathbf b}\beta_{\mathbf b}\mathbf s^{\mathbf b}z^{2|\mathbf b|}.
		\end{equation}
		For example,
		\begin{equation}
			\beta_{(1,0,\ldots)}=\frac13,\qquad
			\beta_{(0,1,0,\ldots)}=\frac1{15},\qquad
			\beta_{(2,0,\ldots)}=\frac7{90}.
		\end{equation}
		
		For each ramification point \(c_i\), define \(\mathbf s^i=(s_1^i,s_2^i,\ldots)\) as the Schur transform of
		\[
		\mathbf u^i=\left(1,\frac{y_3^i}{y_1^i},\frac{y_5^i}{y_1^i},\ldots\right).
		\]
		Equivalently, it can be written as the Laplace-type transformation \cite{eynardIntersectionNumbersSpectral2011,eynardInvariantsSpectralCurves2014}:
		\begin{equation}\label{laplace tranform of y}
			\begin{aligned}
				&\sqrt{\frac w{2\pi}}\int_{x(z_i)\in\mathbb R_+}e^{-w(x(z_i)-x(c_i))}\d y(z_i)
				\\
				=&\sqrt{\frac w{2\pi}}\int_{-\infty}^{\infty}
				e^{-\frac w2z_i^2}\sum_{m=1}^{\infty}m\,y_m^i z_i^{m-1}\d z_i
				\\
				=&\sum_{m=0}^{\infty}(2m+1)!!\,y_{2m+1}^i w^{-m}
				=y_1^i\exp\left(-\sum_{m=1}^{\infty}s_m^i w^{-m}\right).
			\end{aligned}
		\end{equation}
		Here we use the Gaussian integral
		\begin{equation}\label{Gauss integral}
			\int_{-\infty}^{\infty}x^k e^{-\frac12x^2}\d x=
			\begin{cases}
				0, & k\text{ odd},\\
				(k-1)!!\sqrt{2\pi}, & k\text{ even}.
			\end{cases}
		\end{equation}
		Therefore
		\begin{equation}\label{inverse y odd part}
			\frac1{y_1^i+y_3^iz_i^2+y_5^iz_i^4+\cdots}
			=\frac1{y_1^i}\sum_{\mathbf b}\beta_{\mathbf b}(\mathbf s^i)^{\mathbf b}z_i^{2|\mathbf b|}.
		\end{equation}
		
		\subsection{Computation in local Airy coordinate}
		For \(q\in U_i\), expand the Bergman kernel as
		\begin{equation}\label{B expansion by local basis}
			B(p,q)=\sum_{k=0}^{\infty}\d\mathcal B_k^i(p)z_i(q)^k\d z_i(q).
		\end{equation}
		For \(p\in U_j\),
		\begin{equation}\label{define of d B}
			\d\mathcal B_k^i(p)=
			\delta^i_j\frac{k+1}{z_j(p)^{k+2}}\d z_j(p)
			\;+
			\sum_{l=0}^{\infty}B^{j,i}_{l,k}z_j(p)^l\d z_j(p).
		\end{equation}
		Then
		\begin{equation}
			\begin{aligned}
				\int_{q'=\sigma_i(q)}^{q'=q}B(p,q')
				&=\int_{z'=-z_i}^{z'=z_i}\sum_{k=0}^{\infty}\d\mathcal B_k^i(p)(z')^k\d z'
				\\
				&=\sum_{k=0}^{\infty}2\frac{z_i(q)^{2k+1}}{2k+1}\d\mathcal B_{2k}^i(p).
			\end{aligned}
		\end{equation}
		For \(p\in U_i\), the recursion kernel expands as
		\begin{equation}\label{kernel expansion}
			\begin{aligned}
				K(p_0,p)
				&=\frac{\sum_{k=0}^{\infty}\frac{z_i^{2k+1}}{2k+1}\d\mathcal B_{2k}^i(p_0)}
				{\left(\sum_{l=0}^{\infty}y_{2l+1}^iz_i^{2l+2}\right)\d z_i}
				\\
				&=(y_1^i\d z_i)^{-1}
				\sum_{k=0}^{\infty}\frac{z_i^{2k-1}}{2k+1}\d\mathcal B_{2k}^i(p_0)
				\sum_{\mathbf b}\beta_{\mathbf b}(\mathbf s^i)^{\mathbf b}z_i^{2|\mathbf b|}.
			\end{aligned}
		\end{equation}
		Only odd powers of \(z_i\) occur in \(K(p_0,p)\).
		
		Define the normalized basis
		\begin{equation}\label{normalized dxi}
			\d\xi^i_m(p):=(2m-1)!!\,\d\mathcal B_{2m}^i(p),
			\qquad (-1)!!:=1.
		\end{equation}
		The first two stable multi-differentials are
		\begin{equation}
			\begin{aligned}
				W_{0,3}(p_0,p_1,p_2)
				&=-\frac12\sum_{i=1}^N\operatorname{Res}_{p=c_i}K(p_0,p)
				\Big[B(p,p_1)B(\sigma_i(p),p_2)
				\\
				&\qquad\qquad\qquad\qquad\qquad
				+B(p,p_2)B(\sigma_i(p),p_1)\Big]
				\\
				&=\sum_{i=1}^N(y_1^i)^{-1}\d\xi^i_0(p_0)\d\xi^i_0(p_1)\d\xi^i_0(p_2),
			\end{aligned}
		\end{equation}
		and
		\begin{equation}\label{W11}
			\begin{aligned}
				W_{1,1}(p_0)
				&=-\frac12\sum_{i=1}^N\operatorname{Res}_{p=c_i}K(p_0,p)B(p,\sigma_i(p))
				\\
				&=\sum_{i=1}^N(y_1^i)^{-1}
				\left(\frac1{24}\d\xi^i_1(p_0)
				+\frac{s_1^i}{24}\d\xi^i_0(p_0)
				+\frac12B^{i,i}_{0,0}\d\xi^i_0(p_0)\right)
				\\
				&=\sum_{i=1}^N(y_1^i)^{-1}
				\left(\frac1{24}\d\xi^i_1(p_0)
				-\frac18\frac{y_3^i}{y_1^i}\d\xi^i_0(p_0)
				+\frac12B^{i,i}_{0,0}\d\xi^i_0(p_0)\right).
			\end{aligned}
		\end{equation}
		
		The two basic residue operations are
		\begin{equation}\label{two residue objects}
			\begin{aligned}
				&-\frac12\operatorname{Res}_{p=c_i}K(p_0,p)
				\Big[B(p,p_a)\d\xi^j_m(\sigma_i(p))
				+B(\sigma_i(p),p_a)\d\xi^j_m(p)\Big],
				\\
				&-\frac12\operatorname{Res}_{p=c_i}K(p_0,p)
				\Big[\d\xi^{j_1}_{k_1}(p)\d\xi^{j_2}_{k_2}(\sigma_i(p))
				+\d\xi^{j_1}_{k_1}(\sigma_i(p))\d\xi^{j_2}_{k_2}(p)\Big].
			\end{aligned}
		\end{equation}
		First,
		\begin{equation}\label{kernel coefficient}
			\begin{aligned}
				\relax[z_i^{2l-1}]K(p_0,p)\d z_i(p)
				&=(y_1^i)^{-1}
				\sum_{|\mathbf b|\le l}
				\frac{\beta_{\mathbf b}(\mathbf s^i)^{\mathbf b}}
				{(2l+1-2|\mathbf b|)!!}
				\d\xi^i_{l-|\mathbf b|}(p_0),
				\qquad l\ge0 .
			\end{aligned}
		\end{equation}
		
		The first residue is
		\begin{equation}
			\begin{aligned}
				&\operatorname{Res}_{p=c_i}K(p_0,p)B(p,p_a)\d\xi^j_m(\sigma_i(p))
				\\
				=&-(y_1^i)^{-1}(2m-1)!!B^{i,j}_{0,2m}
				\d\xi^i_0(p_0)\d\xi^i_0(p_a)
				\\
				&-(y_1^i)^{-1}\delta^j_i(2m+1)!!
				\sum_{l=0}^{m+1}\sum_{|\mathbf b|\le l}
				\frac{\beta_{\mathbf b}(\mathbf s^i)^{\mathbf b}}
				{(2l+1-2|\mathbf b|)!!(2m+1-2l)!!}
				\d\xi^i_{l-|\mathbf b|}(p_0)
				\d\xi^i_{m+1-l}(p_a).
			\end{aligned}
		\end{equation}
		Equivalently,
		\begin{equation}\label{split}
			B(p_a,-)\d\xi^j_m(-)
			\mapsto
			\sum_{i=1}^{N}\sum_{\substack{k,l\ge0\\ k+l\le m+1}}
			\alpha_{j,m}^{i,k;i,l}\d\xi^i_k(p_0)\d\xi^i_l(p_a),
		\end{equation}
		where
		\begin{equation}\label{alpha}
			\begin{aligned}
				\alpha_{j,m}^{i,k;i,l}
				&=(y_1^i)^{-1}(2m-1)!!B^{i,j}_{0,2m}\delta_{k,0}\delta_{l,0}
				\\
				&\quad
				+(y_1^i)^{-1}\delta^j_i(2m+1)!!
				\sum_{|\mathbf b|\le m+1}
				\frac{\beta_{\mathbf b}(\mathbf s^i)^{\mathbf b}}
				{(2k+1)!!(2l-1)!!}
				\delta_{k+l,m+1-|\mathbf b|}.
			\end{aligned}
		\end{equation}
		
		The second residue gives
		\begin{equation}\label{merge}
			\d\xi^{j_1}_{k_1}(-)\d\xi^{j_2}_{k_2}(-)
			\mapsto
			\sum_{i=1}^{N}\sum_{m=0}^{k_1+k_2+2}
			\mu_{j_1,k_1;j_2,k_2}^{i,m}\d\xi^i_m(p_0),
		\end{equation}
		where
		\begin{equation}\label{mu}
			\begin{aligned}
				\mu_{j_1,k_1;j_2,k_2}^{i,m}
				=&(y_1^i)^{-1}\delta^{j_1}_i\delta^{j_2}_i
				\frac{(2k_1+1)!!(2k_2+1)!!}{(2m+1)!!}
				\sum_{|\mathbf b|=k_1+k_2+2-m}
				\beta_{\mathbf b}(\mathbf s^i)^{\mathbf b}
				\\
				&+(y_1^i)^{-1}\delta^{j_1}_i
				\frac{(2k_1+1)!!(2k_2-1)!!}{(2m+1)!!}
				\sum_{|\mathbf b|\le k_1+1-m}
				\beta_{\mathbf b}(\mathbf s^i)^{\mathbf b}
				B^{i,j_2}_{2k_1+2-2|\mathbf b|-2m,2k_2}
				\\
				&+(y_1^i)^{-1}\delta^{j_2}_i
				\frac{(2k_2+1)!!(2k_1-1)!!}{(2m+1)!!}
				\sum_{|\mathbf b|\le k_2+1-m}
				\beta_{\mathbf b}(\mathbf s^i)^{\mathbf b}
				B^{i,j_1}_{2k_2+2-2|\mathbf b|-2m,2k_1}
				\\
				&+(y_1^i)^{-1}(2k_1-1)!!(2k_2-1)!!
				B^{i,j_1}_{0,2k_1}B^{i,j_2}_{0,2k_2}\delta_{m,0}.
			\end{aligned}
		\end{equation}
		Terms containing a negative index of \(B\) or \(\d\xi\) are understood to be zero.
	
		\begin{remark}
			In \cite{zhouTopologicalRecursionVirasoro2024}, the second-named author called $\alpha$ the splitting coefficients and $\mu$ the merging coefficients. It is natural to guess that they  form the structure constants of a Hopf algebra. We leave this question for future investigation.
		\end{remark}
		
		\subsection{Correlators and their recursion relations}
		In \cite{eynardIntersectionNumbersSpectral2011,eynardInvariantsSpectralCurves2014}, Eynard expanded the multi-differentials $W_{g,n}$ in a suitable basis of differentials and indicated that the correlators with respect to this expansion are related to the intersection numbers on moduli space of Riemann surfaces.
		\begin{definition}
			For $2g-2+n>0$, the correlators $	\left\langle\tau_{i_1,m_1}\cdots\tau_{i_n,m_n}\right\rangle_g^{\mathrm{EO}}$ of the Eynard-Orantin topological recursion for a spectral curve are defined by the expansion of the multi-differentials:
			\begin{equation}\label{EO correlator expansion}
			W_{g,n}(p_1,\ldots,p_n)
			=\sum_{i_a,m_a}
			\left\langle\tau_{i_1,m_1}\cdots\tau_{i_n,m_n}\right\rangle_g^{\mathrm{EO}}
			\prod_{a=1}^{n}\d\xi^{i_a}_{m_a}(p_a).
		\end{equation}
		\end{definition}
		By induction, the summation runs over all tuples satisfying \(\sum_a m_a\le3g-3+n\).
		
		By the computation in the previous subsection, the recursion can be written as
		\begin{proposition}\label{brief expansion of TR}
			The Eynard-Orantin topological recursion is equivalent to the following recursion for the correlators:
			\begin{equation}\label{TR for correlators}
				\begin{aligned}
					&\left\langle
					\tau_{i_0,m_0}\tau_{i_1,m_1}\cdots\tau_{i_n,m_n}
					\right\rangle_g^{\mathrm{EO}} \\
					=&
					\sum_{\substack{1\le j\le n\\ i_j=i_0}}
					\sum_{i,m}
					\alpha^{i_0,m_0;i_j,m_j}_{i,m}
					\left\langle
					\tau_{i,m}
					\prod_{\substack{1\le k\le n\\ k\ne j}}
					\tau_{i_k,m_k}
					\right\rangle_g^{\mathrm{EO}}
					\\
					&+\frac{1}{2}
					\sum_{j_1,k_1,j_2,k_2}
					\mu^{i_0,m_0}_{j_1,k_1;j_2,k_2}
					\Bigg[
					\left\langle
					\tau_{j_1,k_1}\tau_{j_2,k_2}
					\prod_{j=1}^{n}\tau_{i_j,m_j}
					\right\rangle_{g-1}^{\mathrm{EO}}
					\\
					&\qquad\qquad\qquad\qquad
					+
					\sum_{\substack{g_1+g_2=g\\ I_1\sqcup I_2=[n]}}^{\mathrm{st}}
					\left\langle
					\tau_{j_1,k_1}
					\prod_{j\in I_1}\tau_{i_j,m_j}
					\right\rangle_{g_1}^{\mathrm{EO}}
					\left\langle
					\tau_{j_2,k_2}
					\prod_{j\in I_2}\tau_{i_j,m_j}
					\right\rangle_{g_2}^{\mathrm{EO}}
					\Bigg],
				\end{aligned}
			\end{equation}
			where $\sum^{\mathrm{st}}$ indicates that only stable terms are summed. The initial values are
			\begin{equation}
				\begin{aligned}
					\left\langle\tau_{i,0}\tau_{i,0}\tau_{i,0}\right\rangle_0^{\mathrm{EO}}
					&=(y^i_1)^{-1},\\
					\left\langle\tau_{i,1}\right\rangle_1^{\mathrm{EO}}
					&=\frac{1}{24}(y^i_1)^{-1},\\
					\left\langle\tau_{i,0}\right\rangle_1^{\mathrm{EO}}
					&=\frac{1}{24}(y^i_1)^{-1}s^i_1
					+\frac{1}{2}(y^i_1)^{-1}B^{i,i}_{0,0}.
				\end{aligned}
			\end{equation}
		\end{proposition}
		
		Introduce the normalized Bergman coefficients
		\begin{equation}\label{Bhat}
			\widehat B^{i,j}_{k,l}:=(2k-1)!!(2l-1)!!\,B^{i,j}_{2k,2l}.
		\end{equation}
		
		Then the Eynard-Orantin topological recursion is equivalent to the following recursion for the normalized correlators:
		\begin{equation}\label{recursion relation for EO TR}
			\begin{aligned}
				&y_1^{i_0}(2m_0+1)!!
				\left\langle\tau_{i_0,m_0}\tau_{i_1,m_1}\cdots\tau_{i_n,m_n}\right\rangle_g^{\mathrm{EO}}
				\\
				=&\sum_{\substack{1\le j\le n\\ i_j=i_0}}
				\sum_{\mathbf b}\beta_{\mathbf b}(\mathbf s^{i_0})^{\mathbf b}
				\frac{(2(m_0+m_j+|\mathbf b|)-1)!!}{(2m_j-1)!!}
				\left\langle
				\tau_{i_0,m_0+m_j+|\mathbf b|-1}
				\prod_{\substack{k=1\\ k\ne j}}^n\tau_{i_k,m_k}
				\right\rangle_g^{\mathrm{EO}}
				\\
				&+\sum_{\substack{1\le j\le n\\ i_j=i_0}}
				\delta_{m_0,0}\delta_{m_j,0}
				\sum_{i,m}\widehat B^{i,i_0}_{m,0}
				\left\langle\tau_{i,m}
				\prod_{\substack{k=1\\ k\ne j}}^n\tau_{i_k,m_k}
				\right\rangle_g^{\mathrm{EO}}
				\\
				&+\frac12\sum_{\mathbf b}\beta_{\mathbf b}(\mathbf s^{i_0})^{\mathbf b}
				\sum_{k_1+k_2=m_0+|\mathbf b|-2}
				(2k_1+1)!!(2k_2+1)!!
				\\
				&\quad\cdot\left[
				\left\langle\tau_{i_0,k_1}\tau_{i_0,k_2}
				\prod_{j=1}^n\tau_{i_j,m_j}\right\rangle_{g-1}^{\mathrm{EO}}
				+\sum_{\substack{g_1+g_2=g\\I_1\sqcup I_2=[n]}}^s
				\left\langle\tau_{i_0,k_1}\prod_{j\in I_1}\tau_{i_j,m_j}\right\rangle_{g_1}^{\mathrm{EO}}
				\left\langle\tau_{i_0,k_2}\prod_{j\in I_2}\tau_{i_j,m_j}\right\rangle_{g_2}^{\mathrm{EO}}
				\right]
				\\
				&+\sum_{\mathbf b}\beta_{\mathbf b}(\mathbf s^{i_0})^{\mathbf b}
				\sum_{k_1,i,k_2}
				\frac{(2k_1+1)!!\,\widehat B^{i,i_0}_{k_2,k_1+1-|\mathbf b|-m_0}}
				{(2(k_1+1-|\mathbf b|-m_0)-1)!!}
				\\
				&\quad\cdot\left[
				\left\langle\tau_{i_0,k_1}\tau_{i,k_2}
				\prod_{j=1}^{n}\tau_{i_j,m_j}\right\rangle_{g-1}^{\mathrm{EO}}
				+\sum_{\substack{g_1+g_2=g\\I_1\sqcup I_2=[n]}}^s
				\left\langle\tau_{i_0,k_1}\prod_{j\in I_1}\tau_{i_j,m_j}\right\rangle_{g_1}^{\mathrm{EO}}
				\left\langle\tau_{i,k_2}\prod_{j\in I_2}\tau_{i_j,m_j}\right\rangle_{g_2}^{\mathrm{EO}}
				\right]
				\\
				&+\frac12\delta_{m_0,0}
				\sum_{j_1,k_1,j_2,k_2}
				\widehat B^{j_1,i_0}_{k_1,0}\widehat B^{j_2,i_0}_{k_2,0}
				\\
				&\quad\cdot\left[
				\left\langle\tau_{j_1,k_1}\tau_{j_2,k_2}
				\prod_{l=1}^{n}\tau_{i_l,m_l}\right\rangle_{g-1}^{\mathrm{EO}}
				+\sum_{\substack{g_1+g_2=g\\I_1\sqcup I_2=[n]}}^s
				\left\langle\tau_{j_1,k_1}\prod_{l\in I_1}\tau_{i_l,m_l}\right\rangle_{g_1}^{\mathrm{EO}}
				\left\langle\tau_{j_2,k_2}\prod_{l\in I_2}\tau_{i_l,m_l}\right\rangle_{g_2}^{\mathrm{EO}}
				\right].
			\end{aligned}
		\end{equation}
		
		This explicit formula will be compared with the recursion relation obtained from the Virasoro constraints in the next section.

		\vskip 20pt
		
		\section{Intersection numbers on moduli spaces \texorpdfstring{$\overline{\mathcal{M}}_{g,n}$}{Mbar(g,n)} and their Virasoro constraints}\label{section Virasoro constraints for intersection numbers on moduli spaces of Riemann surfaces}
		In this section, we discuss the Virasoro constraints for various intersection numbers on moduli spaces of Riemann surfaces $\overline{\mathcal{M}}_{g,n}$. We first recall the Virasoro constraints for the Witten-Kontsevich tau-function which is the generating series of the intersection numbers of $\psi$-classes \cite{wittenTwodimensionalGravityIntersection1990,kontsevichIntersectionTheoryModuli1992}. Next we recall the Virasoro constraints for the intersection numbers involving both the $\psi$-classes and $\kappa$-classes derived by Liu and Xu \cite{liuRecursionFormulaeHigher2008}, generalizing the work of Mulase and Safnuk \cite{mulaseMirzakhanisRecursionRelations2006}. Then we use an idea in \cite{faberAlgorithmsComputingIntersection1997} to derive Virasoro constraints for the intersection numbers that appear in Eynard's formula, building on the results of \cite{eynardIntersectionNumbersSpectral2011} and \cite{eynardInvariantsSpectralCurves2014}. We also provide a graph representation of this partition function (Subsection \ref{subsection Description as graph summation of intersection numbers}), which is equivalent to Eynard's formula \cite{eynardIntersectionNumbersSpectral2011,eynardInvariantsSpectralCurves2014} and to the DOSS description \cite{dunin-barkowskiIdentificationGiventalFormula2014}. 
		
		\subsection{Witten-Kontsevich tau-function and Virasoro constraints} On the Deligne-Mumford compactified moduli spaces of Riemann surfaces $\overline{\mathcal{M}}_{g,n}$, one can consider the following intersection numbers:
		\begin{equation}
			\langle \tau_{d_1}\tau_{d_2}\cdots\tau_{d_n}\rangle_{g}:=\int_{\overline{\mathcal{M}}_{g,n}} \psi_{1}^{d_1}\psi_2^{d_2}\cdots\psi_n^{d_n}.
		\end{equation}	
		These are defined only under the stability condition $2g - 2 + n > 0$ and satisfy the selection rule
		\begin{equation}\label{selection ruler for tau}
			\sum_{i=1}^{n} d_i=3g-3+n,
		\end{equation}
		otherwise they vanish. Their generating series $F_{\mathrm{WK}}(\mathbf t_{\mathrm{WK}})$ is defined by
		\begin{equation}
			F_{\mathrm{WK}}(\mathbf t_{\mathrm{WK}}):=\sum_{g=0}^{\infty}\lambda^{2g-2}\sum_{d_1,d_2,\cdots,d_n}\langle\tau_{d_1}\tau_{d_2}\cdots\tau_{d_n}\rangle_{g}\frac{t_{\mathrm{WK},d_1}t_{\mathrm{WK},d_2}\cdots t_{\mathrm{WK},d_n}}{n!}.
		\end{equation}
		In a more compact form, it can be written as
		\begin{equation}
			\begin{aligned}
				F_{\mathrm{WK}}(\mathbf t_{\mathrm{WK}})=&\sum_{g=0}^{\infty} \lambda^{2g-2}\sum_{n_0,n_1,\cdots} \langle\tau_{0}^{n_0}\tau_{1}^{n_1}\cdots \tau_{k}^{n_k}\cdots\rangle_{g} \frac{(t_{\mathrm{WK},0})^{n_0}}{n_0!}\frac{(t_{\mathrm{WK},1})^{n_1}}{n_1!}\cdots\frac{(t_{\mathrm{WK},k})^{n_k}}{n_k!}\cdots\\
				=&\sum_{g=0}^{\infty}\lambda^{2g-2} \langle \exp(\sum_{d=0}^{\infty}t_{\mathrm{WK},d} \tau_{d} )\rangle_{g}.
			\end{aligned}
		\end{equation} 
		We will omit the string coupling constant $\lambda$ in this paper, since it can be recovered from the degree count.
		
		By the Dijkgraaf-Verlinde-Verlinde (DVV) relations \cite{dijkgraafLoopEquationsVirasoro1991}, there exists a sequence of quadratic differential operators $\{L_n|n=-1,0,1,\cdots\}$ that annihilate the Witten-Kontsevich tau-function $Z_{\mathrm{WK}}(\mathbf t_{\mathrm{WK}})=\exp(F_{\mathrm{WK}}(\mathbf t_{\mathrm{WK}}))$, i.e.,
		\begin{equation}
			L_m^{\mathrm{WK}}(\mathbf q_{\mathrm{WK}},\partial ) Z_{\mathrm{WK}}(\mathbf q_{\mathrm{WK}})=0.
		\end{equation}
		The explicit expression of these operators is, for $m\geq-1$,
		\begin{equation}\label{Virasoro for WK}
			\begin{aligned}
				&L_{m}^{\mathrm{WK}}(\mathbf q_{\mathrm{WK}},\partial)=\sum_{\substack{r\geq0\\ r+m\geq 0}} \frac{(2r+2m+1)!!}{(2r-1)!!}q_{\mathrm{WK},r}\frac{\partial}{\partial q_{\mathrm{WK},r+m}}\\
				&+\frac{1}{2}\sum_{k_1+k_2=m-1}(2k_1+1)!!(2k_2+1)!!\frac{\partial^2}{\partial q_{\mathrm{WK},k_1}\partial q_{\mathrm{WK},k_2}}+\delta_{m,-1}\frac{(q_{\mathrm{WK},0})^2}{2}+\delta_{m,0}\frac{1}{8},
			\end{aligned}
		\end{equation}
		where the parameters $\mathbf q_{\mathrm{WK}}$ are the dilaton shift of $\mathbf t_{\mathrm{WK}}$, i.e.,
		\[
		q_{\mathrm{WK},r}=t_{\mathrm{WK},r}-\delta_{r,1},
		\]
		and 
		\begin{equation}
			Z_{\mathrm{WK}}(\mathbf q_{\mathrm{WK}})=Z_{\mathrm{WK}}(q_{\mathrm{WK},0},q_{\mathrm{WK},1}+1,q_{\mathrm{WK},2},\cdots).
		\end{equation}
		\begin{remark}
			For convenience, we will use the same notation for $\mathbf t_{\mathrm{WK}}$-parameters and dilaton shift $\mathbf q_{\mathrm{WK}}$-parameters. That is, the partition function $Z_{\mathrm{WK}}(\mathbf q_{\mathrm{WK}})=Z_{\mathrm{WK}}(\mathbf t_{\mathrm{WK}})$, the Virasoro operators $L^{\mathrm{WK}}(\mathbf q_{\mathrm{WK}},\partial)=L^{\mathrm{WK}}(\mathbf t_{\mathrm{WK}},\partial)$,  etc.
		\end{remark}
		
		These operators satisfy the Virasoro commutation relations:
		\begin{equation}
			[\frac{1}{2}L_n^{\mathrm{WK}},\frac{1}{2}L_m^{\mathrm{WK}}]=(n-m)\frac{1}{2}L_{n+m}^{\mathrm{WK}} \quad \forall \ n,m\geq-1.
		\end{equation}
		These operators uniquely determine (up to a multiplicative constant) the function $Z_{\mathrm{WK}}(\mathbf t_{\mathrm{WK}})$, and yield the following recursion for intersection numbers:
		\begin{equation}
			\begin{aligned}
				&\langle \tau_{m_0}\tau_{m_1}\cdots\tau_{m_n}\rangle_{g}\\
				=&\sum_{i=1}^{n}\frac{(2m_0+2m_i-1)!!}{(2m_0+1)!!(2m_i-1)!!}\left\langle \tau_{m_0+m_i-1}\prod_{\substack{j=1\\ j\ne i}}^n\tau_{m_j}\right\rangle_{g}\\
				&+\frac{1}{2}\sum_{k_1+k_2=m_0-2}\frac{(2k_1+1)!!(2k_2+1)!!}{(2m_0+1)!!}\Big[ \langle \tau_{k_1}\tau_{k_2}\prod_{j=1}^{n}\tau_{m_j}\rangle_{g-1}\\
				&+\sum_{\substack{g_1+g_2=g\\I_1\sqcup I_2=[n]}}^{\mathrm{st}}\langle\tau_{k_1}\prod_{j\in I_1}\tau_{m_j}\rangle_{g_1}\langle\tau_{k_2}\prod_{j\in I_2}\tau_{m_j}\rangle_{g_2}
				\Big],
			\end{aligned}
		\end{equation}
		together with the initial values:
		\begin{equation}
			\langle \tau_0\tau_0\tau_0\rangle_0=1, \quad \langle\tau_1\rangle_1=\frac{1}{24}.
		\end{equation}
		
		\subsection{Virasoro constraints for higher Weil-Petersson volumes} Let us now recall the results of Liu and Xu \cite{liuRecursionFormulaeHigher2008} which generalize the work of Mulase and Safnuk \cite{mulaseMirzakhanisRecursionRelations2006} originating from Mirzakhani's work \cite{mirzakhaniWeilPeterssonVolumesIntersection2006}.
		
		Consider the intersection numbers involving both $\psi$-classes and $\kappa$-classes:
		\begin{equation}
			\langle \kappa_{b_1}\kappa_{b_2}\cdots\kappa_{b_m}\tau_{d_1}\tau_{d_2}\cdots\tau_{d_n}\rangle_{g}:=\int_{\overline{\mathcal{M}}_{g,n}}\kappa_{b_1}\kappa_{b_2}\cdots\kappa_{b_m}\psi_1^{d_1}\psi_2^{d_2}\cdots\psi_{n}^{d_n}.
		\end{equation}
		The selection rule is 
		\begin{equation}\label{selection ruler of high weil-petersson}
			\sum_{j=1}^{m}b_j+\sum_{i=1}^{n}d_i=3g-3+n.
		\end{equation}
		
		Their generating series is defined as:
		\begin{equation}\label{free energy with kappa}
			\begin{aligned}
				F(\mathbf{t};\mathbf{s}):=&\sum_{g=0}^{\infty}\sum_{\mathbf{b},\mathbf{n}}\langle\kappa_{1}^{b_1}\kappa_2^{b_2}\cdots\tau_0^{n_0}\tau_1^{n_1}\cdots\rangle_g \frac{\mathbf{s}^{\mathbf{b}}}{\mathbf{b}!}\frac{\mathbf{t}^{\mathbf{n}}}{\mathbf{n}!}\\
				=&\sum_{g=0}^{\infty}\langle \exp(\sum_{i=0}^{\infty}t_i \tau_{i})\exp (\sum_{j=1}^{\infty}s_j\kappa_j)\rangle_{g},
			\end{aligned}
		\end{equation}
		where $\mathbf{b}=(b_1,b_2,\cdots)$, $\mathbf{n}=(n_0,n_1,\cdots)$ are sequences of non-negative integers with only finitely many non-zero terms.
		The partition function $Z(\mathbf{t};\mathbf{s})=\exp F(\mathbf{t};\mathbf{s})$ corresponds to the partition function of higher Weil-Petersson volumes of moduli spaces, and satisfies the generalized constraints \cite{liuRecursionFormulaeHigher2008}:
		\begin{equation}\label{hatL constraint}
			\bar{L}_{n}(\mathbf{t},\partial;\mathbf{s})Z(\mathbf{t};\mathbf{s})=0 \quad n\geq-1,
		\end{equation}
		where
		\begin{equation}\label{LiuXu operator}
			\begin{aligned}
				&\bar{L}_{n}(\mathbf{t},\partial;\mathbf{s})=-(2n+3)!!\frac{\partial}{\partial t_{n+1}}+\sum_{\mathbf{b}}\beta_\mathbf{b}\mathbf{s}^{\mathbf{b}} \sum_{j=0}^{\infty}\frac{(2(j+n+|\mathbf{b}|)+1)!!}{(2j-1)!!}t_j\frac{\partial}{\partial t_{j+n+|\mathbf{b}|}}\\
				&+\frac{1}{2}\sum_{\mathbf{b}}\beta_\mathbf{b}\mathbf{s}^\mathbf{b}\sum_{d_1+d_2=n-1+|\mathbf{b}|}(2d_1+1)!!(2d_2+1)!!\frac{\partial^2}{\partial t_{d_1}\partial t_{d_2}}
				+\delta_{n,-1}(\frac{t_0^2}{2}+\frac{s_1}{24})+\frac{\delta_{n,0}}{8}.
			\end{aligned}
		\end{equation}
		The operators $\bar{L}_k(\mathbf{s})$ satisfy the commutation relations:
		\begin{equation}
			[\frac{1}{2}\bar{L}_{n},\frac{1}{2}\bar{L}_{m}]=(n-m)\sum_{\mathbf{b}}\beta_\mathbf{b}\mathbf{s}^{\mathbf{b}}\cdot\frac{1}{2}\bar{L}_{m+n+|\mathbf{b}|}.
		\end{equation}
		To convert these into Virasoro operators, an idea introduced in \cite{mulaseMirzakhanisRecursionRelations2006}, and was generalized in \cite{liuRecursionFormulaeHigher2008}. Define $L_n(\mathbf{t},\partial;\mathbf{s})$ by:
		\begin{equation}\label{transform of Virasoro to shift}
			L_n(\mathbf{t},\partial;\mathbf{s})=\sum_{\mathbf{L}}\gamma_{\mathbf{L}}\mathbf{s}^\mathbf{L}\cdot \bar{L}_{n+|\mathbf{L}|}(\mathbf{t},\partial;\mathbf{s}),
		\end{equation}
		where the constants $\gamma_{\mathbf{L}}$ are defined by:
		\begin{equation}
			\gamma_{\mathbf{L}}:=\frac{(-1)^{\Vert\mathbf{L}\Vert}}{\mathbf{L}!(2|\mathbf{L}|+1)!!}.
		\end{equation}
		These operators satisfy the Virasoro commutation relations and also annihilate the partition function $Z(\mathbf{t};\mathbf{s})$.
		
		Indeed, these operators $L_n(\mathbf{t},\partial;\mathbf{s})$ are explicitly given by:
		\begin{equation}\label{shift LiuXu}
			\begin{aligned}
				&L_n(\mathbf{t},\partial;\mathbf{s})=\sum_{i=0}^{\infty} \frac{(2n+2i+1)!!}{(2i-1)!!}(\bar{t}_i-\delta_{i,1})\frac{\partial}{\partial \bar{t}_{i+n}}\\
				&+\frac{1}{2}\sum_{k_1+k_2=n-1}(2k_1+1)!!(2k_2+1)!!\frac{\partial^2}{\partial \bar{t}_{k_1}\partial \bar{t}_{k_2}}+\delta_{n,-1}\frac{\bar{t}_0^2}{2}+\delta_{n,0}\frac{1}{8}\\
				&=L_n^{\mathrm{WK}}(\mathbf{q}+\mathbf{p},\partial),
			\end{aligned}
		\end{equation}
		where $\bar{t}_n$ is a shift of $t_n$ beginning from $n\geq2$:
		\begin{equation}
			\bar{t}_n=\begin{cases}
				& t_n, \quad n=0,1\\
				&t_n+p_n, \quad n\geq 2,
			\end{cases}
		\end{equation}
		and the polynomials $p_n$ in $\mathbf{s}$ are given by
		\begin{equation}\label{transform from s to p}
			p_n=\sum_{|\mathbf{L}|=n-1}\frac{(-1)^{\Vert\mathbf{L}\Vert-1}}{\mathbf{L}!}\mathbf{s}^{\mathbf{L}}.
		\end{equation}
		As a consequence, 
		\begin{equation}\label{shift of kappa}
			F(\mathbf{t};\mathbf{s})=F_{\mathrm{WK}}(t_0,t_1,t_2+p_2,t_3+p_3,\cdots).
		\end{equation}
		From the definition (\ref{free energy with kappa}), we obtains
		\begin{equation}
			\begin{aligned}
				\sum_{g=0}^{\infty} \langle \exp(\sum_{i=0}^{\infty} t_i\tau_{i})\exp(\sum_{j=1}^{\infty} s_j\kappa_j)\rangle_{g}=\sum_{g=0}^{\infty} \langle \exp(\sum_{i=0}^{\infty} t_i\tau_{i})\exp(\sum_{k=2}^{\infty} p_k \tau_k)\rangle_g.
			\end{aligned}
		\end{equation}
		Thus, for any operators $\tau_{d_i}$,
		\begin{equation}\label{shift with kappa class in tau class}
			\langle \tau_{d_1}\tau_{d_2}\cdots\tau_{d_n}\exp(\sum_{j=1}^{\infty}s_j\kappa_j)\rangle_g=\langle \tau_{d_1}\tau_{d_2}\cdots\tau_{d_n}\exp(\sum_{k=2}^{\infty}p_k\tau_{k})\rangle_{g}.
		\end{equation}
		
		Notice that
		\begin{equation}
			1-\sum_{k=2}^{\infty}p_k w^{k-1}=\exp(-\sum_{i=1}^{\infty} s_iw^i).
		\end{equation}
		Hence, if $\mathbf{s}$ is the Schur transform of $\mathbf{u} = (u_1 = 1, u_3 = \frac{y_3}{y_1}, \dots, u_{2k+1} = \frac{y_{2k+1}}{y_1}, \dots)$ as in (\ref{Schur transform}), then			
		\begin{equation}
			p_k=-(2k-1)!!u_{2k-1}=-(2k-1)!! \frac{y_{2k-1}}{y_1}, \quad \text{for } k\geq2.
		\end{equation}
		We always assume $\mathbf{p}$ is related to $\mathbf{y}$ by this transformation.
		
		Viewing $\mathbf{s}$ as parameters, we write 
		\begin{equation}
			F(\mathbf{t};\mathbf{s})=\sum_{g=0}^{\infty}\sum_{\mathbf{n}}\langle\tau_0^{n_0}\tau_1^{n_1}\cdots \rangle_{g}(\mathbf{s})\frac{\mathbf{t}^{\mathbf{n}}}{\mathbf{n}!},
		\end{equation}
		and define
		\begin{equation}
			\begin{aligned}
				\langle\tau_{d_1}\tau_{d_2}\cdots\tau_{d_n}\rangle_g(\mathbf{s}):=&\int_{\overline{\mathcal{M}}_{g,n}}\psi_{1}^{d_1}\psi_{2}^{d_2}\cdots\psi_{n}^{d_n}\exp(\sum_{i=1}^{\infty} \kappa_i s_i)\\
				=&\langle \tau_{d_1}\tau_{d_2}\cdots\tau_{d_n}\exp(\sum_{j=1}^{\infty}s_j\kappa_j)\rangle_{g}.
			\end{aligned}
		\end{equation}
		
		These correlators satisfy the recursion relations derived from the constraints (\ref{hatL constraint}): 
		\begin{equation}\label{recursion relation for high weil-petersson}
			\begin{aligned}
				&\langle \tau_{m_0}\tau_{m_1}\cdots\tau_{m_n}\rangle_{g}(\mathbf{s})\\
				=&\sum_{i=1}^{n}\sum_{\mathbf{b}}\beta_\mathbf{b}\mathbf{s}^\mathbf{b}\frac{(2(m_0+m_i+|\mathbf{b}|)-1)!!}{(2m_0+1)!!(2m_i-1)!!}\left\langle \tau_{m_0+m_i-1+|\mathbf{b}|}\prod_{\substack{j=1\\ j\ne i}}^n\tau_{m_j}\right\rangle_{g}(\mathbf{s})\\
				+&\frac{1}{2}\sum_{\mathbf{b}}\beta_\mathbf{b}\mathbf{s}^\mathbf{b}\sum_{k_1+k_2=m_0-2+|\mathbf{b}|}\frac{(2k_1+1)!!(2k_2+1)!!}{(2m_0+1)!!}\Big[ \langle\tau_{k_1}\tau_{k_2}\prod_{j=1}^{n}\tau_{m_j}\rangle_{g-1}(\mathbf{s})\\
				+&\sum_{\substack{g_1+g_2=g\\I_1\sqcup I_2=[n]}}^{s}\langle\tau_{k_1}\prod_{j\in I_1 }\tau_{m_j}\rangle_{g_1}(\mathbf{s})\langle\tau_{k_2}\prod_{j\in I_2}\tau_{m_j}\rangle_{g_2}(\mathbf{s})
				\Big],
			\end{aligned}
		\end{equation} 
		with the initial values:
		\begin{equation}
			\langle\tau_0\tau_0\tau_0\rangle_{0}(\mathbf{s})=1,\quad \langle\tau_{1}\rangle_1(\mathbf{s})=\frac{1}{24}, \quad \langle \tau_0\rangle_1(\mathbf{s})=\frac{s_1}{24}.
		\end{equation}
		
		\subsection{Dilaton transform in intersection numbers} Comparing the recursion relation (\ref{recursion relation for EO TR}) with (\ref{recursion relation for high weil-petersson}), we see that a dilaton transform is needed to introduce a constant factor $y_1$ in the latter.
		
		Define the dilaton rescaling
		\begin{equation}
			\tau_m\mapsto \hat{\tau}_{m}=y_1^{-\frac{2m+1}{3}}\tau_m;\quad \kappa_m\mapsto \hat{\kappa}_{m}= y_1^{-\frac{2m}{3}}\kappa_m,
		\end{equation}
		and the corresponding correlators by
		\begin{equation}
			\begin{aligned}
				&\langle\tau_{d_1}\tau_{d_2}\cdots\tau_{d_n}\rangle_g(\mathbf{s};y_1)\\
				:=&\int_{\overline{\mathcal{M}}_{g,n}}y_1^{-\frac{2d_1+1}{3}}\psi_{1}^{d_1}y_1^{-\frac{2d_2+1}{3}}\psi_{2}^{d_2}\cdots y_1^{-\frac{2d_n+1}{3}}\psi_{n}^{d_n}\exp(\sum_{i=1}^{\infty} y_1^{-\frac{2i}{3}}\kappa_i s_i)\\
				=&\langle \hat{\tau}_{d_1}\hat{\tau}_{d_2}\cdots\hat{\tau}_{d_n}\exp(\sum_{j=1}^{\infty} s_j\hat{\kappa}_j)\rangle_{g}.
			\end{aligned}			
		\end{equation}
		Using the transformation (\ref{transform from s to p}) and relation (\ref{shift with kappa class in tau class}), the correlators can be expressed as
		\begin{equation}\label{rep as hat tau}
			\langle \tau_{d_1}\tau_{d_2}\cdots\tau_{d_n}\rangle_{g}(\mathbf{s};y_1)=\langle \hat{\tau}_{d_1}\hat{\tau}_{d_2}\cdots\hat{\tau}_{d_n}\exp(y_1\sum_{j=2}^{\infty} p_j\hat{\tau}_j)\rangle_{g}.
		\end{equation}
		By the selection rules (\ref{selection ruler for tau}) and (\ref{selection ruler of high weil-petersson}), we have
		\begin{equation}\label{selection ruler for correlator}
			\langle\tau_{d_1}\tau_{d_2}\cdots\tau_{d_n}\rangle_g(\mathbf{s};y_1)=y_1^{-(2g-2+n)}\langle \tau_{d_1}\tau_{d_2}\cdots\tau_{d_n}\rangle_{g}(\mathbf{s}).
		\end{equation}
		Since all the topological recursions (\ref{def of EO TR}), (\ref{recursion relation for EO TR}), (\ref{recursion relation for high weil-petersson}) depend recursively on the Euler characteristic $2g-2+n$, the recursion relation for the dilated correlators is obtained by multiplying the left-hand side of (\ref{recursion relation for high weil-petersson}) by an additional factor $y_1$:
		\begin{equation}
			\begin{aligned}
				&y_1\langle \tau_{m_0}\tau_{m_1}\cdots\tau_{m_n}\rangle_{g}(\mathbf{s};y_1)\\
				=&\sum_{i=1}^{n}\sum_{\mathbf{b}}\beta_\mathbf{b}\mathbf{s}^\mathbf{b}\frac{(2(m_0+m_i+|\mathbf{b}|)-1)!!}{(2m_0+1)!!(2m_i-1)!!}\left\langle \tau_{m_0+m_i-1+|\mathbf{b}|}\prod_{\substack{j=1\\ j\ne i}}^n\tau_{m_j}\right\rangle_{g}(\mathbf{s};y_1)\\
				+&\frac{1}{2}\sum_{\mathbf{b}}\beta_\mathbf{b}\mathbf{s}^\mathbf{b}\sum_{k_1+k_2=m_0-2+|\mathbf{b}|}\frac{(2k_1+1)!!(2k_2+1)!!}{(2m_0+1)!!}\Big[ \langle\tau_{k_1}\tau_{k_2}\prod_{j=1}^{n}\tau_{m_j}\rangle_{g-1}(\mathbf{s};y_1)\\
				+&\sum_{\substack{g_1+g_2=g\\I_1\sqcup I_2=[n]}}^{s}\langle\tau_{k_1}\prod_{j\in I_1 }\tau_{m_j}\rangle_{g_1}(\mathbf{s};y_1)\langle\tau_{k_2}\prod_{j\in I_2}\tau_{m_j}\rangle_{g_2}(\mathbf{s};y_1)
				\Big],
			\end{aligned}
		\end{equation}
		with the initial values
		\begin{equation}
			\langle\tau_0\tau_0\tau_0\rangle_0(\mathbf{s};y_1)=y_1^{-1},\quad \langle\tau_1\rangle_1(\mathbf{s};y_1)=y_{1}^{-1}\frac{1}{24}, \quad \langle\tau_0\rangle_1(\mathbf{s};y_1)=y_1^{-1}\frac{s_1}{24}.
		\end{equation}
		We also consider the generating function
		\begin{equation}\label{relation coupled with kappa and dilated with y}
			\begin{aligned}
				F_{\mathrm{WP}}(\mathbf{t};\mathbf{s};y_1):=&\sum_{g=0}^{\infty}\sum_{\mathbf{n}}\langle \tau_0^{n_0}\tau_{1}^{n_1}\cdots\rangle_{g}(\mathbf{s};y_1)\frac{\mathbf{t}^\mathbf{n}}{\mathbf{n}!}
				=\sum_{g=0}^{\infty} \langle \exp(\sum_{d=0}^{\infty}t_d\tau_d)\rangle_g(\mathbf{s};y_1)\\
				=&\sum_{g=0}^{\infty} \langle \exp(t_0\hat{\tau}_0+t_1\hat{\tau}_1+\sum_{d=2}^{\infty} (t_d+y_1p_d)\hat{\tau}_d)\rangle_{g}\\
				=&F_{\mathrm{WK}}(\hat{t}_0,\hat{t}_1,\hat{t}_2+\hat{p}_{2},\hat{t}_3+\hat{p}_3,\cdots)=F_{\mathrm{WK}}(\hat{\mathbf{t}}+\hat{\mathbf{p}}),
			\end{aligned}
		\end{equation} 
		where $\hat{t}_n=y_1^{-\frac{2n+1}{3}}t_n$ and $\hat{p}_n=y_1^{-\frac{2(n-1)}{3}}p_n$.
		
		By the dilaton equation for the Witten-Kontsevich free energy,
		\begin{equation}
			\frac{\partial}{\partial t_1} F_{\mathrm{WK}}(\mathbf{t})=\sum_{n=0}^{\infty} \frac{2n+1}{3}t_n\frac{\partial}{\partial t_n}F_{\mathrm{WK}}(\mathbf{t})+\frac{1}{24},
		\end{equation}
		one obtains (see \cite{zhouEmergentGeometryUnifications2025}):
		\begin{equation}
			F(t_0,t_1+q,t_2,\cdots)=F(\frac{t_0}{(1-q)^{\frac{1}{3}}},\frac{t_1}{(1-q)^{\frac{3}{3}}},\frac{t_2}{(1-q)^{\frac{5}{3}}},\cdots)+\frac{1}{24}\log \frac{1}{1-q}.
		\end{equation}
		As a result
		\begin{equation}\label{relation coupled with kappa and dilated with y 2}
			\begin{aligned}
				&F_{\mathrm{WP}}(\mathbf t_{\mathrm{WP}};\mathbf{s};y_1)
				=F_{\mathrm{WK}}(q_{\mathrm{WP},0},q_{\mathrm{WP},1}+1,q_{\mathrm{WP},2},\cdots)+\frac{1}{24}\log y_1\\
				&=F_{\mathrm{WK}}(t_{\mathrm{WP},0},t_{\mathrm{WP},1}+1-y_1,t_{\mathrm{WP},2}+y_1p_2,t_{\mathrm{WP},3}+y_1p_3,\cdots)+\frac{1}{24}\log y_1.
			\end{aligned}
		\end{equation}
		Hence we define the dilaton-shift variables of $\mathbf t_{\mathrm{WP}}$ by  \(q_{\mathrm{WP},0}=t_{\mathrm{WP},0}\), $q_{\mathrm{WP},1}=t_{\mathrm{WP},1}-y_1$ and for \(m\geq2\),
		\begin{equation}\label{dilaton shift of WP}
			q_{\mathrm{WP},m}=t_{\mathrm{WP},m}+y_1p_m=t_{\mathrm{WP},m}-(2m-1)!!y_{2m-1}.
		\end{equation}
		
		The partition function $Z_{\mathrm{WP}}=\exp F_{\mathrm{WP}}$ is annihilated and determined by the operators:
		\begin{equation}
			\hat{L}_n(\mathbf{t},\partial;\mathbf{s};y_1)Z_{\mathrm{WP}}(\mathbf{t};\mathbf{s};y_1)=0,
		\end{equation}
		where
		\begin{equation}
			\begin{aligned}
				&\hat{L}_{n}(\mathbf{t},\partial;\mathbf{s};y_1):=y_1^{-\frac{2n}{3}}\bar{L}_{n}(\hat{\mathbf{t}},\partial;\hat{\mathbf{s}})\\
				&=-(2n+3)!!y_1\frac{\partial}{\partial t_{n+1}}+\sum_{\mathbf{b}}\beta_\mathbf{b}\mathbf{s}^{\mathbf{b}} \sum_{j=0}^{\infty}\frac{(2(j+n+|\mathbf{b}|)+1)!!}{(2j-1)!!}t_j\frac{\partial}{\partial t_{j+n+|\mathbf{b}|}}\\
				&+\frac{1}{2}\sum_{\mathbf{b}}\beta_\mathbf{b}\mathbf{s}^\mathbf{b}\sum_{d_1+d_2=n-1+|\mathbf{b}|}(2d_1+1)!!(2d_2+1)!!\frac{\partial^2}{\partial t_{d_1}\partial t_{d_2}}\\
				&+\delta_{n,-1}(\frac{t_0^2}{2}+\frac{s_1}{24})+\frac{\delta_{n,0}}{8}  \qquad n\geq-1.
			\end{aligned}
		\end{equation}
		After the transformation (\ref{transform of Virasoro to shift}), 
		\begin{equation}
			L_n^{\mathrm{WP}}=\sum_{\mathbf{L}}\gamma_{\mathbf{L}}\mathbf{s}^\mathbf{L}\cdot \hat{L}_{n+|\mathbf{L}|},
		\end{equation}
		one obtains
		\begin{equation}
			\begin{aligned}
				L_n^{\mathrm{WP}}(\mathbf t_{\mathrm{WP}},\partial;\mathbf{s};y_1)=L_n^{\mathrm{WK}}(\mathbf q_{\mathrm{WP}},\partial).
			\end{aligned}
		\end{equation}
		It coincides with 
		\begin{equation}
			Z_{\mathrm{WP}}(\mathbf{t}_{\mathrm{WP}};\mathbf{s};y_1)=y_1^{\frac{1}{24}}Z_{\mathrm{WK}}(\mathbf q_{\mathrm{WP}}).
		\end{equation}

		\subsection{Induced Virasoro constraints} 
		To relate the intersection numbers on moduli spaces to the correlators of the Eynard-Orantin topological recursion, we introduce additional formal variables $\widehat B^{i,j}_{k,l}$ for $i, j = 1, 2, \dots, N$ and $k, l = 0, 1, 2, \dots$. We require them to be symmetric, i.e., $\widehat B^{i,j}_{k,l} = \widehat B^{j,i}_{l,k}$. 
		
		Consider several copies of the partition function $Z_{\mathrm{WP}}(\mathbf t^{i}_{\mathrm{TR}}; \mathbf s^{i}; y^{i}_1)$ defined in the previous subsection for $i=1,2,\cdots,N$, and define the \textbf{local glued partition function} by the action of a unitary operator:
		\begin{equation}\label{def of local glued partition function}
			\begin{aligned}
				Z_{\mathrm{loc}}(\mathbf t_{\mathrm{TR}};\mathbf{s};\widehat B;\mathbf{y}):=&\exp(\frac{1}{2}\sum_{i,j=1}^{N}\sum_{k,l\ge0}\widehat B^{i,j}_{k,l}\frac{\partial^2}{\partial t^{i}_{\mathrm{TR},k}\partial t^{j}_{\mathrm{TR},l}})\prod_{m=1}^{N}Z_{\mathrm{WP}}(\mathbf t^{m}_{\mathrm{TR}};\mathbf s^{m};y^{m}_1)\\
				=& e^{D(\widehat B)}\prod_{m=1}^{N}(y^{m}_1)^{\frac{1}{24}}Z_{\mathrm{WK}}(\mathbf q^{m}_{\mathrm{TR}}),
			\end{aligned}
		\end{equation}
		where we write $\mathbf t^{m}_{\mathrm{TR}}=(t^{m}_{\mathrm{TR},0},t^{m}_{\mathrm{TR},1},\cdots)$, $\mathbf s^{m}=(s^{m}_{1},s^{m}_{2},\cdots)$ and $m=1,2,\cdots,N$, $\mathbf{q}^m_{\mathrm{TR}}$ is the dilaton shift of $\mathbf{t}^m_{\mathrm{TR}}$ by (\ref{dilaton shift of WP}); we denote this unitary operator by
		\begin{equation}
			e^{D(\widehat B)}=\exp(\frac{1}{2}\sum_{i,j=1}^{N}\sum_{k,l\ge0}\widehat B^{i,j}_{k,l}\frac{\partial^2}{\partial t^{i}_{\mathrm{TR},k}\partial t^{j}_{\mathrm{TR},l}}).
		\end{equation}
		
		Of course, the partition function $Z_{\mathrm{loc}}(\mathbf t_{\mathrm{TR}};\mathbf{s};\widehat B;\mathbf{y})$ is annihilated and determined by a sequence of operators
		\begin{equation}
			\hat{L}_{i,n}=\hat{L}_{n}(\mathbf t^{i}_{\mathrm{TR}},\partial;\mathbf s^{i};\widehat B;y^{i}_1):=e^{D(\widehat B)}\hat{L}_{n}(\mathbf t^{i}_{\mathrm{TR}},\partial;\mathbf s^{i};y^{i}_1)e^{-D(\widehat B)},
		\end{equation}
		for $i=1,2,\cdots,N$ and $n\geq-1$.
		
		These operators can be computed using the Baker-Campbell-Hausdorff (BCH) formula:
		\begin{equation}
			e^{D}Le^{-D}=L+[D,L]+\frac{1}{2}[D,[D,L]]+\cdots=\sum_{m=0}^{\infty} \frac{1}{m!}\text{ad}(D)^{m}(L),
		\end{equation}
		where $\text{ad}(D)$ is the adjoint action.

		Since $\text{ad}(D)^k(L)=0$ for $k\geq3$, we obtain
		\begin{equation}
			\begin{aligned}
				&\hat{L}_{i,m}=e^{D(\widehat B)}\hat{L}_{m}(\mathbf t^{i}_{\mathrm{TR}},\partial;\mathbf s^{i};y^{i}_1)e^{-D(\widehat B)}\\
				=&\hat{L}_{m}(\mathbf t^{i}_{\mathrm{TR}},\partial;\mathbf s^{i};y^{i}_1)+[D(\widehat B),\hat{L}_m]+\frac{1}{2}[D(\widehat B),[D(\widehat B),\hat{L}_m]]\\
				=&-(2m+3)!!y^{i}_1\frac{\partial}{\partial t^i_{\mathrm{TR},m+1}}+\sum_{\mathbf{b}}\beta_\mathbf{b}(\mathbf s^{i})^{\mathbf{b}} \sum_{j=0}^{\infty}\frac{(2(j+m+|\mathbf{b}|)+1)!!}{(2j-1)!!}t^{i}_{\mathrm{TR},j}\frac{\partial}{\partial t^i_{\mathrm{TR},j+m+|\mathbf{b}|}}\\
				&+\frac{1}{2}\sum_{\mathbf{b}}\beta_\mathbf{b}(\mathbf s^{i})^\mathbf{b}\sum_{d_1+d_2=m-1+|\mathbf{b}|}(2d_1+1)!!(2d_2+1)!!\frac{\partial^2}{\partial t^{i}_{\mathrm{TR},d_1}\partial t^{i}_{\mathrm{TR},d_2}}\\
				&+\sum_{\mathbf{b}}\beta_{\mathbf{b}}(\mathbf s^{i})^{\mathbf{b}}\sum_{j=0}^{\infty}\sum_{a=1}^N\sum_{k\ge0}\widehat B^{a,i}_{k,j}\frac{(2(j+m+|\mathbf{b}|)+1)!!}{(2j-1)!!}\frac{\partial^2}{\partial t^{a}_{\mathrm{TR},k}\partial t^i_{\mathrm{TR},j+m+|\mathbf{b}|}}\\
				&+\delta_{m,-1}\Big(\frac{1}{2}\widehat B^{i,i}_{0,0}+\sum_{(a,k)}\widehat B^{a,i}_{k,0}t^{i}_{\mathrm{TR},0}\frac{\partial}{\partial t^{a}_{\mathrm{TR},k}}+\frac{1}{2} \sum_{(a,k),(b,l)}\widehat B^{a,i}_{k,0}\widehat B^{b,i}_{l,0}\frac{\partial^2}{\partial t^{a}_{\mathrm{TR},k}\partial t^{b}_{\mathrm{TR},l}}     \Big)\\
				&+\delta_{m,-1}(\frac{(t^{i}_{\mathrm{TR},0})^2}{2}+\frac{s^{i}_{1}}{24})+\frac{\delta_{m,0}}{8}.
			\end{aligned}
		\end{equation}
		Moreover, if we define
		\begin{equation}
			L_{m}^{\mathrm{TR}}(\mathbf t_{\mathrm{TR}}^i,\partial):=\sum_{\mathbf{L}}\gamma_\mathbf{L}(\mathbf s^{i})^\mathbf{L}\cdot \hat{L}_{i,m+|\mathbf{L}|},
		\end{equation}
		then these operators can be written as 
		\begin{equation}
			\begin{aligned}
				&L_{m}^{\mathrm{TR}}(\mathbf t_{\mathrm{TR}}^i,\partial)=L_m^{\mathrm{WK}}(\mathbf q^{i}_{\mathrm{TR}},\partial)\\
				&+\sum_{j=0}^{\infty}\sum_{a=1}^N\sum_{k\ge0}\widehat B^{a,i}_{k,j}\frac{(2(j+m)+1)!!}{(2j-1)!!}\frac{\partial^2}{\partial t^{a}_{\mathrm{TR},k}\partial t^i_{\mathrm{TR},j+m}}\\
				&+\delta_{m,-1}\Big(\frac{1}{2}\widehat B^{i,i}_{0,0}+\sum_{(a,k)}\widehat B^{a,i}_{k,0}t^{i}_{\mathrm{TR},0}\frac{\partial}{\partial t^{a}_{\mathrm{TR},k}}+\frac{1}{2} \sum_{(a,k),(b,l)}\widehat B^{a,i}_{k,0}\widehat B^{b,i}_{l,0}\frac{\partial^2}{\partial t^{a}_{\mathrm{TR},k}\partial t^{b}_{\mathrm{TR},l}}     \Big).
			\end{aligned}
		\end{equation}
		They satisfy the Virasoro commutation relations.
		
		\begin{remark}
			These operators are indeed the Virasoro operators for the total ancestor partition function of a semisimple CohFT after the coordinate transformation. Alexandrov derived the $N$ families Virasoro operators for the total ancestor partition function in \cite{alexandrovCutandjoinOperatorsCohomological2024}. Guo and Zhang derived a familiy of summed Virasoro constraints for the semisimple CohFT and the topological recursion in \cite{guoCohomologicalFieldTheory2025,guoVirasoroConstraintsTopological2025}.
		\end{remark}
		
		Expand the local glued free energy of the local glued partition function $Z_{\mathrm{loc}}(\mathbf t_{\mathrm{TR}};\mathbf{s};\widehat B;\mathbf{y})=\exp F_{\mathrm{loc}}(\mathbf t_{\mathrm{TR}};\mathbf{s};\widehat B;\mathbf{y})$ as
		\begin{equation}
			\begin{aligned}
				F_{\mathrm{loc}}(\mathbf t_{\mathrm{TR}};\mathbf{s};\widehat B;\mathbf{y})=\sum_{g=0}^{\infty}\sum_{(i_j,m_j)} \langle \tau_{i_1,m_1}\tau_{i_2,m_2}\cdots\tau_{i_n,m_n}\rangle_{g}^{\mathrm{loc}}\frac{\prod_{j=1}^{n}t^{i_j}_{\mathrm{TR},m_j}}{n!}.
			\end{aligned}
		\end{equation}
		Then the Virasoro constraints give the recursion relations of the \textbf{TR correlators}:
		\begin{equation}\label{recursion relation of TR intersection numbers}
			\begin{aligned}
				&y^{i_0}_1(2m_0+1)!!\langle \tau_{i_0,m_0}\tau_{i_1,m_1}\cdots\tau_{i_n,m_n}\rangle_{g}^{\mathrm{loc}}\\
				=&\sum_{\{j|i_j=i_0,j\in [n]\}}\sum_{\mathbf{b}}\beta_\mathbf{b}(\mathbf s^{i_0})^\mathbf{b}\frac{(2(m_0+m_j+|\mathbf{b}|)-1)!!}{(2m_j-1)!!}\left\langle\tau_{i_0,m_0+m_j+|\mathbf{b}|-1}\prod_{\substack{k=1\\ k\ne j}}^n\tau_{i_k,m_k}\right\rangle_g^{\mathrm{loc}}\\
				&+\sum_{\{j|i_j=i_0,j\in [n]\}}\delta_{m_0,0}\delta_{m_j,0}\sum_{(i,m)}\widehat B^{i,i_0}_{m,0}\left\langle\tau_{i,m}\prod_{\substack{k=1\\ k\ne j}}^n\tau_{i_k,m_k}\right\rangle_{g}^{\mathrm{loc}}\\
				&+\frac{1}{2}\sum_{\mathbf{b}}\beta_\mathbf{b}(\mathbf s^{i_0})^\mathbf{b}\sum_{k_1+k_2=m_0+|\mathbf{b}|-2}(2k_1+1)!!(2k_2+1)!!\Big[\langle \tau_{i_0,k_1}\tau_{i_0,k_2}\prod_{j=1}^{n}\tau_{i_j,m_j}\rangle_{g-1}^{\mathrm{loc}}\\
				&+\sum_{\substack{g_1+g_2=g\\ I_1\sqcup I_2=[n]}}^{s}\langle \tau_{i_0,k_1}\prod_{j\in I_1}\tau_{i_j,m_j}\rangle_{g_1}^{\mathrm{loc}}\langle\tau_{i_0,k_2}\prod_{j\in I_2}\tau_{i_j,m_j}\rangle_{g_2}^{\mathrm{loc}}\Big]\\
				&+\sum_\mathbf{b} \ beta_\mathbf{b}(\mathbf s^{i_0})^\mathbf{b}\sum_{k_1}\sum_{i=1}^N\sum_{k_2\ge0}\frac{(2k_1+1)!!\widehat B^{i,i_0}_{k_2,k_1+1-|\mathbf{b}|-m_0}}{(2(k_1-m_0-|\mathbf{b}|)+1)!!}\Big[\langle \tau_{i_0,k_1}\tau_{i,k_2}\prod_{j=1}^{n}\tau_{i_j,m_j}\rangle_{g-1}^{\mathrm{loc}}\\
				&+\sum_{\substack{g_1+g_2=g\\ I_1\sqcup I_2=[n]}}^{s} \langle \tau_{i_0,k_1}\prod_{j\in I_1}\tau_{i_j,m_j}\rangle_{g_1}^{\mathrm{loc}}\langle\tau_{i,k_2}\prod_{j\in I_2 }\tau_{i_j,m_j}\rangle_{g_2}
				^{\mathrm{loc}}\Big]\\
				&+\frac{1}{2}\delta_{m_0,0}\sum_{(j_1,k_1)(j_2,k_2)}\widehat B^{j_1,i_0}_{k_1,0}\widehat B^{j_2,i_0}_{k_2,0}\Big[  \langle \tau_{j_1,k_1}\tau_{j_2,k_2}\prod_{l=1}^{n} \tau_{i_l,m_l}\rangle_{g-1}^{\mathrm{loc}}\\
				&+\sum_{\substack{g_1+g_2=g\\I_1\sqcup I_2=[n]}}^{s}\langle \tau_{j_1,k_1}\prod_{l\in I_1}\tau_{i_l,m_l}\rangle_{g_1}^{\mathrm{loc}}\langle\tau_{j_2,k_2}\prod_{l\in I_2}\tau_{i_l,m_l}\rangle_{g_2}^{\mathrm{loc}}
				\Big]+\text{initial values}.
			\end{aligned}
		\end{equation}
		The initial values are
		\begin{equation}
			\begin{aligned}
				&\langle\tau_{i,0}\tau_{i,0}\tau_{i,0}\rangle_{0}^{\mathrm{loc}}=(y^{i}_1)^{-1}\\
				&\langle \tau_{i,1}\rangle_1^{\mathrm{loc}}=\frac{1}{24}(y^{i}_1)^{-1}\quad \langle \tau_{i,0}\rangle_1^{\mathrm{loc}}=\frac{1}{24}(y^{i}_1)^{-1}s^{i}_{1}+\frac{1}{2}(y^{i}_1)^{-1}\widehat B^{i,i}_{0,0}.
			\end{aligned}
		\end{equation}

		\subsection{Description as graph summation of intersection numbers} \label{subsection Description as graph summation of intersection numbers}
		In this subsection, we give a description of the local glued partition function as a sum over decorated graphs weighted by intersection numbers on moduli spaces. This is similar to the graph description in \cite{dunin-barkowskiIdentificationGiventalFormula2014}.

		First, we define the data of decorated graphs.
By a decorated graph $\Gamma$, we mean a graph equipped
with the following:

\begin{itemize}[label=--,leftmargin=*,itemsep=0.3em]
  \item A set of vertices $V=V(\Gamma)$.

  \item A set of edges $E=E(\Gamma)$.
  For each edge $e\in E$, $e$ consists of two half-edges
  $e=(he_1,he_2)$ incident to two vertices
  (possibly the same vertex).
  Let $HE=HE(\Gamma)$ denote the set of all half-edges,
  and $HE(v)$ the set of half-edges incident to a vertex $v$.

  \item A set of leaves $L=L(\Gamma)$.
  For each vertex $v$, let $L(v)$ denote the set of leaves
  incident to $v$.
  Set $\mathfrak{n}(v)=|HE(v)|+|L(v)|$.

  \item A vertex coloring
  $\mathfrak{b}:V\to\{1,2,\dots,N\}$.
  The color of a leaf or half-edge is inherited from
  the vertex it is incident to, and is also denoted
  by $\mathfrak{b}$.

  \item A genus assignment
  $\mathfrak{g}:V\to\mathbb{Z}_{\geq 0}$.

  \item A degree assignment
  $\mathfrak{d}:L\sqcup HE\to\mathbb{Z}_{\geq 0}$
  for leaves and half-edges.

  \item The graph is called \textbf{stable} if every vertex
  is stable, i.e.,
  $2\mathfrak{g}(v)-2+\mathfrak{n}(v)>0$
  for all $v\in V$.
\end{itemize}

		The genus of a decorated graph $\Gamma$ is the sum of genus of each vertex and the genus of the underlying graph; that is,
		\begin{equation}
			g(\Gamma):=\sum_{v\in V}\mathfrak{g}(v)+|E|-|V|+1.
		\end{equation}

		Denote by $\operatorname{Aut}(\Gamma)$ the automorphism group of $\Gamma$, i.e., the group of permutations of vertices, half-edges, and leaves that preserve all labels and incidence relations.

		The weight of a stable decorated graph $\Gamma$ is defined as
		\begin{equation}
			w(\Gamma):=\prod_{e\in E(\Gamma)}\widehat B(e) \prod_{v\in V(\Gamma)} \int_{\overline{\mathcal{M}}_{\mathfrak{g}(v),\mathfrak{n}(v)}} \prod_{l\in L(v)}\psi_{l}^{\mathfrak{d}(l)}\prod_{he\in HE(v)} \psi_{he}^{\mathfrak{d}(he)},
		\end{equation}
		where for an edge $e = (he_1, he_2)$ with $he_i \in HE(v_i)$, we set
		\[
		\widehat B(e)=\widehat B^{\mathfrak b(v_1),\mathfrak b(v_2)}_{\mathfrak d(he_1),\mathfrak d(he_2)}.
		\]
		After coupling with the variables $\mathbf{t}$, the weight becomes
		\begin{equation}
			\begin{aligned}
				w(\Gamma;\mathbf{t}):=& \prod_{e\in E(\Gamma)}\widehat B(e) \prod_{v\in V(\Gamma)} \int_{\overline{\mathcal{M}}_{\mathfrak{g}(v),\mathfrak{n}(v)}} \prod_{l\in L(v)}\psi_{l}^{\mathfrak{d}(l)}t^{\mathfrak{b}(v)}_{\mathrm{TR},\mathfrak{d}(l)}\prod_{he\in HE(v)} \psi_{he}^{\mathfrak{d}(he)}\\
				=&w(\Gamma)\prod_{v\in V}\prod_{l\in L(v)} t^{\mathfrak{b}(v)}_{\mathrm{TR},\mathfrak{d}(l)}.
			\end{aligned}
		\end{equation}
		One can think of attaching an operator $\tau_{i,m}$ to each leaf and half-edge, and attaching $\widehat B(e)$ to each edge. Coupling with $\mathbf{t}$ corresponds to attaching $t^{i}_{\mathrm{TR},m}$ to the corresponding leaf.
		\begin{remark}
			We will sum over decorated graphs. The contribution of a specific graph is computed as follows. First, fix a topological configuration $|\Gamma|$ and an ordering of its vertices, leaves, and half-edges. Then assign labels to the vertices, leaves, and half-edges satisfying the required constraints. This produces a decorated graph $\Gamma$. The automorphism group $\operatorname{Aut}(\Gamma)$ consists of those permutations of vertices, leaves, and half-edges that preserve all labels and incidence relations. Two isomorphic decorated graphs count only once.

		\end{remark}

		We have the following lemma.
		\begin{lemma}
			For the product of several Witten-Kontsevich tau-functions, the partition function obtained by applying the operator $\exp(D(\widehat B))$ has the expression
			\begin{equation}
				\begin{aligned}
					Z(\mathbf{t};\mathbf{B}):=&\exp(\frac{1}{2}\sum_{(i,k),(j,l)}\widehat B^{i,j}_{k,l}\frac{\partial^2}{\partial t^{i}_{\mathrm{TR},k}\partial t^{j}_{\mathrm{TR},l}})\prod_{a=1}^{N}Z_{\mathrm{WK}}(\mathbf t^{a}_{\mathrm{TR}})\\
					=&\sum_{\Gamma}\frac{1}{|\operatorname{Aut}(\Gamma)|}w(\Gamma;\mathbf{t}),
				\end{aligned}
			\end{equation}
			where the sum runs over all stable decorated graphs $\Gamma$. The correlators of this partition function are given by
			\begin{equation}
				\langle\frac{\tau_{1,0}^{n_{1,0}}}{n_{1,0}!}\frac{\tau_{2,0}^{n_{2,0}}}{n_{2,0}!}\cdots\frac{\tau_{N,0}^{n_{N,0}}}{n_{N,0}!}\frac{\tau_{1,1}^{n_{1,1}}}{n_{1,1}!}\cdots\rangle_{g}^{c}=\sum\frac{1}{|\operatorname{Aut}(\Gamma)|}w(\Gamma),
			\end{equation}
			where the sum is over all connected stable decorated graphs $\Gamma$ of genus $g$ whose leaves correspond to the given operators.
		\end{lemma}
		\begin{proof}
			It is clear that the product of Witten-Kontsevich tau-functions can be expressed as a sum over stable graphs without edges:
			\begin{equation}
				\begin{aligned}
					\prod_{a=1}^{N}Z(\mathbf t^{a}_{\mathrm{TR}})=\sum_{\Gamma: \text{ no edge}} \frac{1}{|\operatorname{Aut}(\Gamma)|}w(\Gamma;\mathbf{t})\\
				\end{aligned}
			\end{equation}
			The action of the operator $\frac{1}{2}\widehat B^{i,j}_{k,l} \frac{\partial^2}{\partial t^{i}_{\mathrm{TR},k} \partial t^{j}_{\mathrm{TR},l}}$ on the weight of a graph coupled with $\mathbf{t}$ corresponds to joining two leaves (originally attached to $t^{i}_{\mathrm{TR},k}$ and $t^{j}_{\mathrm{TR},l}$) into an edge weighted by $\widehat B^{i,j}_{k,l}$. This glues leaves in all possible ways. When $(i,k) \neq (j,l)$, the symmetry $\widehat B^{i,j}_{k,l} = \widehat B^{j,i}_{l,k}$ absorbs the factor $\frac{1}{2}$. When $(i,k) = (j,l)$, the factor $\frac{1}{2}$ compensates for the ordering of the two half-edges with identical labels. 
			By exponentiating this operator, we glue leaves in all possible ways, thereby generating all graphs with internal edges.

			We now explain the combinatorial factor $\frac{1}{|\operatorname{Aut}(\Gamma)|}$ associated with a decorated graph $\Gamma$. Consider a graph $\Gamma$ formed by a set of edges $E$ and a set of vertices $V$ (each equipped with its incident leaves). Before gluing, the contribution from the edges would carry a factor $\frac{1}{|\operatorname{Aut}(E)|} = \prod_{e} \frac{1}{n_e!}$, where $n_e$ is the multiplicity of the edge type $e$. Similarly, the vertices contribute a factor $\frac{1}{|\operatorname{Aut}(V)|} = \prod_{v} \frac{1}{n_v!  |\operatorname{Aut}(v)|}$, where $n_v$ is the number of vertices of type $v$ and $\operatorname{Aut}(v)$ accounts for the symmetries of the leaves incident to $v$.
			
			When we glue these building blocks into a specific decorated graph $\Gamma$, the symmetries among identical vertices and leaves are partially broken. For instance, a collection of $n_v$ identical vertices may become distributed among several distinct structural roles within $\Gamma$, with $n_{v,1}, n_{v,2}, \dots$ vertices in each role and $\sum_i n_{v,i} = n_v$. The number of ways to assign the original identical vertices to these distinct roles is given by the multinomial coefficient
			\begin{equation*}
				\binom{n_{v}}{n_{v,1}\ n_{v,2}\ \cdots} =\frac{n_{v}!}{n_{v,1}!n_{v,2}!\cdots}.
			\end{equation*}
			A similar phenomenon occurs for the leaves: some leaves are chosen to be glued into edges, while others remain as external leaves. 
			
			The factor $\frac{1}{|\operatorname{Aut}(E)|}$ arising from the edges is canceled by the following mechanism. The gluing operator $\exp(D(\widehat B))$ generates all possible ways of pairing the leaves. When we fix a target decorated graph $\Gamma$, there are generally many different ways to choose which specific leaves are paired to form the edges of $\Gamma$. In fact, the number of such choices is exactly $|\operatorname{Aut}(E)|$, because any permutation of the edges that preserves their types corresponds to a different choice of leaf pairings that produces the same decorated graph $\Gamma$. Since the sum over all graphs counts each distinct decorated graph only once, the overcounting factor $|\operatorname{Aut}(E)|$ from the gluing process precisely cancels the $\frac{1}{|\operatorname{Aut}(E)|}$ prefactor.
			
			After accounting for all combinatorial choices of vertices, leaves, and edge pairings, the combined prefactor reduces to $\frac{1}{|\operatorname{Aut}(\Gamma)|}$.
			
			The expression for the correlators follows immediately by differentiating the free energy and setting $\mathbf{t} = 0$. It is a canonical result for the sum of graphs.
		\end{proof}

		For the local glued partition function that includes $\kappa$-classes and the dilaton shift $\mathbf{y}$, we consider the weight of a graph coupled with $\mathbf{t}$ and $\mathbf{s}$:
		\begin{equation}
			\begin{aligned}
				w(\Gamma;\mathbf{t};\mathbf{s})
				=&\prod_{e\in E(\Gamma)}\widehat B(e) \prod_{v\in V(\Gamma)} \int_{\overline{\mathcal{M}}_{\mathfrak{g}(v),\mathfrak{n}(v)}} \prod_{l\in L(v)}\psi_{l}^{\mathfrak{d}(l)}	t^{\mathfrak{b}(v)}_{\mathrm{TR},\mathfrak{d}(l)}	\\
				&\cdot\prod_{he\in HE(v)} \psi_{he}^{\mathfrak{d}(he)}\cdot \exp(\sum_{j=1}^{\infty} s^{\mathfrak{b}(v)}_j\kappa_{j}).
			\end{aligned}
		\end{equation}
		And the dilated version of it is
		\begin{equation}
			\begin{aligned}
				&\hat{w}(\Gamma;\mathbf{t};\mathbf{s})\\
				=&\prod_{e\in E(\Gamma)}\widehat B(e) \prod_{v\in V(\Gamma)} \int_{\overline{\mathcal{M}}_{\mathfrak{g}(v),\mathfrak{n}(v)}} \prod_{l\in L(v)}\hat{\psi}_{l}^{\mathfrak{d}(l)}	t^{\mathfrak{b}(v)}_{\mathrm{TR},\mathfrak{d}(l)}	\\
				&\cdot\prod_{he\in HE(v)} \hat{\psi}_{he}^{\mathfrak{d}(he)}\cdot \exp(\sum_{j=1}^{\infty} s^{\mathfrak{b}(v)}_j\hat{\kappa}_{j})\\
				=&w(\Gamma;\mathbf{t};\mathbf{s})\prod_{v\in V(\Gamma)} (y^{\mathfrak{b}(v)}_1)^{-(2\mathfrak{g}(v)-2+\mathfrak{n}(v))}.
			\end{aligned}
		\end{equation}
		
		We then have the following theorem.
		\begin{thm}\label{thm of local glued partition function}
			The local glued partition function defined in (\ref{def of local glued partition function}) admits the graph-sum expansion
			\begin{equation}
				\begin{aligned}
					Z_{\mathrm{loc}}(\mathbf t_{\mathrm{TR}};\mathbf{s};\widehat B;\mathbf{y}):=&\exp(\frac{1}{2}\sum_{i,j=1}^{N}\sum_{k,l\ge0}\widehat B^{i,j}_{k,l}\frac{\partial^2}{\partial t^{i}_{\mathrm{TR},k}\partial t^{j}_{\mathrm{TR},l}})\prod_{m=1}^{N}Z_{\mathrm{WP}}(\mathbf t^{m}_{\mathrm{TR}};\mathbf s^{m};y^{m}_1)\\
					=&\sum_{\Gamma}\frac{1}{|\operatorname{Aut}(\Gamma)|} \hat{w}(\Gamma;\mathbf{t};\mathbf{s}),
				\end{aligned}
			\end{equation}
			where the sum runs over all stable decorated graphs $\Gamma$. The correlators are given by
			\begin{equation}
				\langle\frac{\tau_{1,0}^{n_{1,0}}}{n_{1,0}!}\frac{\tau_{2,0}^{n_{2,0}}}{n_{2,0}!}\cdots\frac{\tau_{N,0}^{n_{N,0}}}{n_{N,0}!}\frac{\tau_{1,1}^{n_{1,1}}}{n_{1,1}!}\cdots\rangle_{g}^{\mathrm{loc}}=\sum\frac{1}{|\operatorname{Aut}(\Gamma)|}\hat{w}(\Gamma;\mathbf{s}),
			\end{equation}
			where the sum is over all connected stable decorated graphs $\Gamma$ of genus $g$ whose leaves correspond to the specified operators.
		\end{thm}

		We list the first few non-zero values; one can verify that they coincide with the recursion relation (\ref{recursion relation of TR intersection numbers}):
		\begin{equation}
			\begin{aligned}
				&\langle\tau_{1,0}^3\rangle_{0}^{\mathrm{loc}}=3!\cdot\frac{1}{3!}\cdot 
				\begin{tikzpicture}[baseline=(current bounding box.center)]
					\node [solid black dot, label=right:{(1,0)}] (A) at (0,0) {} ;
					\draw  (A)--(-1,0);
					\draw (A)--(0.5, 0.6);
					\draw (A)--(0.5,-0.6);
				\end{tikzpicture}
				=(y^{1}_1)^{-1},\\[10pt]
				&\langle \tau_{1,0}\rangle_{1}^{\mathrm{loc}}= \frac{1}{2} \cdot
				\begin{tikzpicture}[baseline=(current bounding box.center)]
					\node[solid black dot, label=above:{(1,0)}] (A) at (0,0) {};
					\draw (A)--(-1,0);
					\draw (A) to [loop right, looseness=70] (A);
				\end{tikzpicture}
				+
				\begin{tikzpicture}[baseline=(current bounding box.center)]
					\node [solid black dot, label=right:{(1,1)}] (A) at (0,0) {} ;
					\draw  (A)--(-1,0);
				\end{tikzpicture}
				=\frac{1}{24}(y^{1}_1)^{-1}s^{1}_{1}+\frac{1}{2}(y^{1}_1)^{-1}\widehat B^{1,1}_{0,0},\\[10pt]
				&\langle \tau_{1,1}\rangle_{1}^{\mathrm{loc}}=	
				\begin{tikzpicture}[baseline=(current bounding box.center)]
					\node [solid black dot, label=right:{(1,1)}] (A) at (0,0) {} ;
					\draw  (A)--(-1,0);
					\node  at (-0.5,0.3) {1};
				\end{tikzpicture}
				=\frac{1}{24}(y^{1}_1)^{-1},\\[10pt]
				&\langle \tau_{1,0}^4\rangle_{0}^{\mathrm{loc}}=4!
				\Big(
				\frac{1}{4!}
				\begin{tikzpicture}[baseline=(current bounding box.center)]
					\node [solid black dot, label=right:{(1,0)}] (A) at (0,0) {} ;
					\draw  (A)--(-0.5,-0.6);
					\draw (A)--(0.5, 0.6);
					\draw (A)--(0.5,-0.6);
					\draw (A)--(-0.5,0.6);
				\end{tikzpicture}
				+
				\frac{1}{8}\cdot
				\begin{tikzpicture}[baseline=(current bounding box.center)]
					\node[solid black dot, label=left:{(1,0)}] (A) at (0,0) {};
					\node[solid black dot, label=right:{(1,0)}] (B) at (1.1,0) {};
					\draw (A)--(B);
					\draw (A)--(-0.4,0.5);
					\draw(A) -- (-0.4,-0.5);
					\draw (B)--(1.5,0.6);
					\draw (B)--(1.5,-0.6);
				\end{tikzpicture}
				\Big)
				=(y^{1}_1)^{-2}s^{1}_{1}+3(y^{1}_1)^{-2}\widehat B^{1,1}_{0,0},\\[10pt]
				& \langle \tau_{1,0}^3\tau_{1,1}\rangle_{0}^{\mathrm{loc}}=3! \cdot\frac{1}{3!}\cdot
				\begin{tikzpicture}[baseline=(current bounding box.center)]
					\node [solid black dot, label=right:{(1,0)}] (A) at (0,0) {} ;
					\draw  (A)--(-0.5,-0.6);
					\draw (A)--(0.5, 0.6);
					\node at (0.3,0.6) {1};
					\draw (A)--(0.5,-0.6);
					\draw (A)--(-0.5,0.6);
				\end{tikzpicture}
				=(y^{1}_1)^{-2},\\[10pt]
				& \langle \tau_{1,0}^2\tau_{2,0}^2\rangle_{0}^{\mathrm{loc}}=2!2!\cdot\frac{1}{4}\cdot
				\begin{tikzpicture}[baseline=(current bounding box.center)]
					\node[solid black dot, label=left:{(1,0)}] (A) at (0,0) {};
					\node[solid black dot, label=right:{(2,0)}] (B) at (1.2,0) {};
					\draw (A)--(B);
					\draw (A)--(-0.4,0.5);
					\draw(A) -- (-0.4,-0.5);
					\draw (B)--(1.6,0.5);
					\draw (B)--(1.6,-0.5);
				\end{tikzpicture}
				=(y^{1}_1)^{-1}(y^{2}_1)^{-1}\widehat B^{2,1}_{0,0},\\[10pt]
				&\langle \tau_{1,0}\tau_{2,0}\rangle_1^{\mathrm{loc}}=\frac{1}{2}\cdot 
				\begin{tikzpicture}[baseline=(current bounding box.center)]
					\node [solid black dot, label=above:{(1,0)}] (A) at (0,0) {} ;
					\node [solid black dot, label=above:{(2,0)}] (B) at (1.5,0) {} ;
					\draw  (A)--(-1,0);
					\draw (B)--(2.5,0);
					\draw (A) to [bend right=30](B);
					\draw (A) to [bend left=30] (B);
				\end{tikzpicture}
				=\frac{1}{2}(y^{1}_1)^{-1}(y^{2}_1)^{-1}(\widehat B^{1,2}_{0,0})^2,\\[10pt]
				& \langle\tau_{1,0}^3\tau_{2,0}^2\rangle_{0}^{\mathrm{loc}}=3!2!\cdot \Big( 
				\frac{1}{3!2!}	\cdot 
				\begin{tikzpicture}[baseline=(current bounding box.center)]
					\node[solid black dot, label=left:{(1,0)}] (A) at (0,0) {};
					\node[solid black dot, label=right:{(2,0)}] (B) at (1,0) {};
					\draw (A)--(B);
					\draw (A)--(-0.4,0.5);
					\draw(A) -- (-0.4,-0.5);
					\draw (B)--(1.4,0.5);
					\draw (B)--(1.4,-0.5);
					\draw (A)--(0.4,0.5);
				\end{tikzpicture}
				+
				\frac{1}{3!2!}\cdot 
				\begin{tikzpicture}[baseline=(current bounding box.center)]
					\node[solid black dot, label=left:{(1,0)}] (A) at (0,0) {};
					\node[solid black dot, label=right:{(2,0)}] (B) at (1,0) {};
					\draw (A)--(B);
					\draw (A)--(-0.4,0.5);
					\draw(A) -- (-0.4,-0.5);
					\draw (B)--(1.4,0.5);
					\draw (B)--(1.4,-0.5);
					\draw (A)--(0.4,0.5);
					\node at (0.2,-0.2) {1};
				\end{tikzpicture}	\\[9pt]
				&+\frac{1}{4}\cdot
				\begin{tikzpicture}[baseline=(current bounding box.center)]
					\node[solid black dot, label=left:{(1,0)}] (A) at (0,0) {};
					\node[solid black dot, label=right:{(2,0)}] (B) at (1.6,0) {};
					\node[solid black dot, label=above:{(1,0)}] (C) at (0.8,0) {};
					\draw (A)--(B);
					\draw (A)--(-0.4,0.5);
					\draw(A) -- (-0.4,-0.5);
					\draw (B)--(1.6,0.5);
					\draw (B)--(1.6,-0.5);
					\draw (C)--(0.8,-0.5);
				\end{tikzpicture}
				\Big)
				=(y^{1}_1)^{-2}(y^{2}_1)^{-1}\big(\widehat B^{1,2}_{0,0}s^{1}_{1}+\widehat B^{1,2}_{1,0}+3\widehat B^{1,2}_{0,0}\widehat B^{1,1}_{0,0}\big),
			\end{aligned}
		\end{equation}
		where the factorial coefficients account for repetitions of $\tau$-operators, and the fractions correspond to the orders of automorphism groups. The vertex labels $(b,g)$ indicate the color and genus; we omit the degree labels on leaves and half-edges when they are zero.
		
		\begin{remark}
			In \cite{eynardInvariantsSpectralCurves2014}, Eynard introduced the \textit{combinatorial moduli space} $\overline{\mathcal{M}}^{\mathfrak{b}}_{g,n}$ to describe the TR correlators. The combinatorial moduli space is defined as
			\begin{equation}
				\overline{\mathcal{M}}^{\mathfrak{b}}_{g,n}=\{(\Sigma,p_1,p_2,\cdots,p_n;\sigma)\}/\sim
			\end{equation}
			where $(\Sigma,p_1,p_2,\cdots,p_n)$ is a Riemann surface with $n$ marked points and possibly some nodes. The spherical components with two nodes are allowed. The map $\sigma$ is a stable map from $\Sigma$ to the complete graph with $\mathfrak{b}$ vertices; equivalently, it assigns to each stable component of $\Sigma$ a color corresponding to a ramification point, with the property that contiguous stable components share the same color, meaning that different colors are linked via an unstable component. The local glued partition function and correlators can then be viewed as intersection numbers over $\overline{\mathcal{M}}_{g,n}^{\mathfrak{b}}$. This description is equivalent to the graph-sum formulation presented here.
		\end{remark}

		\vskip 20pt
		
		\section{The equivalence between Eynard-Orantin topological recursion and recursion for Virasoro constraints}\label{section The equivalence between Eynard-Orantin topological recursion and Virasoro constraints}
		In this section, we present our main result concerning the equivalence between the Eynard-Orantin topological recursion for a spectral curve and the Virasoro constraints for the intersection numbers discussed in the preceding sections.
		
		\subsection{Equivalence of two correlators}
		Comparing the recursion relation (\ref{recursion relation for EO TR}) for the correlators generated by the Eynard-Orantin topological recursion with the recursion relation (\ref{recursion relation of TR intersection numbers}) for the TR correlators arising from the Virasoro constraints on moduli spaces, we then obtain the following theorem.
		\begin{thm}\label{main thm: relation}
			For a spectral curve with several ramification points, the Eynard-Orantin topological recursion is equivalent to the recursions satisfied by the correlators of the local glued partition function. More precisely, for $2g-2+n>0$,
			\begin{equation}
				W_{g,n}(p_1,p_2,\cdots,p_n)
				=
				\sum_{(i_j,m_j)}
				\langle \tau_{i_1,m_1}\tau_{i_2,m_2}\cdots\tau_{i_n,m_n}\rangle_g^{\mathrm{loc}}
				\prod_{j=1}^{n}\d\xi^{i_j}_{m_j}(p_j).
			\end{equation}
		\end{thm}
		This also implies that the TR correlators can be non-zero only when
		\[
		\sum_{j=1}^n m_j\le 3g-3+n.
		\]
		
		We can also write this result as
		\begin{equation}
			Z_{\mathrm{loc}}(\mathbf t_{\mathrm{TR}})
			=
			\exp\left(
			\sum_{2g-2+n>0}
			\frac1{n!}
			W_{g,n}\big|_{\d\xi^i_m\to t^i_{\mathrm{TR},m}}
			\right).
		\end{equation}
		
		In terms of the local glued free energy, we have
		\begin{equation}
			\begin{aligned}
				F_{\mathrm{loc},g}(\mathbf t_{\mathrm{TR}};\mathbf s;\widehat B;\mathbf y)
				=&
				\sum_{(i_j,m_j)}
				\langle \tau_{i_1,m_1}\tau_{i_2,m_2}\cdots\tau_{i_n,m_n}\rangle_g^{\mathrm{loc}}
				\frac1{n!}\prod_{j=1}^n t^{i_j}_{\mathrm{TR},m_j}
				\\
				=&
				\sum_{n=1}^{\infty}
				\sum_{\substack{g(\Gamma)=g\\ |L(\Gamma)|=n}}
				\frac1{|\operatorname{Aut}(\Gamma)|}
				\widehat w(\Gamma;\mathbf t_{\mathrm{TR}};\mathbf s).
			\end{aligned}
		\end{equation}
		Consequently, the multi-differentials produced by the Eynard-Orantin topological recursion can be expressed as a sum over graphs:
		\begin{proposition}\label{proposition graph sum of W}
			The multi-differentials of the Eynard-Orantin topological recursion admit the graph-sum expansion
			\begin{equation}
				W_{g,n}(p_1,p_2,\cdots,p_n)
				=
				\sum_{\substack{g(\Gamma)=g\\ |L(\Gamma)|=n}}
				\sum_{\text{orderings of }L}
				\frac1{|\operatorname{Aut}(\Gamma)|}
				\widehat w(\Gamma;\d\xi^i_m(p_-);\mathbf s),
			\end{equation}
			where the leaves of a fixed graph are ordered by $1,2,\dots,n$, and for each leaf $l$ labeled by $k$, we attach the differential $\d\xi^{\mathfrak b(l)}_{\mathfrak d(l)}(p_k)$.
		\end{proposition}
		
		\subsection{Some examples}\label{special situations and examples}
		We now illustrate the equivalence for the three simple examples.
		
		\begin{example}[Airy curve]\label{Airy curve case}
			The simplest example of Eynard-Orantin topological recursion is the Airy curve:
			\begin{equation}\label{Airy curve}
				\Sigma=\mathbb C,\qquad
				x(z)=\frac12 z^2,\qquad
				y(z)=z,\qquad
				B(z_1,z_2)=\frac{\d z_1\d z_2}{(z_1-z_2)^2},
			\end{equation}
			which has a single ramification point at \(z=0\). The standard Bergman kernel on the sphere expands as
			\begin{equation}
				\begin{aligned}
					B(z_1,z_2)
					=&\frac{\d z_1\d z_2}{(z_1-z_2)^2}
					=
					\frac{\d z_1\d z_2}{z_1^2}
					\left(1-\frac{z_2}{z_1}\right)^{-2}
					\\
					=&
					\sum_{n=0}^{\infty}(n+1)\frac{\d z_1}{z_1^{n+2}}z_2^n\d z_2.
				\end{aligned}
			\end{equation}
			Thus
			\begin{equation}
				\d\xi_m(z)=(2m-1)!!\,\d\mathcal B_{2m}(z)
				=
				\frac{(2m+1)!!\,\d z}{z^{2m+2}}.
			\end{equation}
			Theorem~\ref{main thm: relation} then states that for the Airy curve (\ref{Airy curve}), the multi-differentials generated by the Eynard-Orantin topological recursion are given by
			\begin{equation}
				\begin{aligned}
					W_{g,n}(z_1,z_2,\cdots,z_n)
					=&
					\sum_{m_1,m_2,\cdots,m_n}
					\left\langle\tau_{m_1}\tau_{m_2}\cdots\tau_{m_n}\right\rangle_g
					\prod_{i=1}^n
					\frac{(2m_i+1)!!\,\d z_i}{z_i^{2m_i+2}}
					\\
					=&
					\sum_{m_i}
					\int_{\overline{\mathcal M}_{g,n}}
					\psi_1^{m_1}\psi_2^{m_2}\cdots\psi_n^{m_n}
					\prod_{i=1}^n
					\frac{(2m_i+1)!!\,\d z_i}{z_i^{2m_i+2}}.
				\end{aligned}
			\end{equation}
			This result has been established by several works \cite{eynardInvariantsAlgebraicCurves2007,zhouTopologicalRecursionsEynard2013}.
		\end{example}
		
		\begin{example}[Kontsevich curve]\label{Kontsevich curve case}
			The second example is the Kontsevich curve:
			\begin{equation}\label{Kontsevich curve}
				\Sigma=\mathbb C,\qquad
				x(z)=\frac12 z^2,\qquad
				y(z)=\sum_{m=0}^{\infty}y_mz^m,\qquad
				B(z_1,z_2)=\frac{\d z_1\d z_2}{(z_1-z_2)^2},
			\end{equation}
			again with a single ramification point at \(z=0\). The Bergman kernel is the same as for the Airy curve. The multi-differentials given by the Eynard-Orantin topological recursion are
			\begin{equation}
				\begin{aligned}
					&W_{g,n}(z_1,z_2,\cdots,z_n)
					\\
					=&
					y_1^{-(2g-2+n)}
					\sum_{m_1,m_2,\cdots,m_n}
					\left\langle
					\tau_{m_1}\tau_{m_2}\cdots\tau_{m_n}
					e^{\sum_{k=1}^{\infty}s_k\kappa_k}
					\right\rangle_g
					\prod_{i=1}^n
					\frac{(2m_i+1)!!\,\d z_i}{z_i^{2m_i+2}}
					\\
					=&
					y_1^{-(2g-2+n)}
					\sum_{m_i}
					\int_{\overline{\mathcal M}_{g,n}}
					\psi_1^{m_1}\psi_2^{m_2}\cdots\psi_n^{m_n}
					e^{\sum_{k=1}^{\infty}s_k\kappa_k}
					\prod_{i=1}^n
					\frac{(2m_i+1)!!\,\d z_i}{z_i^{2m_i+2}}.
				\end{aligned}
			\end{equation}
			It has been proven by several works \cite{eynardIntersectionNumbersSpectral2011,zhouTopologicalRecursionsEynard2013,zhouEmergentGeometryUnifications2025}.
		\end{example}
		
		\begin{example}[Deformed Bergman kernel]\label{deformation of Bergman kernel}
			The third example is the deformation of the Bergman kernel:
			\begin{equation}
				\begin{aligned}
					&\Sigma=\mathbb C,\qquad
					x(z)=\frac12 z^2,\qquad
					y(z)=z,\\
					&B(z_1,z_2)=\frac{\d z_1\d z_2}{(z_1-z_2)^2}
					+\sum_{k,l\ge0}B_{k,l}z_1^kz_2^l\d z_1\d z_2,
				\end{aligned}
			\end{equation}
			with a single ramification point at \(z=0\). Put
			\[
			\widehat B_{k,l}=(2k-1)!!(2l-1)!!B_{2k,2l}.
			\]
			In \cite{zhouTopologicalRecursionVirasoro2024}, the multi-differentials generated by deforming the Bergman kernel of the Airy curve are expressed as
			\begin{equation}\label{one ramification point W}
				\begin{aligned}
					&W_{g,n}(z_1,z_2,\cdots,z_n)
					\\
					=&
					\sum_{m_1,m_2,\cdots,m_n}
					\int_{\overline{\mathcal M}_{g,n}}
					\exp\left(
					\iota_*\frac12\sum_{k,l}\widehat B_{k,l}\psi^k(\psi')^l
					\right)
					\psi_1^{m_1}\psi_2^{m_2}\cdots\psi_n^{m_n}
					\prod_{i=1}^n\d\xi_{m_i}(z_i),
				\end{aligned}
			\end{equation}
			where \(\iota:\delta\to\overline{\mathcal M}_{g,n}\) is the inclusion map of the boundary divisor in the moduli space. More precisely,
			\begin{equation}\label{iota intersection numbers}
				\begin{aligned}
					&\int_{\overline{\mathcal M}_{g,n}}
					\iota_*\psi^k(\psi')^l
					\prod_{i=1}^n\psi_i^{m_i}
					\\
					=&
					\int_{\overline{\mathcal M}_{g-1,n+2}}
					\psi_{n+1}^k\psi_{n+2}^l
					\prod_{i=1}^n\psi_i^{m_i}
					\\
					&+
					\sum_{\substack{g_1+g_2=g\\ I_1\sqcup I_2=[n]}}^{\mathrm{s}}
					\int_{\overline{\mathcal M}_{g_1,|I_1|+1}}
					\psi_{|I_1|+1}^k\prod_{i\in I_1}\psi_i^{m_i}
					\int_{\overline{\mathcal M}_{g_2,|I_2|+1}}
					\psi_{|I_2|+1}^l\prod_{i\in I_2}\psi_i^{m_i}.
				\end{aligned}
			\end{equation}
			Indeed, in the description of the local glued partition function as a graph sum, if two linked vertices \(v_1\) and \(v_2\) of a graph \(\Gamma\) carry the same color \(\mathfrak b(v_1)=\mathfrak b(v_2)\), they can be contracted into a single vertex, and the intersection number on this new vertex is precisely given by the formula (\ref{iota intersection numbers}).
			
			As a consequence, if the spectral curve has only one ramification point, the correlators of the local glued partition function can be written as intersection numbers on moduli spaces with boundary divisors, as shown in (\ref{one ramification point W}). Further details can be found in \cite{zhouTopologicalRecursionVirasoro2024}.
		\end{example}
		
		\vskip 20pt

		\section{Topological recursion and cohomological field theory}\label{section Topological Recursion and Givental Quantization}
		In this section, under the compatibility and identification assumptions specified below, we compare Eynard--Orantin topological recursion with the ancestor partition function of a semisimple cohomological field theory. 
		This provides a Virasoro-theoretic proof of the result of Dunin-Barkowski, Orantin, Shadrin, and Spitz \cite{dunin-barkowskiIdentificationGiventalFormula2014}.
		Furthermore, keeping the Bergman-kernel/\(R\)-matrix data fixed, we show that varying \(y\) shifts the evaluation point of the corresponding ancestor potential.
		
		\subsection{Bergman kernel and $R$-matrix}
		We begin by discussing certain conditions on the Bergman kernel and deriving some consequences of these conditions.
		
		In \cite{eynardInvariantsSpectralCurves2014}, Eynard gives the following proposition concerning $\d\xi^i_m$ on a global compact spectral curve.
		
		\begin{proposition}[Global recursion for the $\d\xi$-basis]\label{prop:global-dxi-recursion}
			Let $\Sigma$ be a compact connected Riemann surface, and let
			$x:\Sigma\to\mathbb P^1$ be a nonconstant meromorphic function whose
			ramification points $c_1,\ldots,c_N$ are all simple. Let $B$ be the
			Bergman kernel normalized by $\mathcal{A}$-cycles, i.e.,
			\[
			\oint_{\mathcal A_\alpha} B(p,\cdot)=0,
			\qquad \alpha=1,\ldots,g.
			\]
			Choose the local Airy coordinate $z_i$ at $c_i$ such that
			\[
			x=x(c_i)+\frac{z_i^2}{2},
			\]
			and define $\d\xi_m^i$ as in \eqref{normalized dxi}. Then, for every
			$i\in\{1,\ldots,N\}$ and $m\geq 0$, one has the identity of
			meromorphic one-forms on $\Sigma$
			\begin{equation}\label{global condition of d xi}
				-\d\left(\frac{\d\xi_m^i}{\d x}\right)
				=
				\d\xi_{m+1}^i
				+
				\sum_{a=1}^N
				\widehat B_{0,m}^{a,i}\,\d\xi_0^a .
			\end{equation}
			Here $\d\xi_m^i/\d x$ denotes the quotient of the two meromorphic
			one-forms $\d\xi_m^i$ and $\d x$.
		\end{proposition}
		
		We now explain how this condition can be expressed in terms of the local expansion data.
		
		For the following local computation, let
		$\mathcal B_m^i(z_j)$ denote the local primitive of
		$\d\mathcal B_m^i$ whose Laurent expansion in $z_j$ has zero
		constant term. From \eqref{define of d B}, we have
		\begin{equation}
			\mathcal B^i_m(z_j)=-\delta^i_jz_{j}^{-m-1}+\sum_{l=0}^{\infty} B^{i,j}_{m,l}\frac{z_j^{l+1}}{l+1}.
		\end{equation}
		Then
		\begin{equation}
			\begin{aligned}
				-\frac{\d \mathcal B^i_m}{\d x(z_j)}
				=-\frac{\d \mathcal B^i_m(z_j)}{z_j\d z_j}
				=-\delta^i_j\frac{m+1}{z_j^{m+3}}-\sum_{l=0}^{\infty} B^{i,j}_{m,l}z_j^{l-1}.
			\end{aligned}
		\end{equation}
		Substituting the identities
		\begin{equation}
			\begin{aligned}
				&-\delta^i_j\frac{m+1}{z_j^{m+3}}
				=(m+1)\mathcal B^i_{m+2}
				-\sum_{l=0}^{\infty} \frac{m+1}{l+1}B^{i,j}_{m+2,l}z_j^{l+1},\\
				&-B^{i,j}_{m,0}z_j^{-1}
				= -\sum_{a=1}^{N} B^{i,a}_{m,0} \delta^a_jz_j^{-1}
				=\sum_{a=1}^{N}B^{i,a}_{m,0}
				\left(\mathcal B^a_0-\sum_{l=0}^{\infty} B^{a,j}_{0,l}\frac{z_j^{l+1}}{l+1} \right),
			\end{aligned}
		\end{equation}
		we obtain
		\begin{equation}
			\begin{aligned}
				&-\frac{\d \mathcal B^i_m}{\d x(z_j)}
				=(m+1)\mathcal B^i_{m+2}+\sum_{a=1}^{N}B^{i,a}_{m,0}\mathcal B^a_0-B^{i,j}_{m,1}\\
				&-\sum_{l=0}^{\infty}
				\left(
				(l+1) B^{i,j}_{m,l+2}
				+(m+1) B^{i,j}_{m+2,l}
				+\sum_{a=1}^{N} B^{a,i}_{0,m}B^{a,j}_{0,l}
				\right)
				\frac{z_j^{l+1}}{l+1}.
			\end{aligned}
		\end{equation}
		
		Applying the exterior differential to the preceding identity, we obtain
		\begin{equation}
			\begin{aligned}
				-\d\left(
				\frac{\d\mathcal B_m^i}{\d x(z_j)}
				\right)
				=(m+1)\d\mathcal B_{m+2}^i
				+\sum_{a=1}^N B^{i,a}_{m,0}\d\mathcal B_0^a -\sum_{l=0}^{\infty}
				Q^{i,j}_{m,l}z_j^l\d z_j ,
			\end{aligned}
		\end{equation}
		where
		\begin{equation}
			Q^{i,j}_{m,l}
			:=
			(l+1)B^{i,j}_{m,l+2}
			+(m+1)B^{i,j}_{m+2,l}
			+\sum_{a=1}^N B^{a,i}_{0,m}B^{a,j}_{0,l}.
		\end{equation}
		
		If $\Sigma$ is compact and $B$ is normalized by the
		$\mathcal A$-cycles, Proposition \ref{prop:global-dxi-recursion}
		implies
		\[
		Q^{i,j}_{2r,l}=0,
		\qquad
		i,j\in\{1,\ldots,N\},\quad r,l\geq0.
		\]
		Conversely, for local spectral-curve data, it is sufficient to impose
		the differential compatibility condition
		\begin{equation}\label{extra condition on B}
			(l+1)B^{i,j}_{2r,l+2}
			+(2r+1)B^{i,j}_{2r+2,l}
			+\sum_{a=1}^N
			B^{a,i}_{0,2r}B^{a,j}_{0,l}
			=0,
		\end{equation}
		for all $i,j\in\{1,\ldots,N\}$ and $r,l\geq0$. After multiplying the resulting differential identity by
		$(2r-1)!!$, this yields the local formal version of
		\eqref{global condition of d xi}.
		
		Iterating \eqref{global condition of d xi}, one obtains
		\begin{equation}\label{differential basis change}
			\d\xi_m^i
			=
			\sum_{t=0}^m\sum_{a=1}^N
			\left(
			-\widehat B^{a,i}_{0,m-t-1}
			\right)
			(-1)^t\d\xi_{(t)}^a,
		\end{equation}
		where
		\begin{equation}\label{convention on B}
			\widehat B^{a,i}_{0,-1}:=-\delta_a^i,\quad \d \xi^i_{(k)}:=\d \left(\frac{d}{dx}\right)^{k} \xi^i_0.
		\end{equation}
		
		Following \cite{dunin-barkowskiIdentificationGiventalFormula2014}, define a matrix-valued formal series $R(w^{-1})$ by
		\begin{equation}
			\begin{aligned}
				R(-w^{-1})^i{}_j
				=&(R_0)^i{}_j -(R_1)^i{}_jw^{-1}+(R_2)^i{}_jw^{-2}+\cdots\\
				:=&-\frac{1}{\sqrt{2\pi w}}\int_{-\infty}^{\infty}
				\frac{B(z_i,z_j)}{\d z_i}\bigg|_{z_i=0}
				e^{(x(c_j)-x(z_j))w}.
			\end{aligned}
		\end{equation}
		Substituting the expansion (\ref{expansion of B}) into this definition, we obtain
		\begin{equation}\label{definition of R}
			\begin{aligned}
				R(-w^{-1})^i{}_j
				=&-\frac{1}{\sqrt{2\pi w}}\int_{-\infty}^{\infty}
				\left(\frac{\delta^i_j}{z_j^2}+\sum_{l=0}^{\infty}B^{i,j}_{0,l}z_j^l\right)
				e^{-\frac{w}{2}z_j^2}\d z_j\\
				=&\delta^i_j-\sum_{l=0}^{\infty} B^{i,j}_{0,2l} (2l-1)!!w^{-l-1}\\
				=&\delta^i_j-\sum_{l=0}^{\infty} \widehat B^{i,j}_{0,l}w^{-l-1}
				=\sum_{l=0}^{\infty} -\widehat B^{i,j}_{0,l-1}w^{-l},
			\end{aligned}
		\end{equation}
		where we have used the Gaussian integral (\ref{Gauss integral}) in the second line, and for the first term, it is the regularized integral defined by \cite{eynardIntersectionNumbersSpectral2011,dunin-barkowskiIdentificationGiventalFormula2014}
		\begin{equation}
			\int_{-\infty}^{\infty} \frac{1}{z^2}e^{-\frac{\zeta}{2}z^2}\d z:=-\sqrt{2\pi\zeta}.
		\end{equation}
		We also consider the double Laplace-type transform of the Bergman kernel:
		\begin{equation}
			\begin{aligned}
				&-\frac{\sqrt{w_1w_2}}{2\pi }
				\int_{-\infty}^{\infty}\int_{-\infty}^{\infty}
				B(z_i,z_j) e^{w_1(x(c_i)-x(z_i))+w_2(x(c_j)-x(z_j))}\\
				&=\frac{\delta_{ij}}{w_1^{-1}+w_2^{-1}}-\sum_{k,l=0}^{\infty} \widehat B^{i,j}_{k,l}w_{1}^{-k}w_2^{-l},
			\end{aligned}
		\end{equation}
		where we have again used the regularized integral defined by the below for the first term of the Bergman kernel \cite{eynardIntersectionNumbersSpectral2011,dunin-barkowskiIdentificationGiventalFormula2014}
		\begin{equation}
			\int_{-\infty}^\infty\int_{-\infty}^\infty
			\frac{e^{-\frac{\zeta_1}{2}z_1^2-\frac{\zeta_2}{2}z_2^2}}{(z_1-z_2)^2}\d z_1\d z_2
			:=-\frac{2\pi\sqrt{\zeta_1\zeta_2}}{\zeta_1+\zeta_2}.
		\end{equation}
		Assuming condition (\ref{extra condition on B}), one can show that
		\begin{equation}\label{double product formula}
			\sum_{k,l=0}^{\infty} (-1)^{k+l}\widehat B^{i,j}_{k,l}z^kw^l
			=
			\frac{\sum_{a=1}^{N}R(z)^a{}_iR(w)^a{}_j-\delta_{ij}}{z+w}.
		\end{equation}
		This relation has already appeared in \cite{eynardIntersectionNumbersSpectral2011,eynardInvariantsSpectralCurves2014}.
		Indeed, this is equivalent to the identity
		\begin{equation}\label{condition on B 1}
			\widehat B^{i,j}_{k,l+1}+\widehat B^{i,j}_{k+1,l}
			+\sum_{a=1}^{N} \widehat B^{a,i}_{0,k}\widehat B^{a,j}_{0,l}=0,
		\end{equation}
		which is precisely the normalized even part of
		\eqref{extra condition on B}.
		
		Viewing $R(w)$ as a matrix-valued series, and using $*$ to denote transposition, equation (\ref{double product formula}) implies
		\begin{equation}
			R(-w)^{*}R(w)=\mathrm{Id},
		\end{equation}
		i.e., $R(w)^{-1}=R(-w)^{*}$.
		
		\subsection{Base change of multi-differentials}
		In this subsection, we consider a change of basis for the multi-differentials and the corresponding transformation of the partition function.
		
		Recall that the multi-differentials can be written as:
		\begin{equation*}
			W_{g,n}(p_1,p_2,\cdots,p_n)
			=
			\sum_{(i_j,m_j)}
			\langle \tau_{i_1,m_1}\tau_{i_2,m_2}\cdots\tau_{i_n,m_n}\rangle_{g}^{\mathrm{loc}}
			\prod_{j=1}^{n} \d \xi^{i_j}_{m_j}(p_j).
		\end{equation*}
		We may view the set $\{\d\xi^i_m\}$ as a basis of one-forms used to expand the multidifferentials and perform a linear transformation
		\begin{equation}
			\d \xi^i_m=\sum_{j,l}C^{i,j}_{m,l}\d \zeta^j_l,
		\end{equation}
		with some constants $C^{i,j}_{m,l}$. In the new basis, the multi-differentials become
		\begin{equation}
			W_{g,n}=
			\sum_{(j_1,l_1),\cdots,(j_n,l_n)}
			\left\langle
			\prod_{k=1}^{n} \sum_{i,m}C_{m,l_k}^{i,j_k} \tau_{i,m}
			\right\rangle_{g}^{\mathrm{loc}}
			\prod_{k=1}^{n}\d \zeta^{j_k}_{l_k}(p_k).
		\end{equation}
		The partition function generated by these new correlators is
		\begin{equation}
			Z_{\mathrm{new}}(\mathbf t_{\mathrm{new}})
			=
			\exp\left(
			\sum_{2g-2+n>0}\sum_{(j_k,l_k)}
			\left\langle
			\prod_{k=1}^{n} \sum_{i,m}C_{m,l_k}^{i,j_k} \tau_{i,m}
			\right\rangle_{g}^{\mathrm{loc}}
			\frac{1}{n!}\prod_{k=1}^{n}t^{j_k}_{\mathrm{new},l_k}
			\right).
		\end{equation}
		This can be viewed as the transformation
		\begin{equation}
			t^i_{\mathrm{TR},m}=\sum_{j,l} C_{m,l}^{i,j} t^j_{\mathrm{new},l}
		\end{equation}
		applied to the local glued partition function, i.e.,
		\begin{equation}\label{def of new partition}
			Z_{\mathrm{new}}(\mathbf t_{\mathrm{new}})
			=
			Z_{\mathrm{loc}}(\mathbf t_{\mathrm{TR}})\big|_{\mathbf t_{\mathrm{TR}}=\mathbf t_{\mathrm{TR}}(\mathbf t_{\mathrm{new}})}.
		\end{equation}
		Equivalently, the new partition function can be expressed directly in terms of the multi-differentials as
		\begin{equation}
			Z_{\mathrm{new}}(\mathbf t_{\mathrm{new}})
			=
			\exp \sum_{2g-2+n>0}\frac{1}{n!}
			W_{g,n}\big|_{\d \zeta^i_m\to t^i_{\mathrm{new},m}}.
		\end{equation}
		We now consider the specific base change
		\begin{equation}
			\d \xi^i_m=
			\sum_{l=0}^{m}\sum_{a=1}^{N}
			\left(-\widehat B^{a,i}_{0,m-l-1}\right)\cdot
			\left((-1)^{l}\d \xi^a_{(l)}\right),
		\end{equation}
		i.e. we set $\d \zeta^a_l=(-1)^l\d \xi^a_{(l)}$. 
		Denote by $t_{\mathrm{CA}}$ the new variables. The corresponding transformation of $\mathbf t$-variables is
		\begin{equation}\label{transform of t}
			t^i_{\mathrm{TR},m}
			=
			\sum_{l=0}^{m}\sum_{a=1}^{N}
			-\widehat B^{a,i}_{0,m-l-1}t^a_{\mathrm{CA},l}.
		\end{equation}
		Applying \eqref{def of new partition} to this coordinate transformation, we define the transformed topological-recursion partition function
		\begin{equation}
			Z_{\mathrm{CA}}(\mathbf t_{\mathrm{CA}})
			=
			Z_{\mathrm{loc}}
			\bigl(\mathbf t_{\mathrm{TR}}(\mathbf t_{\mathrm{CA}})\bigr).
		\end{equation}
		The notation \(Z_{\mathrm{CA}}\) anticipates its later identification with the canonical ancestor partition function.
		
		We now relate this change of basis to the matrix-valued series $R(w)$. From this point on, repeated Greek and Latin indices are summed according to the Einstein summation convention unless otherwise stated; Greek indices are raised and lowered by $\eta$ (flat metric of the Frobenius manifold), and Latin indices are raised and lowered by $\delta$. For a fixed $m$, we view
		$\mathbf t_{\mathrm{TR},m}=(t^1_{\mathrm{TR},m},t^2_{\mathrm{TR},m},\dots,t^N_{\mathrm{TR},m})$
		as an element of a vector space $V$ with orthonormal basis $(f_i)_{i=1}^N$, i.e.,
		\begin{equation}
			\mathbf t_{\mathrm{TR},m}=f_i t^i_{\mathrm{TR},m}.
		\end{equation}
		We equip $V$ with a symmetric bilinear form $(-, -)$ for which $(f_i)_{i=1}^N$ is an orthonormal basis. The coefficients $R_m$ of the matrix-valued series $R(z) = \sum_{m=0}^{\infty} R_m z^m$ define linear operators on $V$ by
		\begin{equation}
			R_m(f_j a^j)
			=
			f_i (R_m)^i{}_j a^j.
		\end{equation}
		Recall from (\ref{definition of R}) that
		\begin{equation}
			(R_m)^i{}_j=(-1)^{m-1}\widehat B^{i,j}_{0,m-1}.
		\end{equation}
		The transform (\ref{transform of t}) can be written in generating $z$-series form as
		\begin{equation}\label{transform of R on t}
			\begin{aligned}
				\mathbf t_{\mathrm{TR}}(z)
				:=&\sum_{m=0}^{\infty} \mathbf t_{\mathrm{TR},m} z^m
				=\sum_{m=0}^{\infty} f_i t^i_{\mathrm{TR},m}z^m\\
				=&\sum_{m=0}^{\infty}\sum_{l=0}^{m}
				-\widehat B^{a,i}_{0,m-l-1} t^a_{\mathrm{CA},l} f_i z^m\\
				=&\sum_{m=0}^{\infty} R_m^{*}(-z)^m\cdot
				\left( \sum_{l=0}^{\infty} f_a t^a_{\mathrm{CA},l}z^l \right)\\
				=&R(-z)^{*}\cdot \left(\sum_{l=0}^{\infty} \mathbf t_{\mathrm{CA},l} z^l\right)
				=R(z)^{-1}\cdot \mathbf t_{\mathrm{CA}}(z),
			\end{aligned}
		\end{equation}
		where
		$\mathbf t_{\mathrm{CA},l}=(t^1_{\mathrm{CA},l},t^2_{\mathrm{CA},l},\cdots,t^N_{\mathrm{CA},l})
		=f_a t^a_{\mathrm{CA},l}\in V$.
		
		\subsection{Compatibility between $B$ and $y$}
		In DOSS \cite{dunin-barkowskiIdentificationGiventalFormula2014}, the identification of topological recursion with the ancestor theory of a semisimple CohFT requires a compatibility condition between \(B\) and \(y\), which we now state.
		
		The compatibility condition can be written as
		\begin{equation}
			R(-w^{-1})^i{}_{a}y^a_1
			=
			\sqrt{\frac{w}{2\pi }}\int_{-\infty}^{\infty}
			e^{-w(x(z_i)-x(c_i))}\d y(z_i),
		\end{equation}
		or, in terms of the local expansion data,
		\begin{equation}\label{local compatible condition}
			\sum_{a=1}^{N}\widehat B^{a,i}_{0,l}y^a_1
			=p^i_{l+2}y^i_1=-(2l+3)!!y^i_{2l+3},
		\end{equation}
		for all $ i=1,2,\cdots,N$ and $l=0,1,2,\cdots$, where $p^i_{m}$ is related to $\mathbf{s}^i$ via (\ref{transform from s to p}).
		
		Assuming this condition and viewing $\mathbf y_1=(y^1_1,y^2_1,\cdots,y^N_1)$ as a vector in $V$, we obtain
		\begin{equation}
			\begin{aligned}
				R(z)^{-1}\cdot \mathbf y_1z
				=&R(z)^{-1}\cdot\left(f_a y^a_1 z\right)
				=\sum_{m=0}^{\infty}
				-\widehat B^{a,i}_{0,m-1}f_i y^a_1z^m\\
				=&f_iy^i_1z-\sum_{l=2}^{\infty}f_i p^i_ly^i_1z^{l}\\
				=&\sum_{n=1}^{\infty}(2n-1)!!\mathbf y_{2n-1}z^n=:\mathbf y(z),
			\end{aligned}
		\end{equation}
		where $\mathbf y_{2n-1}=f_a y^a_{2n-1}$.
		Combining this with (\ref{transform of R on t}),
		\begin{equation}\label{action of R on shift}
			\mathbf t_{\mathrm{TR}}(z)-\mathbf y(z)
			=
			R(z)^{-1}\cdot \left(\mathbf t_{\mathrm{CA}}(z)-\mathbf y_1z\right).
		\end{equation}
		In Subsection \ref{subsection CohFT}, we set $y^i_1=\psi_{i1}$ corresponding to the elements of transition matrix $\Psi$ of a semisimple Frobenius manifold. For fixed $B$ and hence fixed $R$-matrix, we denote the resulting distinguished choice by
		\begin{equation}\label{specific choice}
			\mathbf y_{B}(z)
			:=
			\sum_{i}\psi_{i1}f_iz
			-\sum_{i,a}\sum_{l=1}^{\infty}\widehat B^{a,i}_{0,l-1}\psi_{a1}f_iz^{l+1}
			=
			R^{-1}(z)\cdot  \sum_{i}\psi_{i1}f_iz.
		\end{equation}
		
		\subsection{Givental Quantization}\label{subsection Givental quantization}
		We now briefly recall the basic definitions of Givental quantization in this subsection.
		
		Let $V$ be a vector space of dimension $N$ equipped with a symmetric non-degenerate bilinear form $(-, -)$. Consider the space $\mathcal{V}$ of vector-valued Laurent series, i.e., $\mathcal{V} = V[[z, z^{-1}]]$. An element $f \in \mathcal{V}$ can be written as
		\begin{equation}\label{formal f}
			\begin{aligned}
				f(z)&=\cdots-p_2z^{-3}+p_1z^{-2}-p_0z^{-1}+q_0+q_1z+q_2z^2+\cdots\\
				&=\sum_{i=0}^{\infty} p_i(-z)^{-(i+1)}+\sum_{j=0}^{\infty} q_jz^j,
			\end{aligned}
		\end{equation}
		where $p_m=p^i_mf_i, q_m=q^i_mf_i$ are in $V$.
		The space $\mathcal{V}$ carries a symplectic bilinear form defined by
		\begin{equation}
			\Omega(f,g)=\frac{1}{2\pi i}\oint (f(-z),g(z)) \d z.
		\end{equation}
		Choose an orthonormal basis $\alpha^1, \alpha^2, \dots, \alpha^N$ of $V$ with respect to $(-, -)$, and write $q_m = \sum_{i=1}^{N} \alpha^i q^i_m$ and $p_m = \sum_{i=1}^{N} \alpha^i p^i_m$. Then the collections $\{q^i_m\}$ and $\{p^i_m\}$ form a Darboux coordinate system with respect to $\Omega(-, -)$.
		
		Given an infinitesimal symplectic transform $L$ of $\mathcal{V}$, i.e.,
		\begin{equation}
			\Omega(Lf,g)+\Omega(f,Lg)=0 \quad \forall f,g\in \mathcal{V},
		\end{equation}
		its Givental quantization $\widehat L$ is a linear differential operator of order $\leq 2$ defined by
		\begin{equation}
			\widehat L:=\frac{1}{2}\widehat{\Omega(Lf,f)},
		\end{equation}
		where $f(z)$ is expressed in the Darboux coordinates (\ref{formal f}), and the quantization is defined by
		\begin{equation}
			\widehat{p^i_mp^j_l}=\frac{\partial^2}{\partial q^i_m\partial q^j_l},\quad
			\widehat{p^i_mq^j_l}=q^j_l\frac{\partial}{\partial q^i_m},\quad
			\widehat{q^i_mq^j_l}=q^i_mq^j_l.
		\end{equation}
		In particular, for a matrix $(r_l)^i{}_j$ with $l \geq 1$, assumed to be symmetric (resp. antisymmetric) when $l$ is odd (resp. even), the transformation $r_l z^l$ is an infinitesimal symplectic transform, and its quantization is given by
		\begin{equation}\label{r operator}
			\begin{aligned}
				-\widehat{r_lz^l}
				=
				\sum_{i,j=1}^{N}\left(
				\sum_{n=0}^{\infty}(r_l)^i{}_j q^j_n \frac{\partial}{\partial q^i_{n+l}}
				+\frac{1}{2}\sum_{n+m=l-1}(r_l)^i{}_j (-1)^{m+1}
				\frac{\partial^2}{\partial q^i_n\partial q^j_m}
				\right).
			\end{aligned}
		\end{equation}
		
		The Fock space of this quantization is the space of functions of the variables in the series $\mathbf q(z)$, i.e., the functions such as $G(q_0,q_1,\cdots)$. The quantized operator acts on such functions. Consider a matrix-valued series
		\begin{equation}
			R(z)=I+R_1z+R_2z^2+\cdots.
		\end{equation}
		If it satisfies the symplectic condition
		\begin{equation}
			R(-z)^*R(z)=I,
		\end{equation}
		then it is a symplectic transform with respect to $\Omega(-,-)$. Its Givental quantization is defined as
		\begin{equation}
			\widehat R:=\exp\bigl(\widehat{\log R}\bigr).
		\end{equation}
		We have the following proposition due to Givental.
		\begin{proposition}[Givental \cite{giventalGromovWittenInvariantsQuantization2001}, Proposition 7.3]\label{Givental proposition}
			The quantized symplectic operator $\widehat R$ acts on a function $G$ as
			\begin{equation}
				(\widehat R \cdot G)(\mathbf q)=[e^{V(\partial_{\mathbf q},\partial_{\mathbf q})/2}G](R^{-1}\mathbf q),
			\end{equation}
			where $R^{-1} \mathbf q$ denotes the product of the $z$-series $R(z)^{-1}$ and $\mathbf q(z)$, and the quadratic form $V = \sum (p_k, V_{k,l} p_l)$ is defined by
			\begin{equation}\label{V and R}
				\sum_{k,l}V_{k,l}(-z)^k(-w)^l:= \frac{R(z)^*R(w)-\mathrm{Id}}{z+w},
			\end{equation}
			as an operator-valued series.
		\end{proposition}
		Also, for a matrix-valued series
		\begin{equation}
			S(1/z)=I+S_1z^{-1}+S_2z^{-2}+\cdots,
		\end{equation}
		if it satisfies the symplectic condition
		\begin{equation}
			S(-1/z)^*S(1/z)=I,
		\end{equation}
		then the following proposition holds:
		\begin{proposition}[Givental \cite{giventalGromovWittenInvariantsQuantization2001}, Proposition 5.3]
			The quantized symplectic operator $\widehat S^{-1}$ acts on a function $G$ as
			\begin{equation}
				(\widehat S^{-1}G)(\mathbf q)=e^{W(\mathbf q,\mathbf q)/2}G([S(1/z)\mathbf q(z)]_{+}),
			\end{equation}
			where $[S(1/z)\mathbf q(z)]_+$ denotes the non-negative part of the $z$-series of the product of $S(1/z)$ and $\mathbf q(z)$, and the quadratic form $W=\sum (W_{kl}q_k,q_l)$ is defined by
			\begin{equation}
				\sum_{k,l\geq 0} W_{kl}w^{-k}z^{-l}:=\frac{S(w^{-1})^*S(z^{-1})-I}{w^{-1}+z^{-1}}.
			\end{equation}
		\end{proposition}
		For a proof, we refer the reader to Section 3.2 of \cite{claderGeometricQuantizationApplications2018}.
		
		\subsection{Frobenius manifolds and semisimple CohFTs}\label{subsection CohFT}
		In this subsection, we briefly review some results about Givental's higher genus formula \cite{giventalGromovWittenInvariantsQuantization2001} and Teleman's reconstruction theorem \cite{telemanStructure2DSemisimple2012} for semisimple CohFTs. The construction starts from a semisimple Frobenius manifold and its monodromy data. For the complete theory of Frobenius manifolds, we refer the reader to \cite{kontsevichGromovWittenClassesQuantum1994} and \cite{dubrovinGeometry2DTopological1996}. We only explain the notation used in this article.
		
		For a semisimple Frobenius manifold $M$ with a flat unit, denote by $\mathbf t=(t^1,t^2,\cdots,t^N)$ the flat coordinates and denote by $\partial/\partial t^1$ the flat unit of the Frobenius algebra $T_{\mathbf t}M$. Then $\{\partial_{t^\alpha}=\partial/\partial t^\alpha, \alpha=1,\cdots,N\}$ is the flat basis of $T_\mathbf tM$, and the matrix $\eta=(\eta_{\alpha\beta})$ is the metric matrix. We assume that the Frobenius manifold has an Euler vector field, denoted as
		\begin{equation}
			E=\sum_{\alpha=1}^{N}d_\alpha t^{\alpha}\partial_{t^\alpha}+\sum_{\alpha:d_\alpha=0} r_\alpha \partial_{t^{\alpha}}.
		\end{equation}
		At a semisimple point, denote by $\mathbf u=(u^1,u^2,\cdots,u^N)$ the canonical coordinates. $\{\partial_{u^i}=\partial/\partial u^i, i=1,\cdots,N\}$ is the canonical basis of $T_{\mathbf u}M$. The inner product of the canonical basis is
		\begin{equation}
			\delta_{ij}\psi_{i1}^2=(\partial_{u^i},\partial_{u^j}).
		\end{equation}
		Denote by $\{f_i=\psi_{i1}^{-1}\partial_{u^i}, i=1,\cdots,N\}$ the normalized canonical basis. Then the transition matrix is $\Psi=(\psi^\alpha{}_i)$ defined by
		\begin{equation}
			\psi^\alpha{}_i=\psi_{i1}^{-1}\frac{\partial t^\alpha}{\partial u^i},
		\end{equation}
		such that
		\begin{equation}
			f_i=\sum_{\alpha=1}^{N} \psi^\alpha{}_i \partial_{t^\alpha}.
		\end{equation}
		
		For any Frobenius manifold $M$, there is a deformed flat connection $\nabla(z)$. A special fundamental solution matrix-valued $z$-series $S(1/z)$ of this flat connection is called a $S$-calibration of $M$. This matrix-valued $z$-series $S(1/z)=\sum_{k=0}^{\infty} S_k z^{-k}$ satisfies the symplectic condition
		\begin{equation}
			S^*(-1/z)S(1/z)=I,
		\end{equation}
		where
		\begin{equation}
			(S_k^*)^\alpha{}_\beta=(S_k)_\beta{}^\alpha=\eta^{\alpha\gamma}(S_k)^{\epsilon}{}_{\gamma}\eta_{\epsilon\beta}.
		\end{equation}
		For a Frobenius manifold defined by a Gromov-Witten theory, there is a natural $S$-calibration,
		\begin{equation}
			S_0=I, \quad (S_k)_{\alpha\beta}=\frac{\partial^2 F_0}{\partial t^{\alpha}_0\partial t^{\beta}_{k-1}}\bigg|_{t^{\gamma}_p=0, p\geq 1} \text{for }k\geq 1,
		\end{equation}
		where $F_0(t^{\alpha}_p)$ is the descendant genus-zero free energy of this Gromov-Witten theory.
		
		When the Frobenius manifold is semisimple, the deformed flat connection has an asymptotic fundamental solution $\Psi R(z)e^{U/z}$ for $z\to 0$, where the matrix $U=\mathrm{diag}(u^1,u^2,\cdots,u^N)$.
		Under the homogeneity condition, the \(R\)-matrix is uniquely determined by the Euler vector field $E$.
		
		In \cite{giventalGromovWittenInvariantsQuantization2001}, Givental gave a famous higher-genus formula for the descendant partition function of a semisimple Frobenius manifold, and Teleman proved the unique reconstruction from a Frobenius manifold to its CohFT in \cite{telemanStructure2DSemisimple2012}.
		\begin{thm}[Givental \cite{giventalGromovWittenInvariantsQuantization2001}, Teleman \cite{telemanStructure2DSemisimple2012}]
			For a homogeneous semisimple CohFT with flat unit, the Givental--Teleman reconstruction expresses its descendant partition function as
			\begin{equation}
				Z_{\mathrm{Des}}=e^{C(u)}\widehat S_{u}^{-1}\widehat\Psi\widehat R_{u}\widehat{e^{U/z}}Z_{\mathrm{WK}}^{\otimes N},
			\end{equation}
			where $u$ is the canonical coordinate of a point in the Frobenius manifold, the function $C(u)=\frac{1}{2}\int^u \sum_i (R_1)^i{}_i\d u^i$ of $u$ is defined up to an additive constant, $\widehat\Psi$ is the coordinate transformation.
		\end{thm}
		Both sides of the equation are independent of the point $u$. Moreover, $\widehat{e^{U/z}}$ is redundant and can be removed \cite{giventalGromovWittenInvariantsQuantization2001}, since it corresponds to the first Virasoro operator of the Witten--Kontsevich partition function.
		
		The relation between the ancestor partition function and the descendant partition function was given by Kontsevich and Manin \cite{kontsevichRelationsCorrelatorsTopological1998} and reformulated by Coates and Givental \cite{coatesQuantumRiemannRoch2007,giventalGromovWittenInvariantsQuantization2001}.
		\begin{thm}
			The relation between the ancestor partition function and the descendant partition function is
			\begin{equation}
				Z_{\mathrm{Des}}(\mathbf t)=e^{F_1(v)}\widehat S^{-1}_{v}Z_{\mathrm{FA}}(\mathbf t;v),
			\end{equation}
			where $v$ is a point of the Frobenius manifold and $F_1(v)$ is the genus-one free energy in the small phase space.
		\end{thm}
		The genus-one free energy in the small phase space $F_1(t)$ of a semisimple Frobenius manifold was conjectured by Givental \cite{giventalEllipticGromovWitten1998} and proved by Dubrovin and Zhang \cite{dubrovinBihamiltonianHierarchies2D1998},
		\begin{equation}
			F_1(u)=C(u)-\frac{1}{24}\sum_{i=1}^{N}\log\psi_{i1}.
		\end{equation}
		Comparing the two theorems, one obtains
		\begin{equation}\label{ancestor partition formula}
			Z_{\mathrm{FA}}=(\prod_{i=1}^N\psi_{i1})^{\frac{1}{24}}\widehat\Psi\widehat RZ^{\otimes N}_{\mathrm{WK}}.
		\end{equation}
		
		\begin{remark}
			All functions acted on by the operators are regarded as functions of dilaton shifted variables $\mathbf{q}$ instead of original variables $\mathbf{t}$.
		\end{remark}
		
		\subsection{Topological recursion and ancestor partition function}
		
		We now return to the setting of topological recursion. Suppose either
		that the spectral curve is compact and its Bergman kernel is
		$\mathcal A$-normalized, or that the local coefficients of $B$
		satisfy the differential compatibility condition
		\eqref{extra condition on B}. Then the matrix-valued series $R(z)$
		defined by \eqref{definition of R} satisfies
		\[
		R(-z)^*R(z)=\mathrm{Id}.
		\]
		
		At this stage, $R(z)$ is a symplectic matrix-valued series associated with the even local expansion data of $B$. To compare topological recursion with Givental reconstruction, we fix a semisimple point $v$ of a Frobenius manifold $M$ and assume that, in the normalized canonical basis, this series agrees with the Givental $R$-matrix at $v$. 
		Let \(\Psi\) and \((\psi_{i1})_{i=1}^N\) denote the corresponding transition and normalization data.
		We choose the distinguished function $y_B$ determined by $R$ and $(\psi_{i1})_{i=1}^N$ through \eqref{specific choice}.
		
		Conversely, starting from the monodromy data \((\Psi,R)\) at a semisimple point, the product formula \eqref{double product formula} determines the normalized even coefficients \(\widehat B^{i,j}_{k,l}\). Equation \eqref{specific choice} determines the odd local expansion of the distinguished function \(y_B\) relevant to topological recursion.
		We call any local spectral curve with these coefficients a local spectral curve associated with \((\Psi,R)\). 
		Its remaining local coefficients are not determined by \((\Psi,R)\): they do not enter the Givental gluing operator, although they remain part of the spectral-curve data and may occur in the basis differentials \(\d\xi_m^i\).
		
		Under this identification, $V=T_vM$, and the basis
		$(f_i)_{i=1}^N$ is the normalized canonical basis of the Frobenius
		algebra at $v$.
		
		The dilaton shift in the canonical ancestor variables is
		\begin{equation}
			q_{\mathrm{CA},m}^i=t_{\mathrm{CA},m}^i-\delta_{1,m}\psi_{i1}.
		\end{equation}
	
		Using (\ref{action of R on shift}), we obtain
		\begin{equation}
			\begin{aligned}
				R(z)^{-1}\cdot \mathbf q_{\mathrm{CA}}(z)=\mathbf t_{\mathrm{TR}}(z)-\mathbf y_{B}(z).
			\end{aligned}
		\end{equation}
		The transformed topological-recursion partition function can be written as
		\begin{equation}
			\begin{aligned}
				Z_{\mathrm{CA}}(\mathbf t_{\mathrm{CA}})
				=
				Z_{\mathrm{loc}}(\mathbf t_{\mathrm{TR}})\big|_{\mathbf t_{\mathrm{TR}}=\mathbf t_{\mathrm{TR}}(\mathbf t_{\mathrm{CA}})}
				=
				[e^{D(\widehat B)}G](R^{-1}\mathbf q_{\mathrm{CA}}),
			\end{aligned}
		\end{equation}
		where, by (\ref{relation coupled with kappa and dilated with y 2}),
		\begin{equation}
			\begin{aligned}
				G(\mathbf q)
				=&G(q^1_0,q^2_0,\cdots,q^N_0,q^1_1,\cdots)\\
				:=&\prod_{i=1}^{N}(y^i_1)^{\frac{1}{24}}
				Z_{\mathrm{WK}}(q^i_0,q^i_1+1,q^i_2,q^i_3,\cdots)\\
				=&\prod_{i=1}^{N}
				Z_{\mathrm{WK}}\left(
				(y^i_1)^{-\frac{1}{3}}q^i_0,
				(y^i_1)^{-\frac{3}{3}}(q^i_1+y^i_1),
				(y^i_1)^{-\frac{5}{3}}q^i_2,\cdots
				\right).
			\end{aligned}
		\end{equation}
		By the relation (\ref{double product formula}), the operator $D(\widehat B)$ coincides with the quadratic operator $V(\partial_{\mathbf q}, \partial_{\mathbf q})$ associated with $R(z)$. Thus, applying Theorem~\ref{main thm: relation} and Proposition~\ref{Givental proposition}, we obtain the following theorem.

		\begin{thm}[DOSS \cite{dunin-barkowskiIdentificationGiventalFormula2014}]\label{Theorem Givental Quantization}
		Suppose either that \(B\) is \(\mathcal A\)-normalized on a compact spectral curve, or that its local coefficients satisfy \eqref{extra condition on B}. 
		Assume moreover that the series \(R(z)\) defined by \eqref{definition of R} agrees, in the normalized canonical basis, with the Givental \(R\)-matrix at a fixed semisimple point, and let \(y=y_B\) be the distinguished choice \eqref{specific choice}.
		
		Then, after the base change \eqref{transform of t}, the transformed partition function \(Z_{\mathrm{CA}}\) coincides with the canonical ancestor partition function and takes the form
			\begin{equation}
				\begin{aligned}
					Z_{\mathrm{CA}}(\mathbf t_{\mathrm{CA}})
					&=\exp\left(
					\sum_{2g-2+n>0}\frac{1}{n!}
					W_{g,n}\big|_{ (-1)^m\d \xi^i_{(m)}\to t^i_{\mathrm{CA},m}}
					\right)\\
					=&\widehat R \prod_{i=1}^{N}
					Z_{\mathrm{WK}}\left(
					(\psi_{i1})^{-\frac{1}{3}}t^i_{\mathrm{CA},0},
					(\psi_{i1})^{-\frac{3}{3}}t^i_{\mathrm{CA},1},
					(\psi_{i1})^{-\frac{5}{3}}t^i_{\mathrm{CA},2},\cdots
					\right)\\
					=&\widehat{\exp\left(\sum_{l=1}^{\infty} r_lz^l\right)}
					\cdot \prod_{i=1}^{N}
					Z_{\mathrm{WK}}\left(
					(\psi_{i1})^{-\frac{1}{3}}t^i_{\mathrm{CA},0},
					(\psi_{i1})^{-\frac{3}{3}}t^i_{\mathrm{CA},1},
					(\psi_{i1})^{-\frac{5}{3}}t^i_{\mathrm{CA},2},\cdots
					\right),
				\end{aligned}
			\end{equation}
			where $R(z)=\sum_{m=0}^{\infty} R_mz^m=\exp(\sum_{l=1}^{\infty} r_l z^l)$ and $(R_m)^i{}_j=(-1)^{m-1}\widehat B^{i,j}_{0,m-1}$, and
			\begin{equation}
				\widehat{r_lz^l}
				=
				\sum_{i,j=1}^{N} (r_l)^i{}_j
				\left(
				\psi_{j1}\frac{\partial}{\partial t^i_{\mathrm{CA},l+1}}
				-\sum_{n=0}^{\infty} t^j_{\mathrm{CA},n}\frac{\partial}{\partial t^i_{\mathrm{CA},n+l}}
				-\frac{1}{2}\sum_{n+m=l-1}(-1)^{m+1}
				\frac{\partial^2}{\partial t^i_{\mathrm{CA},n}\partial t^j_{\mathrm{CA},m}}
				\right).
			\end{equation}
		\end{thm}
		We note that $(-1)^{l+1}(r_l)^i{}_j=(r_l)^j{}_i$ since $r_l$ is symmetric (resp. antisymmetric) when $l$ is odd (resp. even).
		
		\begin{remark}
		In DOSS \cite{dunin-barkowskiIdentificationGiventalFormula2014}, a dilaton operator $\widehat\Delta$ is also introduced. 
		This is equivalent to our dilaton transform $t^i_{\mathrm{CA},m} \mapsto (\psi_{i1})^{-\frac{2m+1}{3}} t^i_{\mathrm{CA},m}$.
		\end{remark}
		
		Apply the \(\Psi\)-change of basis to the dilaton-shifted variables:
		\begin{equation}\label{ancestor coord transfor}
			q^\alpha_{\mathrm{FA},m}=\psi^\alpha{}_i q^i_{\mathrm{CA},m}.
		\end{equation}
		Under this change, the transformed partition function becomes
		\begin{equation}
			Z_{\mathrm{FA}}(\mathbf t_{\mathrm{FA}})
			=
			\widehat\Psi\widehat R(\prod_{i=1}^N \psi_{i1})^{\frac{1}{24}}Z_{\mathrm{WK}}^{\otimes N},
		\end{equation}
		where the dilaton shift for flat ancestor variables is
		\begin{equation}
			q_{\mathrm{FA},m}^{\alpha}=t_{\mathrm{FA},m}^{\alpha}-\delta_{m,1}\delta^{\alpha}_{1}.
		\end{equation}
		This is precisely the flat ancestor partition function of the semisimple CohFT (\ref{ancestor partition formula}). Thus, one obtains the proposition:
		\begin{proposition}
			For a semisimple Frobenius manifold, the canonical ancestor partition function of the local spectral curve associated with its \(R\)-matrix, with \(y=y_B\), coincides with the ancestor partition function of this semisimple CohFT, i.e.
			\begin{equation}
				Z_{\mathrm{CA}}(t^i_{\mathrm{CA},m})
				:=
				\exp\left(
				\sum_{2g-2+n>0}\frac{1}{n!}
				W_{g,n}\big|_{(-1)^m\d \xi^i_{(m)}\to t^i_{\mathrm{CA},m}}
				\right)
				=
				Z_{\mathrm{FA}}(t^\alpha_{\mathrm{FA},m}),
			\end{equation}
			where the two coordinate systems $t^i_{\mathrm{CA},m}$ and $t^\alpha_{\mathrm{FA},m}$ are related to each other by $\Psi$ matrix.
		\end{proposition}
		
		\begin{remark}
			The flat ancestor partition function also depends on a semisimple point \(v\) in the Frobenius manifold. The data \(R\) and \(\Psi\), and hence the associated	coefficients \(\widehat B^{i,j}_{k,l}\) and the distinguished function \(y_B\), depend on \(v\). We suppress this parameter from the notation for convenience.
		\end{remark}
		
		\begin{corollary}\label{origin point EO}
			For the local spectral curve in the preceding proposition, the Eynard--Orantin multidifferentials have the expression
			\begin{equation}
				W_{g,n}
				=
				\sum_{(i_k,m_k)}
				\frac{\partial^n F^{\mathrm{CA}}_g }
				{\partial t^{i_1}_{\mathrm{CA},m_1}\cdots\partial t^{i_n}_{\mathrm{CA},m_n}}
				\bigg|_{\mathbf t_{\mathrm{CA}}=0}
				\prod_{k=1}^{n} (-1)^{m_k}\d \xi^{i_k}_{(m_k)},
			\end{equation}
			where $F^{\mathrm{CA}}_g$ is the canonical ancestor genus-$g$ free energy of this semisimple CohFT, and we write it in the coordinates $t^i_{\mathrm{CA},m}$.
		\end{corollary}
		
		\subsection{Eynard-Orantin topological recursion for a general \(y\)-function}
		In the preceding subsection, we considered a local spectral curve associated with the \((\Psi,R)\)-matrix of a semisimple Frobenius manifold, with \(y=y_B\). This choice corresponds to the evaluation point \(\mathbf t_{\mathrm{CA}}=0\).
		We now keep its Bergman kernel fixed and allow \(y\) to be a general admissible function. 
		
		Denote
		\begin{equation}\label{q'}
			\begin{aligned}
				\mathbf q'(z)
				&=R(z)\cdot (\mathbf t_{\mathrm{TR}}(z)-\mathbf y(z))\\
				&=\mathbf t_{\mathrm{CA}}(z)-R(z)\mathbf y(z)\\
				&=\mathbf q_{\mathrm{CA}}(z)+R(z)\cdot(\mathbf y_B(z)-\mathbf y(z)).
			\end{aligned}
		\end{equation}
		Set $\delta \mathbf y=\mathbf y_B-\mathbf y$.
		Then the transformed topological-recursion partition function associated with the general \(y\)-data is
		\begin{equation}
			Z^{(y)}_{\mathrm{CA}}(\mathbf t_{\mathrm{CA}})
			=
			\widehat R
			\left(\prod_{i=1}^N y^i_1\right)^{\frac{1}{24}}
			Z_{\mathrm{WK}}^{\otimes N}(\mathbf q').
		\end{equation}
		That is,
		\begin{equation}
			Z^{(y)}_{\mathrm{CA}}(\mathbf t_{\mathrm{CA}})
			=
			\left(\prod_{i=1}^N \frac{y^i_1}{\psi_{i1}}\right)^{\frac{1}{24}}
			Z_{\mathrm{FA}}(\mathbf t_{\mathrm{FA}}+\Psi R\cdot\delta\mathbf y).
		\end{equation}
		Thus one has a modification to Corollary \ref{origin point EO}:
		\begin{proposition}\label{generic background EO}
			For such a local spectral curve with a general admissible \(y\)-function, the Eynard--Orantin multidifferentials have the expression
			\begin{equation}
				W_{g,n}
				=
				\sum_{(i_k,m_k)}
				\frac{\partial^n F^{\mathrm{CA}}_g }
				{\partial t^{i_1}_{\mathrm{CA},m_1}\cdots\partial t^{i_n}_{\mathrm{CA},m_n}}
				\bigg|_{\mathbf t_{\mathrm{CA}}=R\cdot\delta \mathbf y}
				\prod_{k=1}^{n} (-1)^{m_k}\d \xi^{i_k}_{(m_k)},
			\end{equation}
			where \(F^{\mathrm{CA}}_g\) is the canonical ancestor genus-\(g\) free energy of this semisimple CohFT and \(\delta\mathbf y=\mathbf y_B-\mathbf y\). Here \(\mathbf y_B\) is the distinguished function associated with \(R\) and \(\psi_{i1}\) through \eqref{specific choice}.
		\end{proposition}
		The prefactor $(\prod_{i=1}^N y^i_1/\psi_{i1})^{1/24}$ is independent of \(\mathbf t_{\mathrm{CA}}\) and represents a genus-one normalization; hence it does not affect the multidifferentials \(W_{g,n}\) for \(n\ge1\).
		
		\begin{remark}
			The shift \(R\cdot\delta\mathbf y\) has no component in the primary directions \(t^i_{\mathrm{CA},0}\); it only shifts the non-primary canonical ancestor variables \(t^i_{\mathrm{CA},m}\) for \(m\ge1\). The displacement of the primary background is expected to be governed by deformations of the \(x\)-function. We leave this issue for future investigation.
		\end{remark}

		\appendix
		
		\section{Topological recursion for Gromov-Witten invariants of \texorpdfstring{$\mathbb P^1$}{P1}}\label{appendix A}
		In this appendix, we discuss the Norbury-Scott spectral curve for $\mathbb{P}^1$. It was conjectured by Norbury and Scott in \cite{norburyGromovWittenInvariants2014} and proved by DOSS in \cite{dunin-barkowskiIdentificationGiventalFormula2014} that the Norbury--Scott spectral curve generates the Gromov--Witten invariants for $\mathbb{P}^1$. In \cite{borotLoopEquationsGromovWitten2019}, Borot and Norbury provided a new method to extract the complete information about Gromov-Witten invariants for $\mathbb{P}^1$ including non-stationary sectors.
		Let
		\[
		\gamma_0=1\in H^0(\mathbb P^1),\qquad
		\gamma_1=\omega\in H^2(\mathbb P^1),\qquad
		\int_{\mathbb P^1}\omega=1 .
		\]
		The descendant Gromov-Witten invariants of $\mathbb P^1$ are
		\begin{equation}
				\left\langle \prod_{j=1}^n \tau_{b_j}(\gamma_{\alpha_j})\right\rangle^{\mathbb P^1}_{g,d}
			:=
			\int_{[\overline{\mathcal M}_{g,n}(\mathbb P^1,d)]^{\mathrm{vir}}}
			\prod_{j=1}^n \psi_j^{b_j}\mathrm{ev}_j^*(\gamma_{\alpha_j}),
			\qquad \alpha_j\in\{0,1\}.
		\end{equation}
		The degree $d$ is determined by the dimension constraint
		\begin{equation}
				\sum_{j=1}^n (b_j+\alpha_j)=2g-2+2d+n;
		\end{equation}
		we omit it from the notation. Insertions of $\gamma_1=\omega$ are called stationary insertions, while insertions of $\gamma_0=1$ are non-stationary insertions.
		
		The Norbury-Scott spectral curve \cite{norburyGromovWittenInvariants2014} for $\mathbb P^1$ is
		\begin{equation}
		\mathcal C_{\mathbb P^1}
		=
		\left(
		\mathbb P^1_z,\quad
		x=z+\frac1z,\quad
		 y=\log z,\quad
		B(z_1,z_2)=\frac{\d z_1\d z_2}{(z_1-z_2)^2}
		\right).
		\end{equation}
		
		The Norbury--Scott spectral curve corresponds to the
		origin $(t^1,t^2)=(0,0)$ of the Frobenius manifold associated with
		the quantum cohomology of $\mathbb P^1$. More generally, put
		\[
		\lambda=e^{t^2/2}.
		\]
		The spectral curve corresponding to the general point 
		$\mathbf t=(t^1,t^2)$ is given by (Appendix A in \cite{dunin-barkowskiDubrovinsSuperpotentialGlobal2019})
		\begin{equation}
			\mathcal C_{\mathbb P^1,\mathbf t}
			=
			\left(
			\mathbb P^1_z,\quad
			x_{\mathbf t}=t^1+\lambda\left(z+\frac1z\right),\quad
			y_{\mathbf t}=\log z,\quad
			B(z_1,z_2)=\frac{\d z_1\d z_2}{(z_1-z_2)^2}
			\right).
		\end{equation}
		The ramification points remain $c_+=1$ and $c_-=-1$, and their
		critical values are
		\begin{equation}
			u^+=t^1+2\lambda,\qquad
			u^-=t^1-2\lambda.
		\end{equation}
		Indeed, the pair $(u^+,u^-)$ forms a system of canonical coordinates
		on the Frobenius manifold.
		
		The local involution is again induced by $z\mapsto 1/z$. The local
		Airy coordinates at the two ramification points are
		\begin{equation}
			x_{\mathbf t}=u^++\frac12\zeta_{+,\mathbf t}^2,
			\qquad
			\zeta_{+,\mathbf t}
			=
			\lambda^{1/2}\sqrt2
			\left(z^{1/2}-z^{-1/2}\right),
		\end{equation}
		and
		\begin{equation}
			x_{\mathbf t}=u^-+\frac12\zeta_{-,\mathbf t}^2,
			\qquad
			\zeta_{-,\mathbf t}
			=
			\lambda^{1/2}\sqrt2
			\left(z^{1/2}+z^{-1/2}\right).
		\end{equation}
		
		Let $y^{\pm}_{2k+1}(\mathbf t)$ and
		$B^{i,j}_{k,l}(\mathbf t)$ denote the local expansion coefficients
		with respect to $\zeta_{\pm,\mathbf t}$. Since
		$y_{\mathbf t}=\log z$ is independent of $\mathbf t$ as a function
		of the global coordinate $z$, the rescaling of the local Airy
		coordinates gives
		\begin{equation}
			y^{\pm}_{2k+1}(\mathbf t)
			=
			\lambda^{-(2k+1)/2}y^{\pm}_{2k+1}(0).
		\end{equation}
		Similarly, the regular coefficients of the standard Bergman kernel
		satisfy
		\begin{equation}
			B^{i,j}_{k,l}(\mathbf t)
			=
			\lambda^{-(k+l+2)/2}B^{i,j}_{k,l}(0),
			\qquad i,j\in\{+,-\}.
		\end{equation}
		Consequently, the normalized Bergman coefficients obey
		\begin{equation}
			\widehat B^{i,j}_{k,l}(\mathbf t)
			=
			\lambda^{-k-l-1}\widehat B^{i,j}_{k,l}(0).
		\end{equation}
		
		Specializing to $(t^1,t^2)=(0,0)$ gives $\lambda=1$ and recovers
		the Norbury--Scott spectral curve and all its local data. We now
		record these special values explicitly.
		
		With the local Airy coordinates
		\begin{equation}
				x=2+\frac12\zeta_+^2,\qquad
			\zeta_+=\sqrt2\left(z^{1/2}-z^{-1/2}\right),
		\end{equation}
		and
		\begin{equation}
				x=-2+\frac12\zeta_-^2,\qquad
			\zeta_-=\sqrt2\left(z^{1/2}+z^{-1/2}\right),
		\end{equation}
		one obtains
		\begin{equation}
			y=\sum_{k=0}^{\infty} y_{2k+1 }^{\pm }\zeta_{\pm }^{2k+1}+\mathrm{const.},
		\end{equation}
		with 
		\begin{equation}
			y_{2k+1}^{\pm }=2(-1)^k\frac{\binom{2k}{k}}{4^k(2k+1)}(2\sqrt{2(\pm 1)})^{-2k-1}.
		\end{equation}
		The expansion of the Bergman kernel in the local Airy coordinate is
		\begin{equation}
			\begin{aligned}
					B(\zeta_{+},\zeta'_{+})&=\frac{\d \zeta_{+}\d \zeta'_{+}}{(\zeta_+-\zeta'_{+})^2}+\sum_{a,b=0}^{\infty} B^{+,+}_{2a,2b} \zeta_{+}^{2a}\zeta_{+}^{\prime 2b}\d \zeta_{+}\d \zeta'_{+}\\
					B(\zeta_{-},\zeta'_{-})&=\frac{\d \zeta_{-}\d \zeta'_{-}}{(\zeta_--\zeta'_{-})^2}+\sum_{a,b=0}^{\infty} B^{-,-}_{2a,2b} \zeta_{-}^{2a}\zeta_{-}^{\prime 2b}\d \zeta_{-}\d \zeta'_{-},
			\end{aligned}
		\end{equation}
		and
		\begin{equation}
			\begin{aligned}
				B(\zeta_{+},\zeta_{-})=&\sum_{a,b=0}^{\infty} B^{+,-}_{2a,2b} \zeta_{+}^{2a}\zeta_{-}^{2b}\d \zeta_{+}\d \zeta_{-}\\
				&+\sum_{a,b=0}^{\infty} B^{+,-}_{2a+1,2b+1}\zeta_{+}^{2a+1}\zeta_{-}^{2b+1}\d\zeta_{+}\d\zeta_{-},
			\end{aligned}
		\end{equation}
		where
		\begin{equation}
			\begin{aligned}
				&B^{+,+}_{2a,2b}=\frac{(-1)^{a+b+1}}{2^{4a+4b+4}}\frac{(2a+1)!!(2b+1)!!}{(a+b+1)a!b!}\\
				&B^{-,-}_{2a,2b}=\frac{1}{2^{4a+4b+4}}\frac{(2a+1)!!(2b+1)!!}{(a+b+1)a!b!}\\
				&B^{+,-}_{2a,2b}=\frac{(-1)^{a+\frac{1}{2}}}{2^{3a+3b+3}}\sum_{l=0}^{\min\{a,b\}}(2l+1)\binom{a+\frac{1}{2}}{a-l}\binom{b+\frac{1}{2}}{b-l}\\
				&B^{+,-}_{2a+1,2b+1}=\frac{(-1)^a}{4}\frac{(a+b+1)!!}{2^{3a+3b+3}a!b!}.
			\end{aligned}
		\end{equation}
		One can check that the Bergman kernel satisfies the condition \eqref{extra condition on B}, and the $y$-function is compatible with the Bergman kernel and $\psi_{\pm 1}$ defined below.
		
		We now describe the Frobenius-manifold data associated with the
		spectral curve $\mathcal C_{\mathbb P^1,\mathbf t}$ at a general
		point $\mathbf t=(t^1,t^2)$.
		Its Frobenius potential is
		\begin{equation}
			F(t^1,t^2)
			=
			\frac12(t^1)^2t^2+e^{t^2}.
		\end{equation}
		Its flat metric is
		\begin{equation}
		\eta=
		\begin{pmatrix}
			0&1\\
			1&0
		\end{pmatrix},
		\end{equation}
		and the canonical coordinates are
		\begin{equation}
			u^+=t^1+2\lambda,\qquad
			u^-=t^1-2\lambda.
		\end{equation}
		Thus
		\begin{equation}
			U_{\mathbf t}
			=
			\begin{pmatrix}
				t^1+2\lambda&0\\
				0&t^1-2\lambda
			\end{pmatrix}.
		\end{equation}
		
		Let $w$ denote the variable of the deformed flat connection and put
		\[
		L=\log w+\gamma,
		\]
		where $\gamma$ is Euler's constant. The deformed flat coordinates of the Dubrovin connection give the following $S$-matrix at the point $\mathbf t$ (Example 3.6.10 in \cite{dubrovinNormalFormsHierarchies2001}):
		\begin{equation}
			S_{\mathbf t}(w)
			=
			e^{wt^1}
			\begin{pmatrix}
				2w\lambda
				\bigl(
				K_1(2\lambda w)-L I_1(2\lambda w)
				\bigr)
				&
				\lambda I_1(2\lambda w)
				\\[4pt]
				-2w
				\bigl(
				K_0(2\lambda w)+L I_0(2\lambda w)
				\bigr)
				&
				I_0(2\lambda w)
			\end{pmatrix},
		\end{equation}
		where $I_\nu$ and $K_\nu$ are the modified Bessel functions.
		Here \(S_{\mathbf t}(w)\) corresponds to \(S_{\mathbf t}(1/z)\) in the notation of Subsection \ref{subsection Givental quantization}, with \(w=z^{-1}\).
		
		Although logarithms occur in this expression, their singular parts cancel, and $S_{\mathbf t}(w)$ is a regular matrix-valued power
		series at $w=0$ satisfying
		\[
		S_{\mathbf t}(0)=I.
		\]
		
		The grading and resonance matrices are independent of
		$\mathbf t$ and are given by
		\begin{equation}
		\mu=
		\begin{pmatrix}
			-\frac12&0\\
			0&\frac12
		\end{pmatrix},
		\qquad
		\rho=
		\begin{pmatrix}
			0&0\\
			2&0
		\end{pmatrix}.
		\end{equation}
		Consequently,
		\begin{equation}
		w^\mu=
		\begin{pmatrix}
			w^{-1/2}&0\\
			0&w^{1/2}
		\end{pmatrix},
		\qquad
		w^\rho=
		\begin{pmatrix}
			1&0\\
			2\log w&1
		\end{pmatrix}.
		\end{equation}
		
		The canonical idempotents are
		\begin{equation}
			\partial_{u^+}
			=
			\frac12\partial_{t^1}
			+\frac1{2\lambda}\partial_{t^2},
			\qquad
			\partial_{u^-}
			=
			\frac12\partial_{t^1}
			-\frac1{2\lambda}\partial_{t^2},
		\end{equation}
		and their squared lengths are
		\begin{equation}
			(\partial_{u^+},\partial_{u^+})
			=
			\frac1{2\lambda},
			\qquad
			(\partial_{u^-},\partial_{u^-})
			=
			-\frac1{2\lambda}.
		\end{equation}
		We choose
		\[
		\psi_{+1}=\frac1{\sqrt{2\lambda}},
		\qquad
		\psi_{-1}=-\frac{i}{\sqrt{2\lambda}},
		\]
		so that the normalized canonical basis is
		\[
		f_+=\sqrt{2\lambda}\,\partial_{u^+},
		\qquad
		f_-=i\sqrt{2\lambda}\,\partial_{u^-}.
		\]
		The transition matrix from the normalized canonical basis to the
		flat basis is therefore
		\begin{equation}
			\Psi_{\mathbf t}
			=
			\frac1{\sqrt2}
			\begin{pmatrix}
				\lambda^{1/2}&i\lambda^{1/2}\\
				\lambda^{-1/2}&-i\lambda^{-1/2}
			\end{pmatrix}.
		\end{equation}
	
		The $R$-matrix at the general point is given by \cite{dunin-barkowskiIdentificationGiventalFormula2014}
		\begin{equation}
			R_{\mathbf t}(z)
			=
			\sum_{k=0}^{\infty}R_{\mathbf t,k}z^k,
		\end{equation}
		where
		\begin{equation}
			R_{\mathbf t,k}
			=
			\lambda^{-k}c_k
			\begin{pmatrix}
				-1&(-1)^{k+1}2ki\\
				2ki&(-1)^{k+1}
			\end{pmatrix},
			\qquad
			c_k=
			\frac{(2k-1)!!(2k-3)!!}{2^{4k}k!}.
		\end{equation}
		We use the conventions \((-1)!!=1,\ (-3)!!=-1.\)
		In particular, $c_0=-1$ and hence
		\[
		R_{\mathbf t,0}=I.
		\]
		Indeed, this agrees with the defining relation
		\[
		(R_{\mathbf t,k})^i{}_j
		=
		(-1)^{k-1}
		\widehat B^{i,j}_{0,k-1}(\mathbf t).
		\]
		
		The central connection matrix is independent of the point
		$\mathbf t$ and is given by
		\begin{equation}
			C=
			\begin{pmatrix}
				0&\dfrac{i}{\sqrt{2\pi}}\\[6pt]
				\sqrt{2\pi}&i\gamma\sqrt{\dfrac2\pi}
			\end{pmatrix}.
		\end{equation}
		Direct multiplication gives the identity
		\begin{equation}
			\begin{aligned}
				S_{\mathbf t}(w)w^\mu w^\rho C
				={}&
				\sqrt{2\pi w}\,e^{wt^1}\\
				&\times
				\begin{pmatrix}
					\lambda I_1(2\lambda w)
					&
					\dfrac{i\lambda}{\pi}K_1(2\lambda w)
					\\[6pt]
					I_0(2\lambda w)
					&
					-\dfrac{i}{\pi}K_0(2\lambda w)
				\end{pmatrix}.
			\end{aligned}
		\end{equation}
		Using the standard asymptotic expansions of the modified Bessel
		functions, one obtains, as $w\to\infty$ with $z=1/w$,
		\begin{equation}
			S_{\mathbf t}(w)w^\mu w^\rho C
			\sim
			\Psi_{\mathbf t}
			R_{\mathbf t}(z)
			e^{U_{\mathbf t}/z}.
		\end{equation}
		Thus the same central connection matrix $C$ gives the asymptotic
		factorization at every semisimple point of the Frobenius manifold.
		
		We now specialize these formulas to the origin
		$(t^1,t^2)=(0,0)$.
		The $S$-matrix becomes
		\begin{equation}
			S_0(w)
			=
			\begin{pmatrix}
				2w\bigl(K_1(2w)-L I_1(2w)\bigr)
				&
				I_1(2w)
				\\[4pt]
				-2w\bigl(K_0(2w)+L I_0(2w)\bigr)
				&
				I_0(2w)
			\end{pmatrix}.
		\end{equation}
		Expanding at $w=0$, one obtains
		\begin{equation}
			S_0(w)=\sum_{\ell=0}^{\infty}S_{0,\ell}w^\ell,
		\end{equation}
		with
		\begin{equation}
			S_{0,2r}
			=
			\frac1{(r!)^2}
			\begin{pmatrix}
				1-2rH_r&0\\
				0&1
			\end{pmatrix},
			\qquad
			S_{0,2r+1}
			=
			\frac1{(r!)^2}
			\begin{pmatrix}
				0&\dfrac1{r+1}\\
				-2H_r&0
			\end{pmatrix},
		\end{equation}
		where
		\[
		H_0=0,\qquad
		H_r=\sum_{j=1}^r\frac1j \quad \text{for }r\geq 1.
		\]
		This is precisely the $S$-matrix appearing in the
		descendant-ancestor transformation \cite{dunin-barkowskiIdentificationGiventalFormula2014}.
		
		Givental's formula for the descendant theory of
		$\mathbb P^1$ can be evaluated at any semisimple point
		$\mathbf t$:
		\begin{equation}
			Z^{\mathbb P^1}_{\mathrm{Des}}
			=
			e^{C(\mathbf t)}
			\widehat S_{\mathbf t}^{-1}
			\widehat\Psi_{\mathbf t}
			\widehat R_{\mathbf t}
			Z_{\mathrm{WK}}^{\otimes 2}.
		\end{equation}
		The right-hand side is independent of the chosen semisimple
		point. 
		
		The primary differentials of the basis for the Norbury-Scott spectral curve are
		\begin{equation}
				\d\xi^+_0=\frac{1}{\sqrt2}\frac{\d z}{(z-1)^2},
			\qquad
			\d\xi^-_0=\frac{i}{\sqrt2}\frac{\d z}{(z+1)^2},
		\end{equation}
		up to the sign determined by the choice of the branch of $\zeta_-$. Borot and Norbury \cite{borotLoopEquationsGromovWitten2019} use the flat basis of differentials
		\begin{equation}
				\Xi^\alpha_{-1}=z^{-\alpha}\d z,\qquad
			\Xi^\alpha_m=-\d\left(\frac{\Xi^\alpha_{m-1}}{\d x}\right),
			\qquad \alpha=0,1,\quad m\geq 0.		
		\end{equation}
		In particular,
		\begin{equation}
				\Xi^0_0=\frac{2z\,\d z}{(z^2-1)^2},
			\qquad
			\Xi^1_0=\frac{(z^2+1)\d z}{(z^2-1)^2}.
		\end{equation}
		This basis is exactly the flat ancestor basis obtained from our canonical ancestor basis at the origin by the $\Psi_0$-transformation:
		\begin{equation}
			\begin{aligned}
					\Xi^0_m
				&=
				\frac{1}{\sqrt2}
				\left((-1)^m\d\xi^+_{(m)}+i(-1)^m\d\xi^-_{(m)}\right),\\
				\Xi^1_m
				&=
				\frac{1}{\sqrt2}
				\left((-1)^m\d\xi^+_{(m)}-i(-1)^m\d\xi^-_{(m)}\right).
			\end{aligned}
		\end{equation}
		Our result, which reproves the result of Dunin-Barkowski, Orantin, Shadrin and Spitz \cite{dunin-barkowskiIdentificationGiventalFormula2014}, is that for the Norbury--Scott spectral curve, the topological recursion partition function in the basis $\Xi^\alpha_m$ coincides with the ancestor partition function of the Gromov-Witten theory of $\mathbb P^1$. Equivalently,
		\begin{equation}
			\exp\left(
			\sum_{2g-2+n>0}
			\frac1{n!}
			W^{\mathbb P^1}_{g,n}
			\big|_{\Xi^\alpha_m\to t_{\mathrm{FA},m}^\alpha}
			\right)
			=
			Z^{\mathbb P^1}_{\mathrm{FA}}(\mathbf{t}_{\mathrm{FA}})
		\end{equation}
		As proved by		Borot and Norbury \cite{borotLoopEquationsGromovWitten2019}, 
  		for $2g-2+n>0$, the general descendant Gromov-Witten invariants of $\mathbb P^1$ are recovered from the Norbury-Scott topological recursion with suitable corrections to the initial values by the formula
		\begin{equation}
				\left\langle
			\prod_{j=1}^n \tau_{b_j}(\gamma_{\alpha_j})
			\right\rangle_g^{\mathbb P^1}
			=
			\left(
			\bigotimes_{j=1}^n I_{b_j}^{\alpha_j}
			\right)
			W^{\mathbb P^1}_{g,n},
		\end{equation}
	where $I_b^\alpha$ ($\alpha= 0,1 $) are the explicitly defined linear functionals on meromorphic one-forms introduced in \cite{borotLoopEquationsGromovWitten2019}.


\end{document}